\documentclass[aps,prd,twocolumn,superscriptaddress,nofootinbib,floatfix]{revtex4-2}
\usepackage{amsmath,amssymb,graphicx,hyperref,xcolor}
\hypersetup{
  pdftitle={The window that survives: detectability of gravitational-wave echoes after horizon backreaction},
  colorlinks=true, linkcolor=blue!60!black, citecolor=blue!60!black, urlcolor=blue!60!black}
\graphicspath{{figures/}}
\newcommand{\Msun}{M_\odot}
\newcommand{\Rwall}{\mathcal{R}_{\rm wall}}
\newcommand{\RBH}{\mathcal{R}_{\rm BH}}
\newcommand{\Tr}{\mathcal{T}}
\newcommand{\code}[1]{\textsf{#1}}

\begin{document}

\title{The window that survives: detectability of gravitational-wave echoes after horizon backreaction}

\author{Jahed Abedi}
\email{jahed.abedi@bimsa.cn}
\affiliation{Beijing Institute of Mathematical Sciences and Applications (BIMSA),
Huairou District, Beijing 101408, P. R. China}

\begin{abstract}
Guo and Mathur argued that radiation falling onto a reflecting surface just outside a horizon forms a trapped surface before its reflection returns, so no echoes escape. We ask what is seen if the backreacted geometry passes a narrow band at the superradiant threshold, $|\omega-m\Omega_H|\lesssim k_c$. We set $k_c=\pi/L_*$, with $L_*$ the tortoise distance from a Planck-distance surface with Boltzmann reflectivity to the equal-area sphere of the backreacted hole. The transmission vanishes at threshold, so the surviving echo has a central dip. A re-formed surface, assumed here, turns it into a few cavity modes. In Cosmic Explorer (CE) a face-on GW150914-like echo at 410~Mpc has SNR $1.8$ and a GW231123-like one at $2.2$~Gpc has $9.4$ ($0.40$ in LIGO at O4). The SNR rises steeply with remnant spin up to $\chi_f\simeq0.84$, that of GW231123, as the ringdown nears $m\Omega_H$ and the transmission steepens. As a detected merger triggers the search, we count signals with SNR $>5$, with source-by-source spins (ranges: common Boltzmann reflectivity phase). In 10~yr CE would see $0.075$--$0.15$ GWTC-5.0 stellar-mass echoes, mostly binaries above $100\,M_\odot$, where only GW231123 lies. The model gives them lower spins; with its spin the count is $2.6$ ($0.035$--$7.2$) and reaches one at SNR $6.5$ ($1.9$--$8.7$). In 4~yr DECIGO would see $0.49$--$14$, $67$--$150$ or $4.6$--$19$ massive-black-hole echoes in three 2016 catalogues and $540$--$900$ in a pulsar-timing-calibrated 2023 one ($0.30$--$9.4$, $43$--$100$, $2.7$--$14$ and $300$--$570$ with SNR $>8$; most under the white-dwarf foreground); LISA at most $2.8$, Taiji $9.4$. Where under one is expected, the loudest has SNR $0.20$--$0.25$ (LIGO A+), $0.79$--$0.97$ (Einstein Telescope), $2.3$--$2.9$ (CE) and at most $3.5$ (TianQin). All numbers are optimistic, conditional on the escape hypothesis, packet energy, background choice and source spins.
\end{abstract}

\maketitle

\section{Introduction}\label{sec:intro}
If a merger remnant had a reflecting surface a Planck length outside its horizon, part of the post-merger radiation falling into the hole would bounce back and leak out as a train of echoes, delayed by $\Delta t_{\rm echo}\simeq(4GM_f/c^3)(1+1/\sqrt{1-\chi^2})\ln(M_f/M_{\rm Pl})$ \cite{Cardoso2016,Cardoso2016b,Abedi2017}. To leading logarithmic order this delay is the scrambling time of the black hole, $\kappa^{-1}\ln(A/4l_p^2)$ \cite{HaydenPreskill2007,Saraswat2020,OshitaAfshordi2023}, and for our remnant the two differ by $1.3\%$. This is one reason why echoes have become a standard probe of quantum structure at horizons \cite{CardosoPani2019,AbediStatus2020,AbediUniverse2020,BertiReview2026}. They have been studied for a wide range of horizon-scale structures and ultracompact objects, among them the 2-2-holes of quadratic gravity \cite{HoldomRen2017,ConklinHoldomRen2018,Ren2019}, firewall-like barriers arising from an action principle \cite{AbediAction2023}, area-quantized horizons \cite{CardosoFoitKleban2019,Agullo2021,Deppe2025}, wormholes \cite{Lai2026}, and long throats near a marginal-trapping threshold \cite{AbediThroat2026}. What the echoes look like, and whether they can be detected, depends strongly on the reflectivity and on how the potential barrier filters the reflected wave \cite{Mark2017,Nakano2017,WangAfshordi2018,TestaPani2018,MaggioTesta2019,Xin2021,LongoMicchi2021,Rosato2025,Rosato2026,ZhangRen2026}. A perfect reflector sends back most of the ringdown energy \cite{Mark2017,Nakano2017,WangAfshordi2018,Xin2021}, but a spinning object with a perfectly reflecting surface is ergoregion-unstable \cite{Maggio2019,Barausse2018}. The Boltzmann reflectivity of a quantum horizon \cite{Oshita2020,WangOshitaAfshordi2020,AbediGW190521,Abedi2023}, by contrast, reflects only within ${\sim}T_H$ of the superradiant threshold $\omega=m\Omega_H$, which lies below the ringdown frequency, and it is free of the ergoregion instability up to the Thorne limit \cite{OshitaTsunaAfshordi2020}. It has been imposed as the surface boundary condition in echo waveforms computed from the Teukolsky equation \cite{Xin2021,SrivastavaChen2021,LongoMicchi2021}, and the same Boltzmann factor, set by $T_H$ alone, has been derived for the reflective horizon of a black mirror \cite{AmaroSeoane2025}. Boltzmann echoes have also been computed from numerical-relativity waveforms by reconstructing the near-horizon spacetime \cite{Ma2022}. The reflectivity has also been discussed for the evaporation of black holes with a reflecting surface, where only the modes with $\omega\sim m\Omega_H$ would contribute \cite{OshitaMotohashiNoda2022}. The Boltzmann echoes of Ref.~\cite{LongoMicchi2021} already show the feature that shapes our signal, a transfer function that vanishes at $\omega=m\Omega_H$ and a double-peaked echo (Sec.~\ref{sec:dip}).

The observational record is mixed. Tentative evidence for echoes was reported for the first binary black-hole events \cite{Abedi2017,AbediHoliday2017,ConklinHoldomRen2018,Holdom2020}, for the binary neutron-star merger GW170817 \cite{AbediAfshordi2019}, and for Boltzmann echoes after GW190521 \cite{AbediGW190521}. Other analyses, with different templates and statistics, found low significance or no evidence \cite{Westerweck2018,AbediComment2018,Nielsen2019,Lo2019,Uchikata2019,Salemi2019,Tsang2018,Tsang2020,LVK2021TGR,Uchikata2023,Miani2023,LVK2025TGR}, and the specific proposal that GW190521 is itself an isolated echo is disfavoured \cite{Lai2026}. Model-agnostic searches for the long-lived cavity modes \cite{RenWu2021,WuGaoRenAfshordi2023,WuZhangHuangRen2026} likewise see no significant signal. A combined search for Boltzmann echoes in 47 events bounds a universal echo amplitude to $A<0.4$ at 90\% confidence, where $A=1$ is the canonical prediction \cite{Abedi2023}. There is general agreement that current data contain no conclusive detection \cite{BertiReview2026}.

Guo and Mathur \cite{GuoMathur2022} raised a basic objection. The infalling energy forms a new trapped surface outside the reflecting one before the reflected wave can come back, and nothing can leave a trapped region. The argument rests on causality, not on a ray picture, and a long wavelength does not by itself allow a disturbance generated inside an event horizon to reach infinity. Partial answers exist. A toy model with a dilaton and a Maxwell field produces, close to extremality, a macroscopic reflecting barrier near the horizon and observable echoes, a counterexample to the trapping argument within that model \cite{AbediAction2023}. Numerical-relativity evolutions with a reflecting boundary close to a dynamical horizon have begun to model the situation directly \cite{Dailey2023,Dailey2025}. Backreaction has also been shown to make the echo train nonperiodic, because the absorbed energy changes the central object \cite{Vellucci2023}. None of these settles the question for a rotating merger remnant hit by its own ringdown.

We do not attempt to overturn this argument. We ask a narrower, conditional question: \emph{if} the backreacted geometry lets a narrow band of frequencies near the superradiant threshold through, what signal results, how does it depend on the reflectivity, and can it be detected? We encode the hypothesis in a phenomenological backreaction filter $S(k)$, $k=\omega-m\Omega_H$, whose width is set by the tortoise thickness of the layer between the reflecting surface and the backreacted horizon. We then test how the conclusions change with the shape, causality and width of $S$, with the background on which the backreaction filter is centred, and with the numerical resolution. A real test of the hypothesis needs a causal evolution on a dynamical background, which is beyond the scope of this paper (Sec.~\ref{sec:conclusions}).

\begin{figure*}
\includegraphics[width=\textwidth]{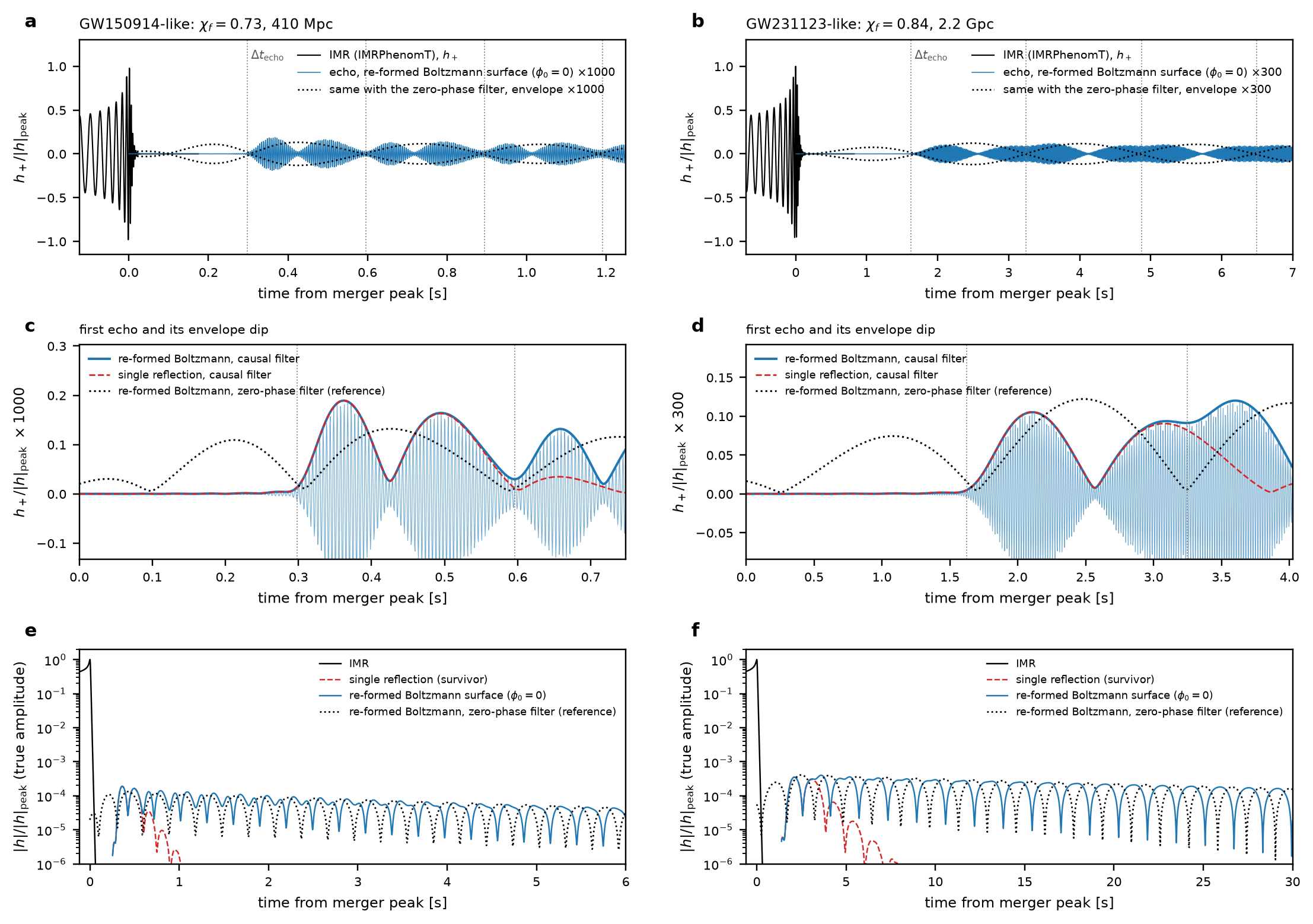}
\caption{IMR signal and the echo that survives the effective horizon backreaction for the two reference sources (IMRPhenomT, face-on, time from the strain peak, strain in units of the merger peak). Left: GW150914-like source ($M_f=67.21\,\Msun$, $\chi_f=0.7268$, 410~Mpc, $|h|_{\rm peak}=2.0\times10^{-21}$). Right: GW231123-like source ($M_f^{\rm src}=224\,\Msun$ in the source frame, $\chi_f=0.84$, $z=0.39$, $D_L=2.2$~Gpc, $|h|_{\rm peak}=1.8\times10^{-21}$). (a, b) $h_+$ of the IMR signal (black) and of the echo with a re-formed Boltzmann surface (default; Sec.~\ref{sec:reformed}; $N\to\infty$, $\phi_0=0$; blue), multiplied by $10^3$ and $300$; vertical dotted lines mark multiples of the geometric echo time $u_e=\Delta t_{\rm echo}$. The blue echo is drawn with the causal (minimum-phase) Butterworth filter of the same width as the adopted raised cosine, so that it does not precede the unfiltered reflection; its optimal SNR equals that of the zero-phase Butterworth filter of the same magnitude and differs from that of the adopted raised cosine (Appendix~\ref{app:robust}). The black dotted curves are the envelope $\pm|h|$ of the same echo for the adopted zero-phase filter, with the same scaling. It begins before $u_e$; Appendix~\ref{app:robust} discusses how far this is excluded on causal grounds. In all panels it is drawn only from the start of the ingoing packet, $4M_f$ before the strain peak, because the small part before that time cannot come from a causal response. (c, d) Zoom on the first echo: $h_+$ and envelope of the re-formed signal, envelope of the single-reflection survivor (dashed), and the envelope for the adopted zero-phase filter (dotted), whose central dip lies at $u_e$; with the causal filter the envelope minimum between the two main lobes of the first echo (single reflection) lies $384M_f$ ($617M_f$) after $u_e$ (Appendix~\ref{app:robust}). Before $2u_e$ the re-formed signal differs from the single-reflection survivor by at most $13\%$ ($9\%$) of the survivor peak on the finite frequency support of the figure; the two are not identical there, because the constant-phase Boltzmann factor is not a strictly retarded response. (e, f) True envelopes $|h|$ on a logarithmic scale: IMR, single-reflection survivor, re-formed signal, and the re-formed signal for the zero-phase filter (dotted).}
\label{fig:imrecho}
\end{figure*}

Figure~\ref{fig:imrecho} shows the outcome for two sources. The inspiral--merger--ringdown (IMR) signal is followed, after a delay of $\Delta t_{\rm echo}=0.298$~s for the GW150914-like source and $1.62$~s for the GW231123-like one, by the part of the first reflection that survives the effective horizon backreaction. Its peak strain is $2\times10^{-4}$ and $4\times10^{-4}$ of the merger peak, respectively, so it is magnified by $10^3$ and $300$ in the figure, and its envelope shows the central dip derived in Sec.~\ref{sec:dip}. The GW231123-like echo is stronger relative to the merger for two reasons (Sec.~\ref{sec:model}). The escape band sits at the superradiant threshold, where the transmission through the potential barrier has a zero, and at the higher remnant spin the transmission rises more steeply from that zero. Its ringdown also carries a larger fraction of the remnant mass. Our default assumption is that the reflecting surface re-forms after the absorption (Sec.~\ref{sec:reformed}). The first echo is then followed, every $\Delta t_{\rm echo}$, by the slowly decaying ringing of the cavity modes.

The paper is organised as follows. Section~\ref{sec:model} summarises the sources, the ingoing packet, the barrier scattering data and the first reflection. Section~\ref{sec:backreaction} sets up the backreaction-filter model, and Sec.~\ref{sec:dip} derives the central envelope dip and shows how it depends on the backreaction filter. Section~\ref{sec:reformed} treats a surface that re-forms after the absorption, the cavity modes it supports and its stability. Section~\ref{sec:detect} gives the detectability in eleven detectors, including source-frame masses and astrophysical populations with source-dependent remnant spins, and Sec.~\ref{sec:modes} adds the other $(\ell,m)$ modes of the ringdown, $m=0$ included. We conclude in Sec.~\ref{sec:conclusions}. The appendices collect the source parameters and scattering data, a step-by-step derivation of the backreaction-filter model, the post-merger time window, the noise curves, the time-domain Teukolsky construction, numerical checks and the code. All codes and data are public~\cite{EchoRepo}, and Appendix~\ref{app:code} lists the codes with the abbreviated names used in the text.

\section{Echo model}\label{sec:model}
\subsection{Source and ingoing packet}\label{sec:source}
We use two reference sources, both described by the $\ell=|m|=2$ mode of the aligned-spin IMRPhenomT waveform \cite{IMRPhenomT}. The first is a GW150914-like binary \cite{GW150914}: component masses $36+29\,\Msun$ in the source frame, $z=0.09$ and $D_L=410$~Mpc. Its remnant has $M_f^{\rm src}=61.66\,\Msun$ in the source frame ($M_f=(1+z)M_f^{\rm src}=67.21\,\Msun$ in the detector frame) and $\chi_f=0.7268$. The second is modelled on GW231123, the most massive binary observed so far \cite{GW231123}: component masses $137+103\,\Msun$ in the source frame, aligned spins chosen to reproduce the published remnant spin, $z=0.39$ and $D_L=2.2$~Gpc. Its remnant has $M_f^{\rm src}=224\,\Msun$ in the source frame ($M_f=(1+z)M_f^{\rm src}=311\,\Msun$ in the detector frame) and $\chi_f=0.84$. Appendix~\ref{app:setup} lists all parameters. Unless stated otherwise the numbers in the text refer to the GW150914-like source, and the GW231123-like values are given alongside. We use $G=c=1$ and write $h=h_+-ih_\times$, $r\psi_4=\tfrac12\partial_t^2(rh)$ (Kinnersley tetrad) and $\tilde F(\omega)=\int F e^{i\omega t}dt$. Throughout, $M_f$ is the detector-frame (redshifted) remnant mass and $M_f^{\rm src}=M_f/(1+z)$ the source-frame mass. Every step of the calculation uses $M_f$: the waveform, the Kerr quantities ($r_\pm$, $\Omega_H$, $T_H$, the tortoise coordinate and $k$), the Boltzmann reflectivity, the backreaction and the location of the surface (Sec.~\ref{sec:wall}). Times, lengths and frequencies are given in units of $M_f$, and values in seconds and hertz are observed values. The source-frame mass enters only where sources and populations are specified. Both reference sources are specified by their source-frame masses and redshift, and every calculation uses their detector-frame masses. The other multipoles of the source, $m=0$ included, are treated in Sec.~\ref{sec:modes}; together they contribute less than $1.8\times10^{-4}$ of the echo SNR$^2$.

We select the post-merger signal, $F(u)=\tfrac12\ddot h_{22}(u)\,w(u)$, with a raised-cosine ramp centred on the strain peak (Appendix~\ref{app:window}). We inject its ingoing counterpart inside the light ring and evolve it to the horizon with a $1{+}1$ time-domain Teukolsky equation (Appendix~\ref{app:td}). Its spectrum $|\tilde\Phi_{\rm in}|^2$ has a long tail toward low frequencies. On the remnant it peaks at $M_f\omega=0.521$ for the GW150914-like source, $2.8\,T_H$ above the threshold $mM_f\Omega_H=0.4309$. For the GW231123-like source it peaks at $0.580$, $1.3\,T_H$ above $mM_f\Omega_H=0.545$ [Fig.~\ref{fig:walls}(a)]. Its $|\tilde\Phi_{\rm in}|^2$-weighted mean frequency lies lower, $M_f\bar\omega=0.494$ and $0.541$, and $19\%$ and $45\%$ of the power lie below the threshold. The hole that absorbs the packet is not the remnant but the hole before the absorption (Sec.~\ref{sec:backreaction}). Its threshold is lower than that of the remnant for the GW150914-like source, $0.4068/M_f$. For the GW231123-like source it is higher, $0.641/M_f$, and lies above the ringdown frequency. As a result $94\%$ of the GW231123-like power lies below the threshold of the hole that absorbs it, against $13\%$ for the GW150914-like source.

\subsection{Barrier scattering and energy normalisation}
In the Teukolsky variable, outgoing waves grow exponentially toward the horizon. On a time-domain grid that reaches the surface, the reflected wave would therefore be swamped by the ingoing one. We use the time-domain evolution only to obtain the ingoing $\psi_4$ spectrum $\tilde\Phi_{\rm in}(\omega)$ at a reference point $r_*^p$ on the horizon side, and treat the near-horizon cavity in the frequency domain with exact Kerr barrier data. Three functions are needed. The first is the reflection coefficient from inside, $\RBH(\omega)$. The second is the transmission $\Tr(\omega)$ from a unit outgoing horizon wave to $\psi_4$ at infinity. The third is the factor $A_{\rm trans}(\omega)$, which converts the ingoing $\psi_4$ spectrum into a horizon plane-wave amplitude. We obtain them from the Teukolsky ``up'' solution in Sasaki--Nakamura variables \cite{SasakiNakamura1982,Hughes2000}, with energy weights from the Teukolsky--Press horizon flux \cite{TeukolskyPress1974}. Appendix~\ref{app:setup} describes the construction, the frequency grids, the reciprocity test of the greybody factor and the phase convention. As expected, the energy-weighted $|\RBH|^2$ equals $1-\Gamma$ and exceeds unity for $\omega<m\Omega_H$, where the hole is superradiant. At high frequency the same coefficient falls exponentially, on a scale that is not set by $T_H$~\cite{Abedi:2026nonthermal}.

\subsection{Boltzmann surface and the first reflection}\label{sec:wall}
Following Ref.~\cite{Abedi2017}, we place the reflecting surface one Planck proper length $l_p$ outside the would-be horizon,
\begin{equation}
\begin{aligned}
&\int_{r_+}^{r_++\Delta r}\sqrt{g_{rr}}\,dr\Big|_{\theta=0}\sim l_p,\\
&\Delta r=\frac{\sqrt{1-\chi^2}\,l_p^2}{4M_f\left(1+\sqrt{1-\chi^2}\right)},
\end{aligned}
\label{eq:planck}
\end{equation}
The echo delay is twice the tortoise distance between the barrier and the surface, $\Delta t_{\rm echo}\simeq4M_f(1+1/\sqrt{1-\chi^2})\ln(M_f/l_p)$, with the detector-frame mass both in the prefactor and in the logarithm; Appendix~\ref{app:setup} gives its value for our remnant. We place the surface at $r_*^{\rm wall}=-\Delta t_{\rm echo}/2$ with this leading-logarithm formula, which drops constants of order $M_f$. In the full Kerr tortoise coordinate used for the barrier this puts it $3.0\,l_p$ (proper distance on the axis) outside the horizon, within the order-unity freedom of Eq.~\eqref{eq:planck}. The echo delay, the cavity mode spacing and the layer thickness $L_*$ of Sec.~\ref{sec:backreaction} all use this same $r_*^{\rm wall}$. A surface at exactly $l_p$ would sit at $r_*=-455.5M_f$, with $\Delta t_{\rm echo}=911M_f$, $L_*=423.6M_f$ and a $k_c$ smaller by $1.2\%$. We adopt the Boltzmann amplitude reflectivity, defined on energy-normalised waves,
\begin{equation}
|\Rwall|=e^{-|k|/2T_H},\qquad k\equiv\omega-m\Omega_H ,
\label{eq:refl}
\end{equation}
with a constant phase $\phi_0$, the Boltzmann reflectivity phase. A frequency-dependent phase is tested in Appendix~\ref{app:robust}. This is the reflectivity obtained for quantum black-hole horizons \cite{Oshita2020,WangOshitaAfshordi2020,AmaroSeoane2025}, used in echo waveforms computed from black-hole perturbation theory \cite{Xin2021,SrivastavaChen2021,LongoMicchi2021,Ma2022} and in echo searches \cite{AbediGW190521,Abedi2023}, and discussed for evaporating black holes \cite{OshitaMotohashiNoda2022}. We do not consider a perfectly reflecting surface, $|\Rwall|=1$. On a spinning background its cavity modes with $\omega<m\Omega_H$ grow, because the round-trip gain $\sqrt{1-\Gamma}$ exceeds unity there. This is the ergoregion instability \cite{Maggio2019}, and such objects would already have spun down or produced a stochastic background \cite{Barausse2018}. The Boltzmann surface is stable, because its absorption $1-|\Rwall|^2\simeq|k|/T_H$ outweighs the superradiant amplification near threshold (Sec.~\ref{sec:reformed}). With $W=\Rwall e^{i\phi_0}e^{-2ikr_*^{\rm wall}}\sqrt{f_{\rm in}/f_{\rm out}}$ and $a_{\rm in}=A_{\rm trans}\tilde\Phi_{\rm in}e^{ikr_*^p}$, the first reflection that reaches infinity is
\begin{equation}
r\tilde h_1(\omega)=-\frac{2}{\omega^2}\,\Tr\,W\,a_{\rm in},
\label{eq:h1}
\end{equation}
Here $f_{\rm in}$ and $f_{\rm out}$ are the horizon energy fluxes per unit squared amplitude of the ingoing and outgoing waves (Appendix~\ref{app:setup}), so $\sqrt{f_{\rm in}/f_{\rm out}}$ maps an ingoing onto an outgoing energy-normalised amplitude at the surface. Each later echo picks up a further factor of $\RBH W$, so the $n$-th carries $(\RBH W)^{n-1}$.

For the GW150914-like source the first reflection reaches $4.01\%$ of the merger peak and carries $0.83\%$ of the energy delivered to the horizon. For the GW231123-like source it reaches $4.91\%$. Both are small because the ringdown lies well above the threshold. The ringdown frequency of the GW150914-like source, $M_f\omega_{220}=0.545$, lies $3.5\,T_H$ above the threshold $mM_f\Omega_H=0.4309$, where the Boltzmann reflectivity is only $|\Rwall|^2=0.029$. For the GW231123-like source $M_f\omega_{220}=0.615$ lies $2.5\,T_H$ above $mM_f\Omega_H=0.545$, where $|\Rwall|^2=0.082$ [Fig.~\ref{fig:walls}(a)]. The GW231123-like ringdown also carries more energy, $2.70\%$ of the remnant mass against $1.86\%$. The escape band of Sec.~\ref{sec:backreaction} contains $6.4\%$ of the field power $|\tilde\Phi_{\rm in}(\omega)|^2$ of the ingoing packet, against $5.4\%$ for GW150914. Because the horizon flux $dE_H/d\omega\propto k\,|\tilde\Phi_{\rm in}|^2$ vanishes at the threshold, the gross absorbed energy inside the band is much smaller, $0.23\%$ and $0.14\%$. The survivor is set by the field amplitude near threshold, not by the energy deposited there. For the GW231123-like source a large part of the packet is superradiant ($\omega<m\Omega_H$), on the remnant and even more on the hole that absorbs it, which matters for the energy bookkeeping below.

\section{A model for the echo that survives the effective horizon backreaction}\label{sec:backreaction}
\paragraph*{The objection and the hypothesis.} Guo and Mathur \cite{GuoMathur2022} consider the part of the merger wave that falls onto a surface at a Planckian distance and is reflected. The round trip takes $\Delta t_{\rm echo}\simeq900M_f$, and long before the reflected wave can return, the infalling energy has formed a closed trapped surface outside the reflecting one. Nothing inside a trapped region can influence infinity, so no echo escapes. This is a statement about causality; it does not rely on treating the reflected wave as a ray. Nor does a wave description get around it. An extended packet can have support outside a newly forming horizon, but whether any part of the reflected signal does can only be settled by following its spacetime support in the evolving geometry. The counterexamples and numerical studies mentioned in Sec.~\ref{sec:intro} \cite{AbediAction2023,Dailey2023,Dailey2025} show that the outcome depends on the near-horizon physics, but they do not yet cover a Kerr remnant with an infalling ringdown packet.

\begin{figure*}
\includegraphics[width=\textwidth]{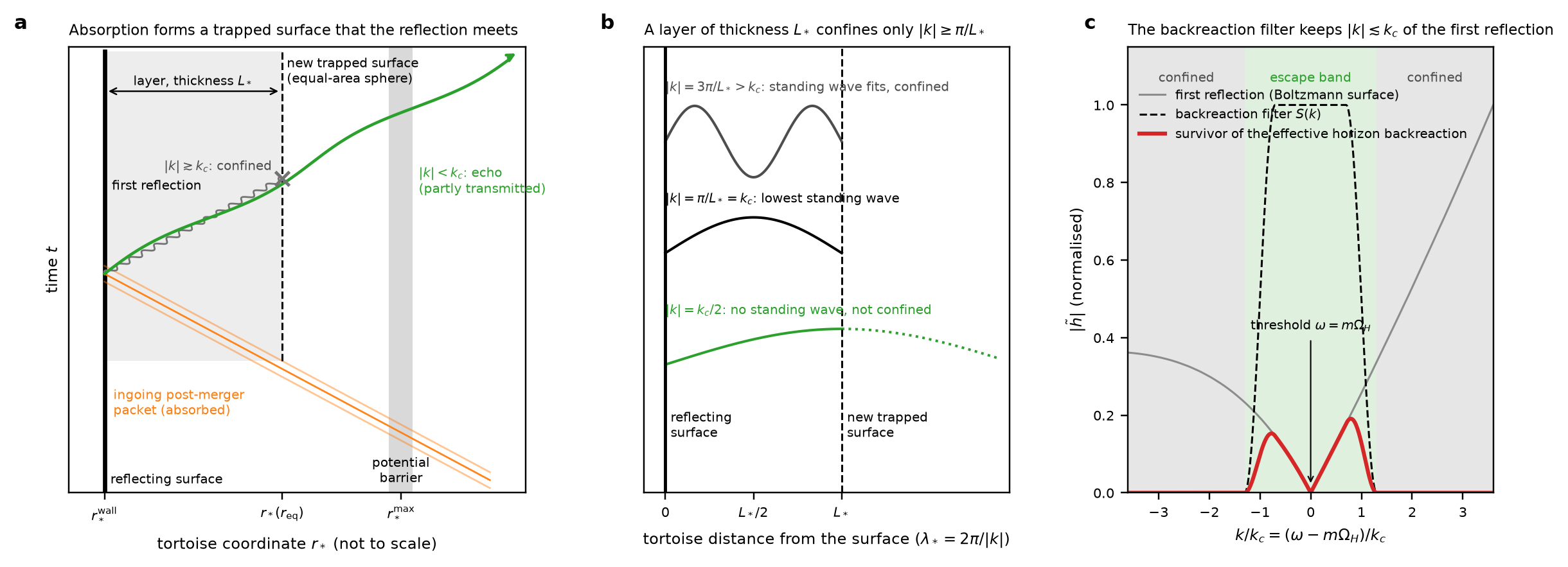}
\caption{Schematic of the backreaction-filter hypothesis of Eq.~\eqref{eq:filter}. Panels (a) and (b) are not to scale and illustrate the assumed picture, not a computed solution. (a) The ingoing post-merger packet (orange) is absorbed, and its energy is assumed to form a new trapped surface, which we place at the equal-area sphere $r_*(r_{\rm eq})$ (dashed, Eq.~\eqref{eq:area}), well before the first reflection from the surface at $r_*^{\rm wall}$ can return ($\Delta t_{\rm echo}\simeq900M_f$). The reflection reaches it after crossing the layer of tortoise thickness $L_*$. Under the hypothesis its components with $|k|\gtrsim k_c$ (grey) stay confined, while those with $|k|<k_c$ (green) are not and reach the potential barrier, where they are partly transmitted as the echo. Here $k=\omega-m\Omega_H$ is the frequency in the horizon frame, so ``low frequency'' means close to the superradiant threshold $\omega=m\Omega_H$ ($207$~Hz for the GW150914-like remnant), not small $\omega$. (b) The confinement heuristic: near the horizon a wave has a fixed tortoise wavelength $\lambda_*=2\pi/|k|$, and a layer of thickness $L_*$ supports standing waves only for $|k|\ge\pi/L_*\equiv k_c$ (Eq.~\eqref{eq:kc}); for $|k|<k_c$ none fits (dotted continuation). (c) The adopted backreaction filter $S(k)$ of Eq.~\eqref{eq:survivor} (dashed) applied to the computed first-reflection spectrum of the Boltzmann surface (grey; reference source, $\ell=m=2$) gives the survivor (red). The green band is the support $|k|<1.3k_c$ ($\pm4.7$~Hz); the grey regions are removed. The first reflection vanishes at $k=0$ with the barrier transmission, which produces the central dip of Sec.~\ref{sec:dip}. All curves are normalised to the maximum of $|\tilde h_1|$ in the plotted range.}
\label{fig:schematic}
\end{figure*}

We therefore \emph{assume}, as a phenomenological hypothesis, that the backreaction acts on the first reflection as a linear filter in the horizon wavenumber,
\begin{equation}
\tilde h_{\rm surv}(\omega)=S(k)\,\tilde h_1(\omega),\qquad |S(k)|\le1,
\label{eq:filter}
\end{equation}
which passes a narrow band around the superradiant threshold and blocks the rest, and we work out what follows. We call $S$ the backreaction filter. Its width is set, through a confinement argument, by the layer between the reflecting surface and the horizon grown by the absorbed energy. Near the horizon the Teukolsky equation in $(t,r_*)$ is translation invariant, so waves have a fixed tortoise wavelength $2\pi/|k|$, and a layer of tortoise thickness $L_*$ cannot hold a standing wave with $|k|<\pi/L_*$ (Fig.~\ref{fig:schematic}). The threshold is also the real part of the horizon modes $\omega=m\Omega_H-ij\kappa$ ($j=1,2,\dots$) that horizon regularity singles out among the outgoing solutions~\cite{AbediAfshordi2026}. We take this as the definition of the width of the backreaction filter, $k_c\equiv\pi/L_*$, without claiming that it follows from the causal structure. The hypothesis may simply be wrong, in which case no echo survives. Testing it requires a causal evolution of the reflected wave on a dynamical, accreting background (Sec.~\ref{sec:conclusions}). Appendix~\ref{app:derivation} derives the relations of this section step by step. For the reflecting surface we assume only that it does not re-form outside the new horizon within $\Delta t_{\rm echo}$. If it does, later echoes reappear with the same band limit, and we treat that case in Sec.~\ref{sec:reformed}.

\begin{figure*}
\includegraphics[width=\textwidth]{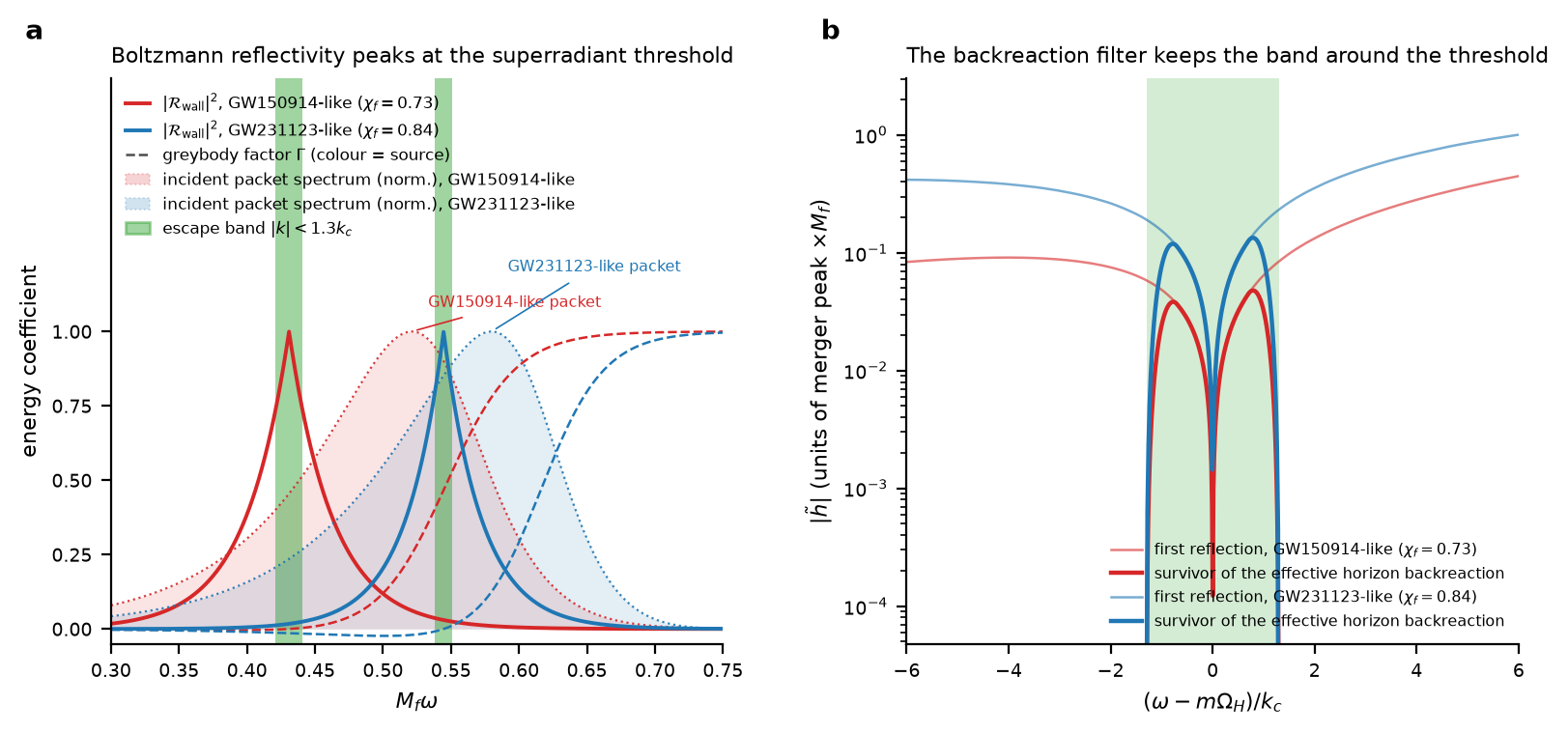}
\caption{Boltzmann surface, escape band and first reflection for the GW150914-like (red, $\chi_f=0.7268$) and GW231123-like (blue, $\chi_f=0.84$) remnants. All curves are those of the remnants, on which the first reflection is evaluated (Sec.~\ref{sec:backreaction}). (a) Boltzmann energy reflectivity $|\Rwall|^2$ (solid), greybody factor $\Gamma$ (dashed) and incident packet spectrum (shaded, dotted outline, normalised to its maximum), each in the colour of its source; the GW150914-like packet (red) peaks near the GW231123-like threshold, and the labels mark the peak of each packet. The green bands are the escape bands $|k|<1.3k_c$ of the backreaction-filter model (Sec.~\ref{sec:backreaction}). The packets peak $2.8\,T_H$ (GW150914-like) and $1.3\,T_H$ (GW231123-like) above their thresholds (Sec.~\ref{sec:model}). (b) Spectrum of the first reflection (thin) and of the survivor, the part of it that survives the effective horizon backreaction (thick), versus $k/k_c$, in units of the merger peak of each source.}
\label{fig:walls}
\end{figure*}

\paragraph*{Energy and angular momentum of the absorbed packet.} We normalise the ingoing packet by the energy it delivers to the horizon of the hole that absorbs it. That hole is not the remnant $(M_f,\chi_f)$ but the hole before the absorption, $(M_i,\chi_i)$, which is determined in the next paragraph. The horizon fluxes of this paragraph therefore carry its Teukolsky--Press weights and its threshold, $k_i=\omega-m\Omega_H^{(i)}$. Components below that threshold are superradiantly amplified and carry negative horizon flux, so the net absorbed energy can be much smaller than the gross, or negative. We therefore fix the gross absorbed energy, $\int|dE_H/d\omega|\,d\omega=\epsilon_E E_{\rm RD}$, with $E_{\rm RD}=1.25\,\Msun=1.86\times10^{-2}M_f$ the energy of the post-merger signal and $\epsilon_E=1$ by default. The net absorbed energy is then $E_{\rm in}=\int(dE_H/d\omega)\,d\omega=0.955\,\epsilon_EE_{\rm RD}$ for the GW150914-like packet. For the GW231123-like packet it is negative, $E_{\rm in}=-0.986\,\epsilon_EE_{\rm RD}$: the threshold of the hole that absorbs it lies above the ringdown frequency, so almost the whole packet is superradiant and extracts energy from the hole. (A normalisation of the net energy is discussed in Sec.~\ref{sec:detect}.) Equal $\psi_4$ amplitude corresponds to $\epsilon_E=0.0229$ and $0.0088$, respectively. An $m=2$ wave deposits angular momentum $(m/\omega)\,dE_H$ per frequency, so
\begin{equation}
\Delta J=\int\frac{m}{\omega}\frac{dE_H}{d\omega}\,d\omega=3.490\,M_f\,E_{\rm in},
\label{eq:dJ}
\end{equation}
where we use the horizon energy spectrum $dE_H/d\omega$ of the computed packet. Using the mean frequency instead, $m/\bar\omega=4.048M_f$, would overestimate $\Delta J$ by $16\%$. Combining this modal relation with the first law gives
\begin{equation}
\frac{\kappa}{8\pi}\,dA=dE_H-\Omega_H\,dJ_H=\frac{k}{\omega}\,dE_H .
\label{eq:firstlaw}
\end{equation}
with $\kappa$, $\Omega_H$ and $k$ those of the absorbing hole. Because $dE_H\propto\alpha\propto k$, the integrand $(k/\omega)\,dE_H$ is never negative. Superradiant modes ($k<0$) \emph{lower} the mass of the hole---they carry $2.3\%$ of the gross flux of the GW150914-like packet and $99.3\%$ of the GW231123-like one---but they still \emph{increase} its area. Integrated over the GW150914-like packet, $(\kappa/8\pi)\Delta A=0.290\,E_{\rm in}$ at linear order. The exact Kerr areas for $\epsilon_E=1$ give $0.950$ times this value; the difference is the second-order change of $\kappa$ and $\Omega_H$ during the absorption. For the GW231123-like packet $\Delta J=4.81\,M_f\,E_{\rm in}$ with $E_{\rm in}<0$: the absorption lowers the mass of the hole by $2.7\%$ and spins it down from $\chi_i=0.919$ to $\chi_f=0.84$, while its area still grows.

\paragraph*{Mass bookkeeping and the equal-area mapping.} The remnant mass and spin of the IMR waveform are fixed by what has been radiated to infinity, $M_f=M_{\rm ADM}-E_{\rm rad}$ and likewise for the angular momentum. The ingoing packet is the inward counterpart of the post-merger signal. It is emitted at the light ring toward the hole and never reaches infinity, so its energy is contained in $M_f$. The same is true of a remnant measured from the ringdown. The quasinormal modes are oscillations of the potential barrier at the light ring. When they are emitted, some $10M_f$ after the peak, the packet is already inside the light ring but still far from the horizon in tortoise distance. It reaches the reflecting surface only about $\Delta t_{\rm echo}/2\simeq450M_f$ later. The barrier responds to the mass and angular momentum it encloses (in spherical symmetry this is Birkhoff's theorem), whether or not they have crossed the horizon. We therefore take $(M_f,\chi_f)$ as the hole \emph{after} the absorption, and the hole that absorbs the packet as its progenitor,
\begin{equation}
M_i+E_{\rm in}=M_f,\qquad \chi_iM_i^2+\Delta J=\chi_fM_f^2 .
\label{eq:progenitor}
\end{equation}
Because $E_{\rm in}$ and $\Delta J$ are fluxes onto the horizon of $(M_i,\chi_i)$, with the gross flux fixed at $\epsilon_EE_{\rm RD}$, Eq.~\eqref{eq:progenitor} has to be solved self-consistently. For the GW150914-like source the solution is unique, $M_i=0.9823\,M_f=66.02\,\Msun$ and $\chi_i=0.6892$, and the area grows by $\Delta A/A=1.4\times10^{-2}$. For the GW231123-like source $M_i=1.0266\,M_f$ and $\chi_i=0.919$: the hole before absorption is heavier and spins faster than the remnant. In the ringdown of GW250114, a fit that includes the first horizon mode, $\omega=2\Omega_H-i\kappa$, likewise returns a remnant that is heavier and spins faster than the inspiral predicts, which Ref.~\cite{AbediAfshordi2026} attributes to a horizon that has not yet settled. Table~\ref{tab:progenitor} lists the progenitors of all spin templates. The net absorbed energy changes sign between $\chi_f=0.81$ and $0.825$. Its sign is set by where the threshold of the absorbing hole lies relative to the packet. For the $\chi_f=0.81$ template it lies at $m\Omega_H^{(i)}=0.506/M_f$, below the ringdown frequency $0.593/M_f$, and the net flux is $+0.54$ of the gross. For the GW231123-like source it lies above it ($0.641/M_f$ against $0.615/M_f$). The change is discontinuous. Up to $\chi_f=0.81$ the spin of the hole before absorption rises smoothly, to $\chi_i=0.799$. At $\chi_f=0.825$ it is $0.901$, and the net flux is $-0.94$ of the gross. In between the mapping folds over: in a narrow interval of remnant spins inside $0.805<\chi_f<0.8125$, three holes satisfy the two conservation laws ($\chi_i=0.799, 0.827\ {\rm and}\ 0.875$ for the $\chi_f=0.81$ template), and we adopt the one nearest to the remnant spin. Above this interval only the energy-extracting branch remains, so the net flux changes sign without passing through zero. A packet that extracts energy needs an absorbing hole that is heavier and spins faster than the remnant, and this moves its threshold further above the packet. For $\chi_f=0.90$ the progenitor is already close to extremal, $\chi_i=0.989$. For $\chi_f=0.955$ no sub-extremal progenitor exists at $\epsilon_E=1$. A packet with the full ringdown energy cannot have been absorbed by a hole that ends at this spin. For such remnants we lower $\epsilon_E$ to the largest value that admits a progenitor with $\chi_i\le0.998$. This value is the spin limit of thin-disc accretion \cite{Thorne1974}, which we adopt as a cap; the mapping itself requires only $\chi_i<1$. It gives $\epsilon_E=0.45$ at $\chi_f=0.955$. We compute the fluxes from the packet field evolved on the remnant and the flux weights of the absorbing hole. Evolving the packet on the progenitor itself changes nothing of consequence. The hole then settles at a mass of $1.00004\,M_f$ and $\chi=0.7269$ ($1.00006\,M_f$ and $0.8389$ for the GW231123-like source), and $k_c$ below changes by $1\times10^{-5}$ ($1\times10^{-4}$) in relative terms. The map counts the full signed ingoing flux as the energy that forms the new trapped region, i.e.\ the energy enclosed before the reflection. One could instead count only the part absorbed by the surface on the first pass, $(1-|\Rwall|^2)$ of the incident flux. The net-to-gross ratio would then be $0.930$ instead of $0.955$ ($-0.977$ instead of $-0.986$). The same change of mass makes the echo spacing drift in models where the reflecting object follows the absorbed energy \cite{Vellucci2023}. Here it enters only through the location of the new horizon. To place the new horizon relative to the reflecting surface we need a rule that compares the two holes. We use the Boyer--Lindquist sphere of the Kerr geometry before absorption, $(M_i,a_i)$, whose exact area
\begin{equation}
A(r)=2\pi\!\int_0^\pi\!\sin\theta\,\sqrt{(r^2+a_i^2)^2-\Delta_i\,a_i^2\sin^2\theta}\;d\theta
\label{eq:area}
\end{equation}
equals the area $4\pi(r_+^2+a^2)$ of the remnant horizon. Note that Eq.~\eqref{eq:area} reduces to $4\pi(r^2+a_i^2)$ only at $\Delta_i=0$. This sphere lies at $r_{\rm eq}-r_+^{(i)}=1.4\times10^{-2}M_f$, or $r_*(r_{\rm eq})=-9.79M_f$ in the tortoise coordinate of that geometry. Its reflecting surface sits at $r_*^{{\rm wall},i}=-428.3M_f$, so the layer between the surface and the mapped horizon has tortoise thickness
\begin{equation}
\begin{aligned}
L_*&=r_*(r_{\rm eq})-r_*^{{\rm wall},i}=418.5\,M_f,\\
k_c&=\pi/L_*=7.51\times10^{-3}/M_f ,
\end{aligned}
\label{eq:kc}
\end{equation}
where all lengths are in units of $M_f$. With $M_f=3.310\times10^{-4}$~s this corresponds to $f_c=k_c/2\pi=3.61$~Hz around the threshold of the remnant, $f_H=207$~Hz (the threshold of the hole before absorption is $196$~Hz; the first reflection is evaluated on the remnant, see below). The proper thickness of the layer is $0.333M_f$ in the equatorial plane and $0.358M_f$ on the axis. For the GW231123-like remnant, $L_*=665M_f$ and $f_c=0.49$~Hz around $f_H=56.6$~Hz.

We use the area because, for a quasi-stationary horizon, it is the one geometric quantity that does not depend on the choice of coordinates. It equals $16\pi M_{\rm irr}^2$, grows monotonically during absorption (second law), and is the same for the apparent and event horizons once the hole has settled. Boyer--Lindquist radii of the two holes refer to different coordinate systems and cannot be compared directly, and a Misner--Sharp-type quasi-local mass exists only in spherical symmetry. Still, $L_*$ and $r_{\rm eq}$ come from a prescription. They do not locate a marginally trapped surface, which would require the outgoing null expansion to vanish in the dynamical spacetime.

Fortunately $L_*$ is insensitive to the details, because $r_*$ is logarithmic in $r-r_+$. The mean-frequency $\Delta J$ with the approximate area $4\pi(r^2+a_i^2)$ gives $414.08M_f$, the spectral $\Delta J$ with the approximate area $418.47M_f$, and our adopted choice, the spectral $\Delta J$ with the exact area, $418.51M_f$. Lowering the packet energy changes little: $L_*=430M_f$ for $\epsilon_E=0.0229$ and $423M_f$ for $\epsilon_E=10^{-3}$, while at $10^6\,\Msun$ $L_*=463M_f$. The width $k_c$ is close to the cavity mode spacing $\pi/|r_*^{\rm wall}|$, so the escape band $|k|<1.3k_c$ spans $2.8$ mode spacings. Where the resonances fall inside it is set by the total round-trip phase, that is, by the Boltzmann reflectivity phase $\phi_0$ and the phase of $\RBH$; there is no general reason for one of them to sit at $k=0$.

\paragraph*{Shape of the backreaction filter.} For definiteness we adopt the even, zero-phase raised cosine
\begin{equation}
S(k)=\begin{cases}1, & x<0.7,\\[2pt] \tfrac12+\tfrac12\cos\!\Big[\dfrac{\pi(x-0.7)}{0.6}\Big], & 0.7\le x\le1.3,\\[6pt] 0, & x>1.3,\end{cases}
\label{eq:survivor}
\end{equation}
with $x\equiv|k|/k_c$. With this choice $k_c$ is both the half-width at half-maximum of $S$ and its equivalent rectangular half-width ($\int S\,dk=2k_c$), while $S$ is non-zero only for $|k|<1.3k_c$, i.e.\ within $\pm4.7$~Hz (Fig.~\ref{fig:schematic}c). In Appendix~\ref{app:robust} we replace it with seven other shapes, one of them causal, and vary $k_c$ by a factor of two.

\paragraph*{The surviving signal.} We call the part of the first reflection that survives the effective horizon backreaction, $\tilde h_{\rm surv}$ of Eq.~\eqref{eq:filter}, the survivor. For the GW150914-like source it peaks at $1.21\times10^{-4}$ of the merger strain (the amplitude scales as $\sqrt{\epsilon_E}$) and carries $1.0\times10^{-4}$ of the first-reflection energy; for the GW231123-like source the peak is $2.2\times10^{-4}$, $1.8$ times larger. Inside the band $|k|<1.3k_c$ the Boltzmann reflectivity is $e^{-|k|/2T_H}\ge e^{-1.3k_c/2T_H}=0.861$ ($0.896$ for GW231123), and $0.911$ at the lobe centres $|k|\simeq0.8k_c$ (Sec.~\ref{sec:dip}). The backreaction filter therefore keeps the band in which the surface reflects almost perfectly, and the survivor is set mainly by the barrier transmission and by how much of the packet lies near threshold [Fig.~\ref{fig:walls}(b)]. A reflectivity whose magnitude departs from unity within $\pm k_c$ of $m\Omega_H$ would change the survivor amplitude accordingly, and a frequency-dependent phase would change its waveform.

\begin{figure*}
\includegraphics[width=\textwidth]{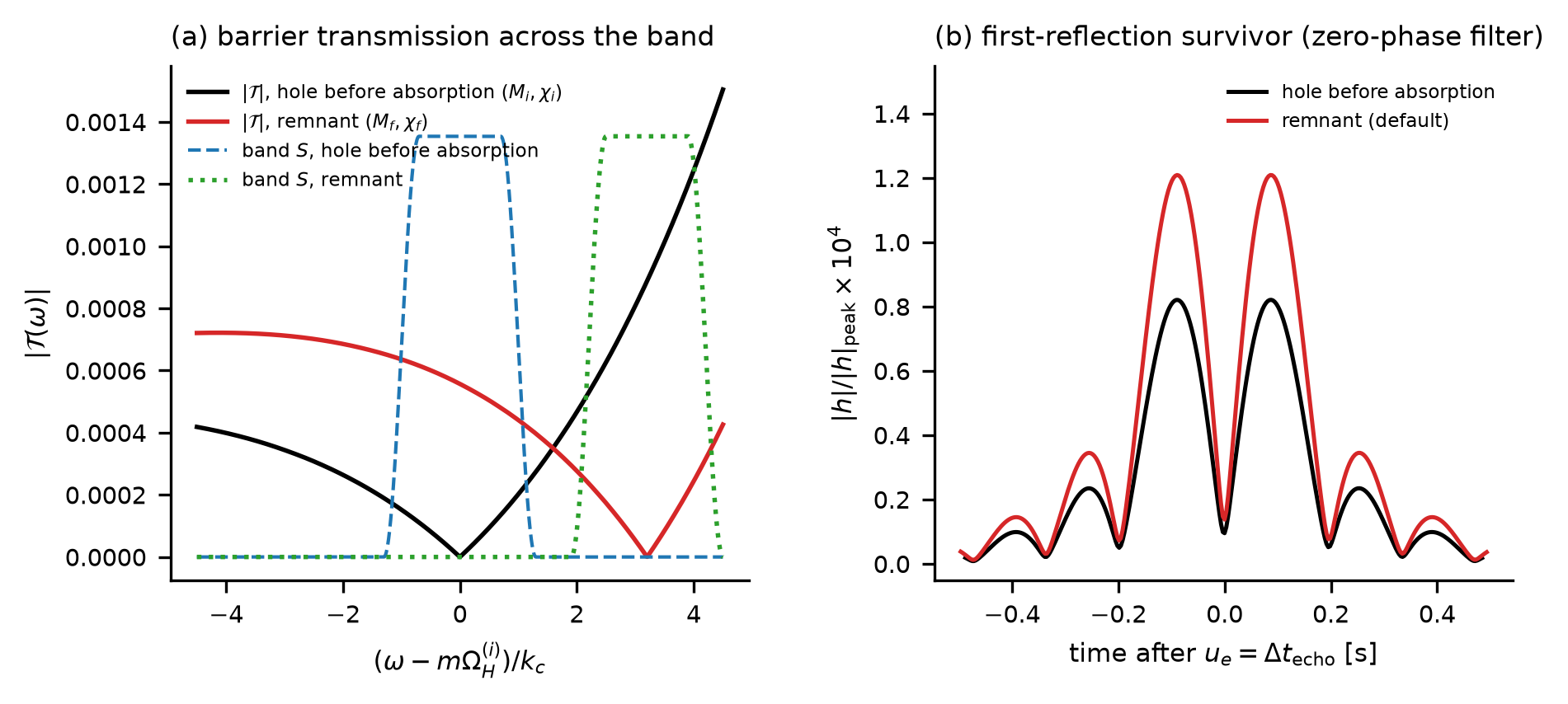}
\caption{Which Kerr background carries the first reflection (GW150914-like source). (a) Barrier transmission $|\Tr(\omega)|$ of the hole before absorption (black; zero at $m\Omega_H^{(i)}$) and of the remnant (red; zero at $m\Omega_H$, $3.2\,k_c$ higher), against the horizon wavenumber of the hole before absorption; the dashed and dotted curves are the escape band $S$ of the two evaluations, centred on the threshold of the hole before absorption and on that of the remnant (scaled). (b) Envelope of the survivor, the part of the first reflection that survives the effective horizon backreaction (zero-phase filter, $\epsilon_E=1$): everything on the hole before absorption (black, the bracket) and everything on the remnant with $k_c$ from the mapping (red, our default). Code \code{MCF}~\cite{EchoRepo}.}
\label{fig:metric}
\end{figure*}

\paragraph*{Horizon-frequency shift: which hole carries the first reflection.} The absorption changes the hole from $(M_i,\chi_i)$ to the remnant $(M_f,\chi_f)$, and its horizon frequency from $m\Omega_H^{(i)}=0.4068/M_f$ to $m\Omega_H=0.4309/M_f$ (GW231123-like: from $0.641/M_f$ to $0.545/M_f$). The shift, $0.0240/M_f=3.2\,k_c$ ($21\,k_c$, the other way), is larger than the bandwidth of the backreaction filter, so it matters on which Kerr background the reflected wave propagates. The absorption is complete long before the reflected wave gets back to the barrier (next paragraph), so the wave that returns from the surface climbs the barrier of the remnant. The transmission $\Tr$, its zero and the threshold are therefore those of $(M_f,\chi_f)$. We take the Boltzmann factor and the Planck-distance location of the surface from the same hole, and give a variant below. Our default evaluation is therefore done on the remnant. A run denotes one complete calculation with our pipeline code \code{KEP}~\cite{EchoRepo} (packet, barrier data, first reflection, backreaction filter and surface; Appendix~\ref{app:code}), and the default runs are \texttt{V0} for the GW150914-like source and \texttt{GW231123} for the GW231123-like one. The time-domain packet, the barrier data, the surface ($r_*^{\rm wall}=-450.0M_f$, $\Delta t_{\rm echo}=900M_f=0.298$~s) and the centre of the escape band are those of $(M_f,\chi_f)$, while $L_*$ and $k_c$ come from the equal-area mapping above. The amplitude of the packet is fixed by its gross flux onto the hole before absorption. It is $0.852$ ($0.864$) times the amplitude that would deliver the same gross flux to the remnant. As a bracket we evaluate the same model entirely on the hole before absorption (runs \texttt{I0} and \texttt{IG}, the counterparts of \texttt{V0} and \texttt{GW231123}; Fig.~\ref{fig:metric}). Packet, barrier, surface, Boltzmann factor and band are then those of $(M_i,\chi_i)$, with $\Delta t_{\rm echo}=0.284$~s. Relative to that evaluation the survivor amplitude of the default changes by $+47\%$ for the GW150914-like source and by $+27\%$ for the GW231123-like source, and its SNR is larger by factors $1.49$ and $1.16$. The dip depth and hump separation are $0.114$ and $543M_i$ there against $0.113$ and $533M_f$ here. Across the spin templates the threshold shift runs from $-182\,k_c$ to $+3.2\,k_c$ (Table~\ref{tab:progenitor}). For $\chi_f\gtrsim0.825$ the packet is superradiant on the hole that absorbs it. That hole is heavier and spins faster than the remnant, and its threshold lies far above the band of the remnant. Neither static evaluation bounds a non-adiabatic evolution.

The sweep is far from adiabatic. At the horizon probe of the time-domain evolution, $80\%$ of the cumulative $|\partial_t\Phi|^2$, from 10 to 90\%, crosses within $\Delta t_{\rm abs}=22M_f$. We use this quantity as a proxy for the energy flux, not as the signed Teukolsky flux. That is short compared with both the coherence time of the escape band, $1/k_c=133M_f$, and the echo delay, $900M_f$. The adiabaticity parameter, $(d\,m\Omega_H/dt)/k_c^2\simeq(\Delta m\Omega_H/k_c)/(k_c\Delta t_{\rm abs})\approx19$, is large, so a backreaction filter that tracks an instantaneous threshold makes no sense. In the sudden limit the hole has settled into the remnant before the reflected wave returns, which is why we evaluate on it. The layer is $8.4$ times longer than the packet, which takes $51M_f$ between the arrival of $1\%$ and $99\%$ of its energy. A wave reflected at the surface therefore meets the last in-going part of the packet within $18$--$25M_f$ of tortoise distance from the surface. It then crosses the remaining layer, more than $401M_f$, and the barrier in the geometry of the remnant. For the GW231123-like source the corresponding distances are $12$--$26M_f$ and $621M_f$. For the reflection itself the choice of hole is less clear. The arrival of the front at the surface and the passage of the tail through the final equal-area sphere are spacelike separated ($375M_f$ of Boyer--Lindquist time against $426M_f$ of tortoise distance), so their order depends on the time slicing. Along in-going slices the front reaches the surface $51M_f$ before the tail. As a variant we therefore reflect each in-going slice with the Boltzmann factor and at the Planck-distance surface of the hole it finds on arrival, and keep the escape band and the barrier of the remnant (\code{ARC}~\cite{EchoRepo}). The Boltzmann factor inside the band then falls from $0.86$--$1$ to $0.71$--$0.79$, and the survivor peak and SNR become $0.80$ ($0.48$) times those of the default; the central dip remains (depth $0.087$ and $0.030$, against $0.113$ and $0.060$). This is within the other uncertainties of the amplitude, and we keep the remnant for the reflection as well. How the transmitted band responds to the sudden shift is a question only the dynamical calculation can answer.

\paragraph*{Robustness to the backreaction filter.} Because the form of the backreaction filter is a hypothesis, we have tested how much the results depend on it (Appendix~\ref{app:robust}, Table~\ref{tab:robust}, Fig.~\ref{fig:robust}). We replaced the adopted raised cosine by seven other shapes, one of them causal, varied its width by a factor of two and added an in-band phase. Across shapes the survivor peak changes by up to $57\%$ and the LIGO O4 SNR by $-11$ to $+23\%$. The width matters more: $k_c\to(0.5$--$2)\,k_c$ multiplies the SNR by $0.37$--$2.7$, i.e.\ $\rho\propto k_c^{1.44}$, and an in-band phase changes the waveform but not the SNR. The existence of a narrow-band survivor and its central dip are therefore not artefacts of the shape of the backreaction filter, while its absolute amplitude is uncertain by factors of order unity.

\section{The central envelope dip}\label{sec:dip}
\begin{figure*}
\includegraphics[width=\textwidth]{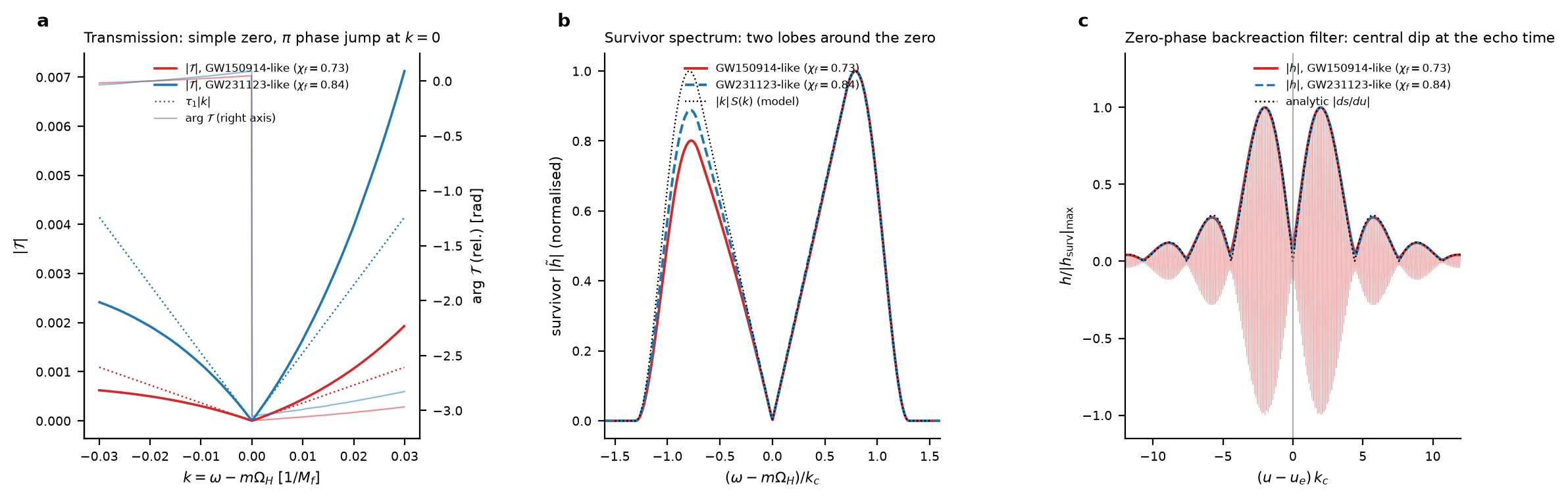}
\caption{Origin of the central envelope dip of the survivor, the echo that survives the effective horizon backreaction (adopted backreaction filter), for the GW150914-like (red) and GW231123-like (blue) remnants. (a) Horizon-to-infinity transmission $|\Tr|$ near the superradiant threshold with its linear approximation $\tau_1|k|$ (dotted) and its phase (thin, right axis): a simple zero with a $\pi$ phase jump. (b) Survivor spectra (normalised) versus $k/k_c$ and the model $|k|S(k)$. (c) Survivor waveform ($h_+$ and envelope) versus $(u-u_e)k_c$, normalised to the peak, with the analytic envelope $|s'(u-u_e)|$ of Eq.~\eqref{eq:rc}; the geometric echo time is $u_e=0.298$~s (GW150914-like) and $1.623$~s (GW231123-like).}
\label{fig:dip}
\end{figure*}

The envelope of the survivor has a minimum close to the geometric echo time [Fig.~\ref{fig:dip}(c)]. Times and $L_*$ are in units of $M_f$ ($L_*=418.5M_f$ for the GW150914-like source). The same effect was seen by Longo Micchi, Afshordi and Chirenti \cite{LongoMicchi2021} in Boltzmann echoes computed from perturbative inspiral--merger--ringdown waveforms of a spinning compact object. Their echo transfer function goes to zero at the superradiant bound frequency $\omega_{\rm SR}=m\Omega_H$, which gives it two humps, and the echoes of their rapidly spinning case ($a=0.9M_f$) have a double peak that they attribute to this double-hump structure. There the humps are as wide as the Boltzmann factor, ${\sim}T_H$, and unequal, the superradiant side being in general the weaker. Here the backreaction filter keeps only $|k|\lesssim k_c\ll T_H$, placed symmetrically about the zero, so the dip is nearly symmetric and is present already in the first reflection. We first derive it for the adopted backreaction filter and then sort out which parts of the result depend on its form.

\paragraph*{(i) Linear zero of the transmission.}
Let $a_{\rm out}$ be the energy-normalised amplitude of a wave leaving the cavity toward the barrier. Energy conservation gives $|\Tr|^2|a_{\rm out}|^2/\omega^2=\Gamma(\omega)\,f_{\rm out}(\omega)|a_{\rm out}|^2$, where $f_{\rm out}$ is the horizon-side flux per unit SN amplitude. Both horizon fluxes are proportional to $k$: $f_{\rm in}\propto\alpha\propto k(k^2+4\epsilon^2)(k^2+16\epsilon^2)$, and $f_{\rm out}/f_{\rm in}\to1$ as $k\to0$ (numerically $1.000$ at $k=0$). The greybody factor vanishes linearly at threshold, $\Gamma\simeq\gamma_1 k+\gamma_2k^2$, with $\gamma_1\simeq0.29M_f$ and $\gamma_2\simeq7.5M_f^2$ from a fit to the dense grid ($|k|\le0.012/M_f$). Hence
\begin{equation}
\begin{aligned}
|\Tr(\omega)|^2&=\omega^2 f_{\rm out}\Gamma\;\propto\;k^2\\
\Rightarrow\quad \Tr(\omega)&=\tau_1\,(\omega-m\Omega_H)+O(k^2),
\end{aligned}
\label{eq:Tzero}
\end{equation}
The dense grid confirms this: $|\Tr(0)|=2\times10^{-9}$, the phase jumps by $3.14$~rad across $k=0$, and $|\Tr|\propto|k|^{0.88}$ for $k<0$ and $|k|^{1.13}$ for $k>0$ over $0.0005<|k|<0.02$, the exponents differing from 1 through the $O(k^2)$ terms. Since $\Tr$ is the response function of a causal system and is analytic near the real axis, a simple zero is a sign change, not a cusp. Physically, at $\omega=m\Omega_H$ the mode co-rotates with the horizon, the horizon flux vanishes, and nothing gets through.

\paragraph*{(ii) Survivor spectrum as an odd factor times a continuous function.}
The survivor of Eqs.~\eqref{eq:h1} and \eqref{eq:filter} can be written as
\begin{equation}
\begin{aligned}
\tilde h_{\rm surv}(\omega)&=-\frac{2}{\omega^2}\,S(k)\,\Tr\,W\,a_{\rm in}=k\,G(k),\\
G(k)&=-\frac{2\tau_1}{\omega^2}\,S(k)\,W\,a_{\rm in}\,[1+O(k)],
\end{aligned}
\end{equation}
with $G$ continuous and confined to $|k|<1.3k_c$. It is smooth except at $k=0$, where the Boltzmann factor $|\Rwall|=e^{-|k|/2T_H}$ has a cusp: $G$ is continuous there, and its slope jumps by $-G(0)/T_H$. Across the band this factor falls only to $0.86$ ($0.90$ for the GW231123-like source), so the cusp is a correction to the leading odd factor $k$ and does not remove the zero at threshold. In the time domain the factor acts as a convolution with the Lorentzian $b/[\pi(b^2+u^2)]$, $b=1/(2T_H)=15.4M_f$. This kernel is two-sided: a reflectivity with a constant phase, like the zero-phase filter, is not a retarded response, although its width is far below the hump separation. In particular $W$ is regular at $k=0$: $f_{\rm in}$ and $f_{\rm out}$ both vanish linearly there, their ratio tends to unity, and $\sqrt{f_{\rm in}/f_{\rm out}}$ departs from 1 by at most $1.7\%$ in the band. Within this band $\omega^{-2}$, $a_{\rm in}$ and $\sqrt{f_{\rm in}/f_{\rm out}}$ vary by only a few percent, $\Rwall$ is even in $k$, and $e^{-2ikr_*^{\rm wall}}$ is a pure delay by the echo time $u_e=-2r_*^{\rm wall}$.

\paragraph*{(iii) Time domain: derivative of a smooth packet.}
With $\tilde h(\omega)=\int h(u)e^{i\omega u}du$, multiplying by $k$ amounts to acting with $i\,\partial_u$ on the demodulated signal:
\begin{equation}
\begin{aligned}
h_{\rm surv}(u)&=e^{-im\Omega_H u}\;i\,\frac{d}{du}\,g(u),\\
g(u)&\equiv\frac{1}{2\pi}\int G(k)\,e^{-iku}\,dk .
\end{aligned}
\label{eq:deriv}
\end{equation}

\paragraph*{(iv) Closed-form envelope for the adopted backreaction filter.}
To leading order $G(k)\propto S(k)\,e^{-2ikr_*^{\rm wall}}$, so $g(u)\propto s(u-u_e)$, where $s$ is the Fourier transform of the backreaction filter~\eqref{eq:survivor}. For a flat top up to $0.7k_c$ and a cosine roll-off to $1.3k_c$, this is the raised-cosine kernel
\begin{equation}
s(u)=\frac{k_c}{\pi}\,\frac{\sin x}{x}\,\frac{\cos\beta x}{1-(2\beta x/\pi)^2},\qquad x=k_cu,\quad\beta=0.3 ,
\label{eq:rc}
\end{equation}
which is real and even. The envelope $|h_{\rm surv}|\propto|s'(u-u_e)|$ therefore vanishes at $u=u_e$ and has two main humps at $x=\pm x_1$, where $x_1=1.9874$ is the first extremum of $s'$. The hump separation is
\begin{equation}
\Delta u_{\rm hump}=\frac{2x_1}{k_c}=\frac{2x_1}{\pi}L_*=530M_f\;(0.175~{\rm s}),
\label{eq:sep}
\end{equation}
and weaker side humps follow from the oscillating tail of $s$. The computed survivors agree with this picture. For the GW150914-like source the minimum lies $2M_f$ before $u_e$, the hump separation is $533M_f$, and the envelope follows $|s'(u-u_e)|$ with correlation coefficient $0.9991$. For the GW231123-like source the hump separation is $846M_f$ ($1.30$~s, analytic $841M_f$) and the dip depth is $0.060$. Plotted against $(u-u_e)k_c$, the two envelopes coincide [Fig.~\ref{fig:dip}(c)]. This confirms the derivation for the adopted backreaction filter, but it is not an independent test of the backreaction. The dip is not complete. Because of the $O(k^2)$ term of $\Tr$, the two spectral lobes carry $39\%$ ($k<0$) and $61\%$ ($k>0$) of the survivor energy, and this odd part of $G$ fills the dip to $0.113$ of the hump height.

\paragraph*{(v) What depends on the backreaction filter.} Two ingredients produce the dip: the zero of $\Tr$ at threshold, which belongs to the Kerr barrier, and the symmetry of the backreaction filter. For every even, zero-phase filter in Table~\ref{tab:robust} the minimum stays at $u_e$ to within $3M_f$, while its depth ($0.10$--$0.24$) and the hump separation ($325$--$558M_f$) depend on the shape. The causal Butterworth filter of the same width keeps a dip of depth $0.16$, but shifts it by $384M_f$ ($0.127$~s) and makes the humps unequal (ratio $1.16$) [Fig.~\ref{fig:robust}(a)]. The shift is not the group delay at threshold, which is $2.61/k_c=348M_f$ for this filter. The group delay grows to $3.91/k_c=521M_f$ at $|k|\simeq0.89k_c$, and the survivor has no weight at $k=0$ and most of it in the lobes, so its node moves by an intermediate amount. Because zero-phase filters are acausal, the timing of their dip diagnoses the transmission zero rather than predicting when a dip would be observed. A response that begins before the geometric echo time is not excluded outright, because the retained waves are longer than the cavity itself (Appendix~\ref{app:robust}), but nothing in the model selects the zero-phase response. The feature is best described as a central envelope dip predicted by the adopted backreaction filter. Its existence follows from the transmission zero for an even filter and a packet spectrum that is smooth and nearly symmetric about the threshold; the cusp of the even Boltzmann factor does not change this. A frequency-dependent escape phase can fill it (Table~\ref{tab:robust}), and its timing, depth and hump separation measure the unknown escape response. What survives is a pair of wave packets at $m\Omega_H$ with opposite phase, reflecting the $\pi$ phase jump of $\Tr$, separated by ${\sim}L_*$. With zero-phase filters the pair is centred on the echo time.

\section{Re-formed surface and cavity modes}\label{sec:reformed}
\paragraph*{Cavity modes in the echo spectra.} The echo transfer function $\Tr W a_{\rm in}/(1-\RBH W)$ has poles at the modes of the cavity between the potential barrier and the surface. These are the low-frequency modes of horizonless objects studied by Maggio \emph{et al.}~\cite{Maggio2019}. For $s=-2$ their Eqs.~(29)--(30) give $\omega_R\simeq m\Omega_H-\pi(q+1)/2|r_*^0|$ and $\omega_I\propto(\omega_R-m\Omega_H)$. The factor $\omega_R-m\Omega_H$ comes from the superradiant amplification $Z=|\RBH|^2-1=-\Gamma$, that is, from the same linear zero of the greybody factor that produces the central dip (Sec.~\ref{sec:dip}). The modes lie on both sides of $m\Omega_H$; for a perfect surface those below it grow, because $|\RBH|>1$ there, and those above it decay. We evaluate the modes with the exact barrier data (\code{CMS}~\cite{EchoRepo}), recomputing the spectra on a grid $\Delta\omega=2\times10^{-6}/M_f$ fine enough to resolve the lines. The comb spacing is $\pi/|r_*^{\rm wall}|=0.93\,k_c$, since $L_*\simeq|r_*^{\rm wall}|$, and its position depends on the Boltzmann reflectivity phase: $\phi_0=0$ puts the modes at odd multiples of $\pi/2|r_*^{\rm wall}|$, their axial family, and $\phi_0=\pi$ gives the polar family. The damping or growth rate is $\omega_I\simeq\ln|\RBH \Rwall|/2|r_*^{\rm wall}|$. Both expressions are the leading terms of a weak-damping expansion. The exact denominator is the derivative of the total round-trip phase, $\tau=\partial_\omega\arg(\RBH W)$, and the spacing is $2\pi/\tau$. They reduce to the geometric values when the phase of $\RBH$ varies slowly: across the escape band $\tau$ is shorter than $2|r_*^{\rm wall}|$ by $0.6$--$0.7\%$ for both reference sources. The mode parameters quoted below use the derivative of the full round-trip factor. Our case, $M_f\omega\simeq0.42$ and $\chi_f=0.7268$, lies outside the range $M_f\Omega_H\ll1$ of the low-frequency formula, and the growth rate of the perfect-surface modes, computed with the exact barrier data, exceeds their Eq.~(30) by a factor $67$--$81$ for $q=1$--$5$ ($318$--$383$ for the GW231123-like remnant, at $M_f\omega\simeq0.53$). With the Boltzmann surface the in-band modes are damped, because the absorption $1-|\Rwall|^2\simeq|k|/T_H$ exceeds $Z$ (the quenched case of their Eq.~(48)). A perfect surface would let the modes with $k<0$ grow; this is the ergoregion instability, and the reason why we do not consider it.

\begin{figure*}
\includegraphics[width=\textwidth]{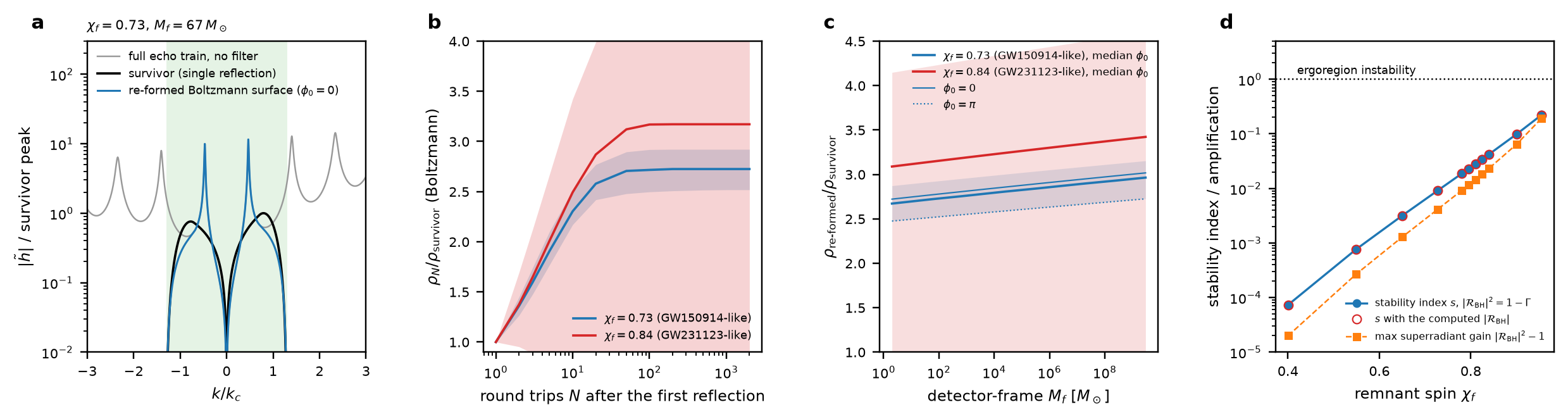}
\caption{Re-formed Boltzmann surface and cavity modes. (a) Spectra near threshold for the GW150914-like remnant ($M_f=67.21\,\Msun$), normalised to the survivor peak: full unfiltered Boltzmann echo train (grey), survivor of a single reflection (black) and survivor with a re-formed surface, Eq.~\eqref{eq:reformed} with $N\to\infty$ and $\phi_0=0$ (blue); green: escape band. (b) SNR after $N$ round trips relative to the survivor for the GW150914-like (blue) and GW231123-like (red) remnants (median and range over $\phi_0$). (c) Gain $\rho_\infty/\rho_{\rm surv}$ versus detector-frame remnant mass, median over $\phi_0$ and range (band) for both remnants, and $\phi_0=0,\pi$ for the GW150914-like one. (d) Stability: largest value over the superradiant range of the index $s=2T_H\ln|\RBH|/|k|$, with $|\RBH|^2=1-\Gamma$ (filled) and with the computed barrier reflection coefficient (open), and the largest superradiant amplification $|\RBH|^2-1$, versus remnant spin for the population templates; $s<1$ means no ergoregion instability.}
\label{fig:reformed}
\end{figure*}

In the full, unfiltered Boltzmann echo train of earlier work \cite{AbediGW190521,Abedi2023} most of the SNR comes from the first echo near the spectral peak of the packet [Fig.~\ref{fig:reformed}(a)]. This agrees with the small excitation factors of long-lived cavity modes \cite{Rosato2026}, and with the finding that the early echoes are carried by high-frequency, nearly reflectionless scattering modes rather than by the low-frequency resonances \cite{Rosato2025}. The backreaction filter reverses this hierarchy: it removes the high-frequency part and keeps only the band near threshold, where the resonances sit. The survivor contains no modes at all: a single reflection has no poles, and its band spans only $2.8$ mode spacings.

\paragraph*{Re-formed surface.} Up to now we have assumed that the reflecting surface does not re-form within $\Delta t_{\rm echo}$ (Sec.~\ref{sec:backreaction}). From Sec.~\ref{sec:detect} on we take the re-formed surface as the default. If the reflecting structure is a property of horizons, for example Planck-scale structure at a fixed proper distance outside whatever the current horizon is, then the hole has a surface of its own after the absorption. The in-band wave reflected back by the barrier keeps bouncing, and after $N$ round trips
\begin{equation}
\tilde h_N=\tilde h_{\rm surv}\,\frac{1-(\RBH W)^N}{1-\RBH W}.
\label{eq:reformed}
\end{equation}
Later bounces do not pass through the backreaction filter again, since the survivor carries only $1.0\times10^{-4}$ of the first-reflection energy. In the sudden limit (Sec.~\ref{sec:backreaction}) the cavity is that of the remnant, on which we evaluate Eq.~\eqref{eq:reformed} (\code{RFC}~\cite{EchoRepo}).

For the Boltzmann surface the series converges within ${\sim}20$--$50$ round trips [Fig.~\ref{fig:reformed}(b)]. After 10 round trips the SNR is already $2.2$--$2.5$ times that of the survivor. The final gain $\rho_\infty/\rho_{\rm surv}$ is $2.47$--$3.15$ over all masses from $2$ to $3\times10^9\,\Msun$ and all $\phi_0$ (median $2.82$). It is $2.21$--$3.35$ when the model is evaluated on the hole before absorption, and $0.68$--$4.61$ (median $3.25$) for the GW231123-like remnant. A gain below one means that the re-formed signal is weaker than the single reflection. Of the two reference sources it occurs only for the GW231123-like one, and only near $\phi_0=\pi$ (for $5.3\%$ of the phases at its own mass). The generic templates on the energy-extracting branch ($\chi_f\ge0.825$, Table~\ref{tab:progenitor}) behave in the same way, with smallest gains of $0.79$ ($\chi_f=0.825$), $0.67$ ($0.84$), $0.79$ ($0.90$) and $0.62$ ($0.955$). All statistics of the gain over $\phi_0$ use $2048$ equally spaced phases. This matters at high spin, where the gain is sharply peaked in $\phi_0$: with 16 phases the median at the native mass of the template would be off by $11\%$ for $\chi_f=0.955$ and by $3.9\%$ for $0.90$, against at most $0.4\%$ for the other templates. For this source the layer is thicker than the cavity, so the mode spacing $\pi/|r_*^{\rm wall}|$ is $1.26\,k_c$, against $0.93\,k_c$ for the GW150914-like source. At $\phi_0=\pi$ one mode sits at the threshold, where the transmission vanishes, and its neighbours lie at $\pm1.26\,k_c$, at the edge of the band. The two lobes of the survivor, at $\pm0.8k_c$, then fall between resonances, where the later bounces interfere destructively and $|1-\RBH W|^{-1}=0.57$. For the GW150914-like source a mode always lies close to one of the lobes and the gain stays above 2.47. The range widens with spin, to $0.79$--$6.7$ at $\chi_f=0.90$ and $0.62$--$12$ at $\chi_f=0.955$. The layer is thicker there ($L_*=1453M_f$ and $3132M_f$, Table~\ref{tab:progenitor}), so the band $|k|<1.3k_c$ spans only $1.1$ and $0.7$ mode spacings, and whether a mode falls inside it depends on $\phi_0$. Because the band is narrow, the gain is the same in every detector to within $0.2\%$. It also depends only weakly on mass [Fig.~\ref{fig:reformed}(c)], since $k_c$ and the mode spacing $\pi/|r_*^{\rm wall}|$ scale together. Two or three in-band modes carry the gain. The sum of their individual powers is $101$--$108\%$ of $\rho_\infty^2$, and their interference terms add only $0.4$--$1.0\%$. A coherent reconstruction of the waveform from the local residues leaves a residual of $24$--$32\%$ of the signal norm, so the lines account for the power but not for the full waveform. Their quality factors are $Q=\omega_R/2|\omega_I|=1310$--$8817$. These are real-axis resonance estimates, one Newton step from the phase crossing of the round-trip factor, not complex-frequency roots of a causal wall response. For each $\phi_0$ the strongest one or two modes are individually louder than the survivor. Their SNRs are $13$--$16$ against $6.2$ in CE ($300\,\Msun$, 410~Mpc). Modes at the band edge or at $k=0$ are much weaker ($1.8$--$2.5$). For $\phi_0=\pi$ the phase condition is also met at $k=0$, where $\Tr$ and $1-\RBH W$ vanish together and the Boltzmann factor has a cusp. The one-step real-axis estimate is not reliable there (its result changes with the sampling around the threshold), so we quote no lifetime for this point. Whatever is trapped at the threshold radiates little, because $\Tr\to0$ there.

\paragraph*{Stability of the re-formed Boltzmann surface.} A re-formed surface makes sense only if the cavity it forms with the barrier is stable, i.e.\ if the round-trip gain $|\RBH\Rwall|$ stays below unity at every frequency $k\neq0$. At the threshold itself both factors tend to one, so the gain tends to unity, which is the marginal point of the geometric series. Near threshold $|\RBH|^2=1-\Gamma>1$ for $k<0$, while $|\Rwall|=e^{-|k|/2T_H}$, so the condition is $s(k)\equiv2T_H\ln|\RBH(k)|/|k|<1$ for all $k<0$. Using $|\RBH|^2=1-\Gamma$ from the barrier data, the largest value of $s$ over the superradiant range of our grid is $0.0090$ for the GW150914-like remnant, $0.042$ for the GW231123-like one and $0.22$ at $\chi_f=0.955$, the highest remnant spin among the population templates of Sec.~\ref{sec:detect} [Fig.~\ref{fig:reformed}(d)]. The margin shrinks with spin, because the superradiant amplification grows while $T_H$ falls, but it stays large. The re-formed Boltzmann surface is therefore ergoregion-stable for all remnants considered here. This agrees with Oshita \emph{et al.}, who find the Boltzmann reflectivity stable up to the Thorne limit \cite{OshitaTsunaAfshordi2020}. Every round trip absorbs a fraction ${\approx}|k|/T_H$ of the in-band energy, so the cavity rings down within a few tens of round trips.

\paragraph*{Effect on the first echo.} The re-formed signal stays close to the survivor until the first in-band bounce returns, one round trip ($2|r_*^{\rm wall}|\simeq\Delta t_{\rm echo}$) after the first echo [Fig.~\ref{fig:imrecho}(c, d)]. Before $2u_e$ the two differ by at most $13\%$ of the survivor peak for the GW150914-like source and $9\%$ for the GW231123-like one. They would be identical up to the first return only for a retarded transfer function, which neither the zero-phase filter nor the constant-phase Boltzmann factor is. With the zero-phase reference filter, whose response is not causal, the leading tail of the second passage reaches back to $u_e$, but the central dip of the first echo survives. For the Boltzmann surface its depth (the envelope minimum relative to the lower neighbouring hump) changes from $0.113$ to $0.105$, and its position moves by $+30M_f$. The cavity ringing thus adds to the signal after the first echo but leaves the dip intact.

\section{Detectability}\label{sec:detect}
\paragraph*{Default signal.} Unless stated otherwise, the SNRs, reaches and expected numbers in this section refer to a Boltzmann surface that re-forms after the absorption (Sec.~\ref{sec:reformed}). We sum all round trips ($N\to\infty$) and quote the range over the unknown Boltzmann reflectivity phase $\phi_0$, or its median where a single number is given. For a single source these refer to the gain at its mass and spin; for the expected numbers they refer to the counts at a phase common to all sources (see below). All quoted SNRs of single sources are optimal and face-on. Since the band is narrow, the re-formed SNR is simply $\rho=G(M_f,\phi_0)\,\rho_{\rm surv}$, with a gain $G$ that does not depend on the detector but only on the remnant mass (through the surface logarithm), its spin and $\phi_0$ [Fig.~\ref{fig:reformed}(c)]. We therefore compute the survivor SNR $\rho_{\rm surv}$ for every detector, mass and redshift and multiply it by $G$. The tables and the waterfall plots also give the single-reflection values, for a surface that does not re-form during the observation. All SNRs use the $\ell=m=2$ mode. The other multipoles (Sec.~\ref{sec:modes}) change the face-on SNRs by less than $1.1\times10^{-4}$ and the SNR at any inclination by less than $2.1\times10^{-4}$, and hence the expected numbers by less than $6.2\times10^{-4}$.

\paragraph*{Conventions.} We rescale the survivor with mass ($h\propto M_f$, $t\propto M_f$, with $r_*^{\rm wall}$ and $L_*$ recomputed for each mass) and compute optimal (matched-filter) face-on SNRs in seven ground-based and four space-based detectors. The ground-based ones are the LIGO Hanford--Livingston network at three sensitivities, namely the O4 ``high'' projection (BNS range 190~Mpc per interferometer), the Advanced LIGO design and A+, together with Advanced Virgo, KAGRA, ET-D (single interferometer) and CE. Appendix~\ref{app:psd} lists the source of each curve \cite{T2000012,Shoemaker2010,T1800042,ObsScen2018,T1600593,Hild2011,P1600143}. The space-based ones are LISA \cite{Robson2019}, Taiji \cite{Taiji2017,Taiji2020,TaijiOverview2021,TaijiConfusion2023}, TianQin \cite{TianQin2016} and DECIGO \cite{DECIGO2011}. The two reference sources are stellar-mass binaries, and we do not use them as models of the massive binaries of the space band. The space-detector numbers of this section are for generic templates with the same remnant spins ($\chi_f=0.7268$ and $0.84$; mass ratio $0.8$, equal aligned spins). Since the model is scale-free and the spins are the same, their SNRs differ from those of the rescaled reference sources by at most $0.6\%$. Where a LISA or DECIGO column appears in the numerical checks of Sec.~\ref{sec:modes} and of the appendices, it weights the reference signal shape with a second noise curve, as a numerical check in another band, and is not a source model. The design of Taiji is close to that of LISA: the same acceleration noise, with longer arms ($3\times10^6$ against $2.5\times10^6$~km) and lower optical-metrology noise (8 against 15~pm/$\sqrt{\rm Hz}$). We use the analytic sky-averaged curve of Ref.~\cite{TaijiConfusion2023}, which has the same form as the LISA curve, with its Galactic confusion noise for four years of data. Since the signal is a narrow line at $f_H$, the ratio of the Taiji and LISA SNRs of one source is $[S_n^{\rm LISA}(f_H)/S_n^{\rm Taiji}(f_H)]^{1/2}$. For remnants of $10^6$--$10^8\,\Msun$ at 3~Gpc this ratio is $0.9$--$2.1$. It is largest at $10^6\,\Msun$, it is $1.8$ at $3\times10^6\,\Msun$, and it falls below one only near $10^7\,\Msun$ ($0.89$). There $f_H\simeq1.4$~mHz, where the fit of the confusion noise of Ref.~\cite{TaijiConfusion2023} lies above that of Ref.~\cite{Robson2019}.
 For a single ground interferometer with $F_+=1$ we use $\rho^2=4\int|\tilde h_+|^2/S_n\,df$, with $|\tilde h_+|^2=|\tilde h|^2/4$ face-on. For the IMR source this agrees with \texttt{pycbc.filter.sigma} to $0.2\%$. Advanced Virgo, KAGRA, ET-D and CE are treated in this way. Every LIGO number in this paper, by contrast, is for the Hanford--Livingston network: the two interferometers at their actual positions and orientations, with the same noise curve, and their SNRs added in quadrature. We do not quote single-interferometer values for LIGO. For a face-on source the response of each interferometer, $F_+^2+F_\times^2$, does not depend on the polarisation angle, and the optimal network SNR is taken at the sky position that maximises the sum over the two detectors. It is $W=1.37$ times the SNR of one optimally oriented interferometer. Here $W^2=1.89$ rather than 2, because the detectors are $27^\circ$ apart and cannot both have the source overhead (Appendix~\ref{app:psd}). Averaging over sky position, polarisation and inclination divides the SNRs by 2.5 for a single interferometer and by $2.43$ for the LIGO network. The space curves are already sky- and polarisation-averaged, and there $\rho^2=4\int(|\tilde h_+|^2+|\tilde h_\times|^2)/S_n\,df$, with $|\tilde h_+|^2+|\tilde h_\times|^2=|\tilde h|^2/2$ face-on. For DECIGO $S_n=\tfrac52S_h^{\rm YS}$. The noise curves are validated in Appendix~\ref{app:psd}. All SNRs are for $\epsilon_E=1$ (gross absorbed energy equal to $E_{\rm RD}$) and scale as $\sqrt{\epsilon_E}$.

In Table~\ref{tab:survivor} the masses are detector-frame masses, $M_f$, at the quoted luminosity distance, with Euclidean scaling $\rho\propto1/D_L$; the echo depends on the source only through $M_f$, $\chi_f$, the mass ratio and $D_L$, so no redshift has to be specified there. For a source of fixed source-frame mass we use $M_f=(1+z)M_f^{\rm src}$, $f_{\rm obs}=f_{\rm src}/(1+z)$ and the Planck18 luminosity distance $D_L(z)$ \cite{Planck2018,FinnChernoff1993}; the surface location $r_*^{\rm wall}\propto\ln(M_f/M_{\rm Pl})$ and the gain $G$ are then set by the detector-frame mass (Table~\ref{tab:cosmo}). These are optimal matched-filter SNRs for a known template. We quote two nominal thresholds. SNR 8 is the conventional one, and we use it for the reach in distance and redshift. For the expected numbers we also use SNR 5, because an echo search is triggered by a detected merger. The time of the merger, the sky position and, through the remnant mass and spin, the frequency $f_H$ are then known in advance, so the search covers few independent templates. For a template of unknown phase in Gaussian noise the false-alarm probability per independent trial is $e^{-\rho^2/2}=3.7\times10^{-6}$ at $\rho=5$, so $10^2$ trials still leave it below $10^{-3}$. Neither threshold is a demonstrated search sensitivity.

\begin{table*}
\caption{Optimal face-on SNR in ET-D and CE (single interferometer with $F_+=1$; $\epsilon_E=1$, adopted backreaction filter) for the GW150914-like ($\chi_f=0.7268$) and GW231123-like ($\chi_f=0.84$) signal shapes, rescaled to the detector-frame remnant mass $M_f$ at $D_L=410$~Mpc. Upper row of each mass: re-formed Boltzmann surface ($N\to\infty$), median over $\phi_0$, with the range over $\phi_0$ in parentheses (Fig.~\ref{fig:reformed}). Lower row (``single''): survivor of a single reflection. The $67\,\Msun$ entries of the first shape are the GW150914-like reference source. Last rows: the GW231123-like reference source at its own distance, where its detector-frame mass is $M_f=$$311\,\Msun$. The IMR SNRs of the two reference sources are 1077 and 1023 in ET-D and 3825 and 3405 in CE. The other detectors are shown in Fig.~\ref{fig:det}(b).}
\label{tab:survivor}
\begin{ruledtabular}
\begin{tabular}{lcccc}
 & \multicolumn{2}{c}{GW150914-like shape ($\chi_f=0.7268$)} & \multicolumn{2}{c}{GW231123-like shape ($\chi_f=0.84$)}\\
$M_f$ [$\Msun$] & ET-D & CE & ET-D & CE\\
\hline
30 & $0.26$ ($0.24$--$0.27$) & $0.42$ ($0.39$--$0.45$) & $0.64$ ($0.14$--$0.85$) & $1.0$ ($0.22$--$1.4$)\\
{\scriptsize single} & {\scriptsize $0.094$} & {\scriptsize $0.15$} & {\scriptsize $0.20$} & {\scriptsize $0.33$}\\[1pt]
67 & $0.92$ ($0.85$--$0.99$) & $1.8$ ($1.7$--$1.9$) & $2.6$ ($0.56$--$3.5$) & $4.7$ ($1.0$--$6.4$)\\
{\scriptsize single} & {\scriptsize $0.34$} & {\scriptsize $0.66$} & {\scriptsize $0.83$} & {\scriptsize $1.5$}\\[1pt]
130 & $2.1$ ($1.9$--$2.2$) & $5.0$ ($4.7$--$5.4$) & $6.2$ ($1.3$--$8.4$) & $14$ ($3.0$--$19$)\\
{\scriptsize single} & {\scriptsize $0.77$} & {\scriptsize $1.8$} & {\scriptsize $2.0$} & {\scriptsize $4.4$}\\[1pt]
300 & $4.3$ ($4.0$--$4.6$) & $17$ ($16$--$18$) & $15$ ($3.2$--$20$) & $48$ ($10$--$65$)\\
{\scriptsize single} & {\scriptsize $1.6$} & {\scriptsize $6.2$} & {\scriptsize $4.7$} & {\scriptsize $15$}\\[1pt]
500 & $6.9$ ($6.4$--$7.4$) & $34$ ($31$--$36$) & $19$ ($4.0$--$25$) & $98$ ($21$--$130$)\\
{\scriptsize single} & {\scriptsize $2.5$} & {\scriptsize $12$} & {\scriptsize $5.8$} & {\scriptsize $31$}\\[1pt]
800 & $12$ ($11$--$12$) & $53$ ($49$--$57$) & $38$ ($8.2$--$52$) & $170$ ($37$--$230$)\\
{\scriptsize single} & {\scriptsize $4.2$} & {\scriptsize $19$} & {\scriptsize $12$} & {\scriptsize $54$}\\[1pt]
1200 & $19$ ($17$--$20$) & $46$ ($42$--$49$) & $53$ ($11$--$71$) & $210$ ($46$--$290$)\\
{\scriptsize single} & {\scriptsize $6.8$} & {\scriptsize $17$} & {\scriptsize $17$} & {\scriptsize $67$}\\[1pt]
2000 & $35$ ($32$--$37$) & $13$ ($12$--$14$) & $110$ ($24$--$150$) & $130$ ($27$--$170$)\\
{\scriptsize single} & {\scriptsize $13$} & {\scriptsize $4.7$} & {\scriptsize $35$} & {\scriptsize $39$}\\[1pt]
\hline
\multicolumn{5}{c}{GW231123-like source at its own distance, $D_L=2.2$~Gpc}\\
311 & & & $2.9$ ($0.61$--$3.8$) & $9.4$ ($2.0$--$13$)\\
{\scriptsize single} & & & {\scriptsize $0.90$} & {\scriptsize $3.0$}\\
\end{tabular}
\end{ruledtabular}
\end{table*}

\begin{figure*}
\includegraphics[width=\textwidth]{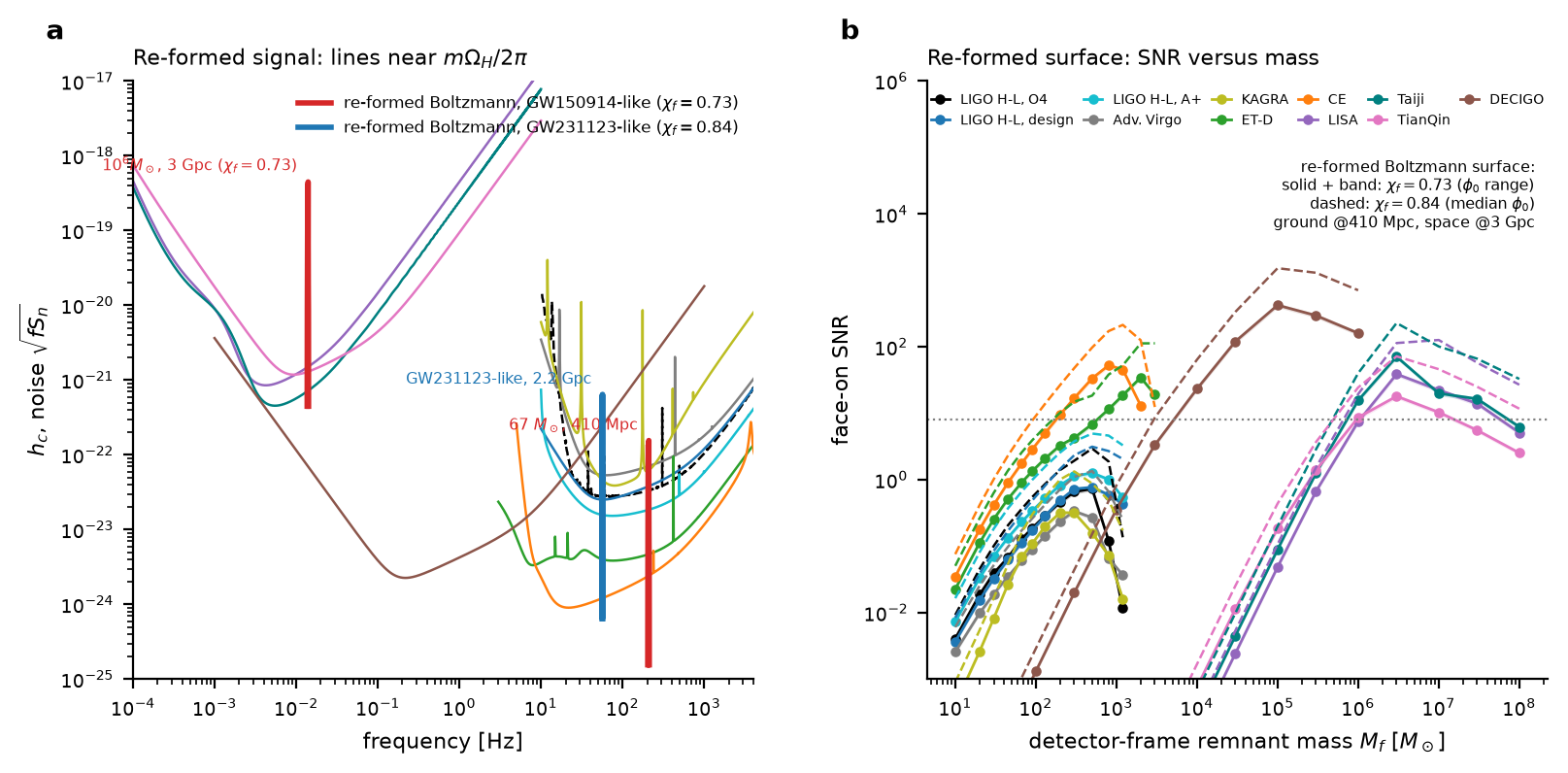}
\caption{Detectability of the re-formed Boltzmann signal (default scenario). (a) Characteristic strain at $\phi_0=0$ for the $67\,\Msun$ GW150914-like remnant at 410~Mpc, the GW231123-like remnant at $2.2$~Gpc and a $10^6\,\Msun$ remnant ($\chi_f=0.7268$) at 3~Gpc, against the characteristic noise amplitude $\sqrt{fS_n}$ of the eleven detectors (for LIGO the effective noise of the Hanford--Livingston network, $\sqrt{fS_n}/W$; colours as in panel b; ground: $2f|\tilde h_+|$; space: $2f(|\tilde h_+|^2+|\tilde h_\times|^2)^{1/2}$; Appendix~\ref{app:psd}). The signal is a few narrow cavity lines within $|k|<1.3k_c$ of $m\Omega_H$; for such lines the height of $h_c$ above $\sqrt{fS_n}$ is not a direct measure of the SNR, which is given in panel (b). (b) Optimal face-on SNR versus detector-frame remnant mass $M_f$ for the $\chi_f=0.7268$ signal shape (filled symbols, median over $\phi_0$; band: range over $\phi_0$) and the $\chi_f=0.84$ shape of the GW231123-like run (dashed, median); ground detectors at 410~Mpc, space detectors at 3~Gpc. The space-detector curves of this panel, and the $10^6\,\Msun$ line of panel (a), are the reference shapes rescaled in mass; the generic templates used for the space detectors in the text differ from them by at most $0.6\%$.}
\label{fig:det}
\end{figure*}

\paragraph*{Results for the GW150914-like source.} Table~\ref{tab:survivor} (ET-D and CE) and Fig.~\ref{fig:det} (all detectors) collect the results. For the GW150914-like remnant at 410~Mpc, face-on and optimally oriented, the re-formed signal at the median $\phi_0$ has SNR $0.11$ in the LIGO network at design sensitivity, $0.13$ at O4 sensitivity and $0.23$ at A+ sensitivity, $0.061$ in Advanced Virgo, $0.070$ in KAGRA, $0.92$ in ET-D and $1.8$ in CE. That is a factor 2.72 (2.52--2.92 over $\phi_0$) above the single-reflection survivor, which gives $0.042$ in LIGO at design sensitivity and $0.66$ in CE. Reaching SNR 8 would take a source at $5.8$~Mpc for LIGO at design sensitivity or $92$~Mpc for CE ($2.1$ and $34$~Mpc for a single reflection). For reference, the single-reflection survivor keeps $0.2$--$1.7\%$ of the SNR of the unfiltered Boltzmann echo train (all echoes, over the full frequency range of each noise curve, for the same detector and source) and $2\times10^{-5}$--$7\times10^{-4}$ of the IMR SNR.

\paragraph*{The GW231123-like source.} The second reference source is more favourable, although it lies $5.4$ times farther away. At $D_L=2.2$~Gpc the re-formed signal has face-on SNR $0.40$ (0.086--0.54 over $\phi_0$) in the LIGO network at O4 sensitivity, $0.45$ at design sensitivity and $0.73$ at A+ sensitivity, $0.21$ in Advanced Virgo, $0.25$ in KAGRA, $2.9$ in ET-D and $9.4$ in CE (last rows of Table~\ref{tab:survivor}). A single reflection gives $0.13$ in LIGO O4 and $3.0$ in CE. The re-formed value in LIGO O4 is $3.1$ times that of the GW150914-like source. Relative to the IMR signal of the same source the echo carries $0.47\%$ of the SNR ($0.09\%$ for GW150914-like). Scaled to the observed network SNR of GW231123, $20.7$ \cite{GW231123}, this corresponds to an echo SNR of ${\approx}0.097$ (single reflection ${\approx}0.03$) in the O4 network, well below a detection threshold. In CE the same event would exceed SNR 8 for most values of $\phi_0$ (2.0--13). Two effects combine. First, at the higher spin the barrier transmission rises $3.8$ times more steeply from its zero and the ringdown carries a larger share of the remnant mass, so the survivor is $1.8$ times stronger relative to the merger (Sec.~\ref{sec:model}). Second, the heavier remnant moves the threshold, $f_H=56.6$~Hz, into the most sensitive band of the ground detectors. The first effect is a property of the spin, not of the event: the GW231123-like template rescaled to $67\,\Msun$ at 410~Mpc gives $2.4$ times the GW150914-like SNR. Because of this strong spin dependence, the population estimates below assign each source its own remnant spin.

\paragraph*{Net-energy normalisation for GW231123.} The normalisation matters most for this source, because its packet lies largely below the threshold. One could ask instead that the \emph{net} absorbed energy equal the ringdown energy. With the bookkeeping of Sec.~\ref{sec:backreaction} this is not possible: a packet that delivers a net energy $E_{\rm RD}$ would have to be so large that no hole can absorb it and end at $\chi_f=0.84$, and Eq.~\eqref{eq:progenitor} has no solution. The largest net energy for which a progenitor exists is $0.34\%$ of $E_{\rm RD}$. The hole before absorption then has $\chi_i=0.841$, close to the remnant, the net-to-gross ratio on it is $0.069$, and the gross flux is $0.13\%$ of the remnant mass ($0.071\%$ absorbed and $0.062\%$ extracted). The packet amplitude is $0.26$ times that of the default and the layer is thinner, $L_*=512M_f$. The re-formed signal then has SNR $0.14$ (0.11--0.14 over $\phi_0$) in the LIGO network at the projected O4 sensitivity for an optimally oriented source at $2.2$~Gpc, $0.34$ times the default. With the noise measured in the two detectors around the event \cite{GWOSC2023,GWOSCO4a} the default gives $0.31$ and the net normalisation $0.10$; scaled to the observed network SNR these are ${\approx}0.10$ and ${\approx}0.03$. The net normalisation therefore gives a weaker echo than the gross one, and with either the echo of GW231123 stays well below the threshold of the O4 network (\code{NNG}~\cite{EchoRepo}).

Larger SNRs need more massive remnants. In CE a $300\,\Msun$ remnant at 410~Mpc (GW150914-like shape) gives 16--18 for the re-formed surface, against $6.2$ for a single reflection, and 31--36 at $500\,\Msun$. ET-D reaches 17--20 at $1200\,\Msun$ and 32--37 at $2000\,\Msun$. At 3~Gpc, for the generic template of the same spin, LISA reaches 36--42 and Taiji 66--76 at $3\times10^6\,\Msun$, and DECIGO exceeds SNR 8 for $10^{4}$--$10^{6}$$\,\Msun$, with 390--460 at $10^5\,\Msun$. TianQin stays below 8 for a single reflection and reaches it for $10^{6}$--$10^{7}$$\,\Msun$ with the re-formed surface (maximum 18). In detail, the sky-averaged SNRs at 3~Gpc for the re-formed surface [single reflection] are: LISA 7.5 [2.6], 39 [14], 23 [7.8], 14 [4.9] and Taiji 16 [5.5], 72 [25], 20 [7.0], 17 [5.7] for $M_f$$=10^6$, $3\times10^6$, $10^7$ and $3\times10^7\,\Msun$; TianQin 8.7 [3.1], 18 [6.3], 10 [3.6] for $10^6$, $3\times10^6$ and $10^7\,\Msun$; DECIGO 23 [8.4], 120 [43], 430 [150], 300 [100], 160 [57] for $10^4$, $3\times10^4$, $10^5$, $3\times10^5$ and $10^6\,\Msun$. The ground detectors have an upper mass limit because $f_H=207\,{\rm Hz}\,(67.21\,\Msun/M_f)$ eventually leaves their band.

\begin{table*}[t]
\caption{Fixed source-frame remnant mass, Planck18 cosmology, $\chi_f=0.7268$ signal shape (generic template for the space detectors): SNR at $z=0.1$, and the redshift $z_8$ and luminosity distance $D_{L,8}$ at which SNR $=8$ (face-on, $\epsilon_E=1$), for the re-formed surface (range over $\phi_0$) and for a single reflection.}
\label{tab:cosmo}
\begin{ruledtabular}
\footnotesize
\begin{tabular}{lccccc}
& \multicolumn{2}{c}{$\rho(z{=}0.1)$} & \multicolumn{2}{c}{$z_8$} & $D_{L,8}$ [Gpc]\\
det., $M_f^{\rm src}$ [$\Msun$] & re-formed & single & re-formed & single & re-formed\\
\hline
CE, $300$ & 15--18 & 6.1 & 0.21--0.24 & 0.075 & 1.0--1.3\\
CE, $600$ & 36--42 & 14 & 0.45--0.51 & 0.18 & 2.6--3.0\\
ET-D, $1200$ & 17--20 & 6.8 & 0.24--0.29 & 0.084 & 1.2--1.5\\
LISA, $3\times10^6$ & 250--290 & 95 & 1.6--1.7 & 0.96 & 12--13\\
LISA, $10^7$ & 120--140 & 44 & 0.82--0.92 & 0.39 & 5.4--6.1\\
Taiji, $3\times10^6$ & 430--500 & 160 & 1.4--1.5 & 0.90 & 10--11\\
DECIGO, $10^5$ & 2500--2900 & 960 & 6.5--7.1 & 3.5 & 65--72\\
\end{tabular}
\end{ruledtabular}
\end{table*}

\paragraph*{Source-frame masses and cosmological reach.} Table~\ref{tab:cosmo} considers a remnant of fixed source-frame mass at redshift $z$, and Fig.~\ref{fig:robust}(c) shows the single-reflection SNR versus $z$. Redshift lowers $f_{\rm obs}$, which helps the ground detectors for light remnants and hurts them for heavy ones, and the luminosity distance grows faster than the comoving one. With the re-formed surface and the $\chi_f=0.7268$ shape (generic template for the space detectors), SNR 8 is reached out to $z=$0.21--0.24 for a $300\,\Msun$ remnant in CE, $z=$1.6--1.7 for $3\times10^6\,\Msun$ in LISA and $z=$1.4--1.5 in Taiji, and $z=$6.5--7.1 ($D_L=$65--72~Gpc) for $10^5\,\Msun$ in DECIGO. A single reflection gives 0.075, 0.96, 0.90 and 3.5. At this mass the Taiji reach is smaller than that of LISA although its SNR at low redshift is larger, because the redshifted line moves to $1.9$~mHz, where its confusion-noise fit is the higher one [Fig.~\ref{fig:robust}(c)].

\begin{figure*}
\includegraphics[width=\textwidth]{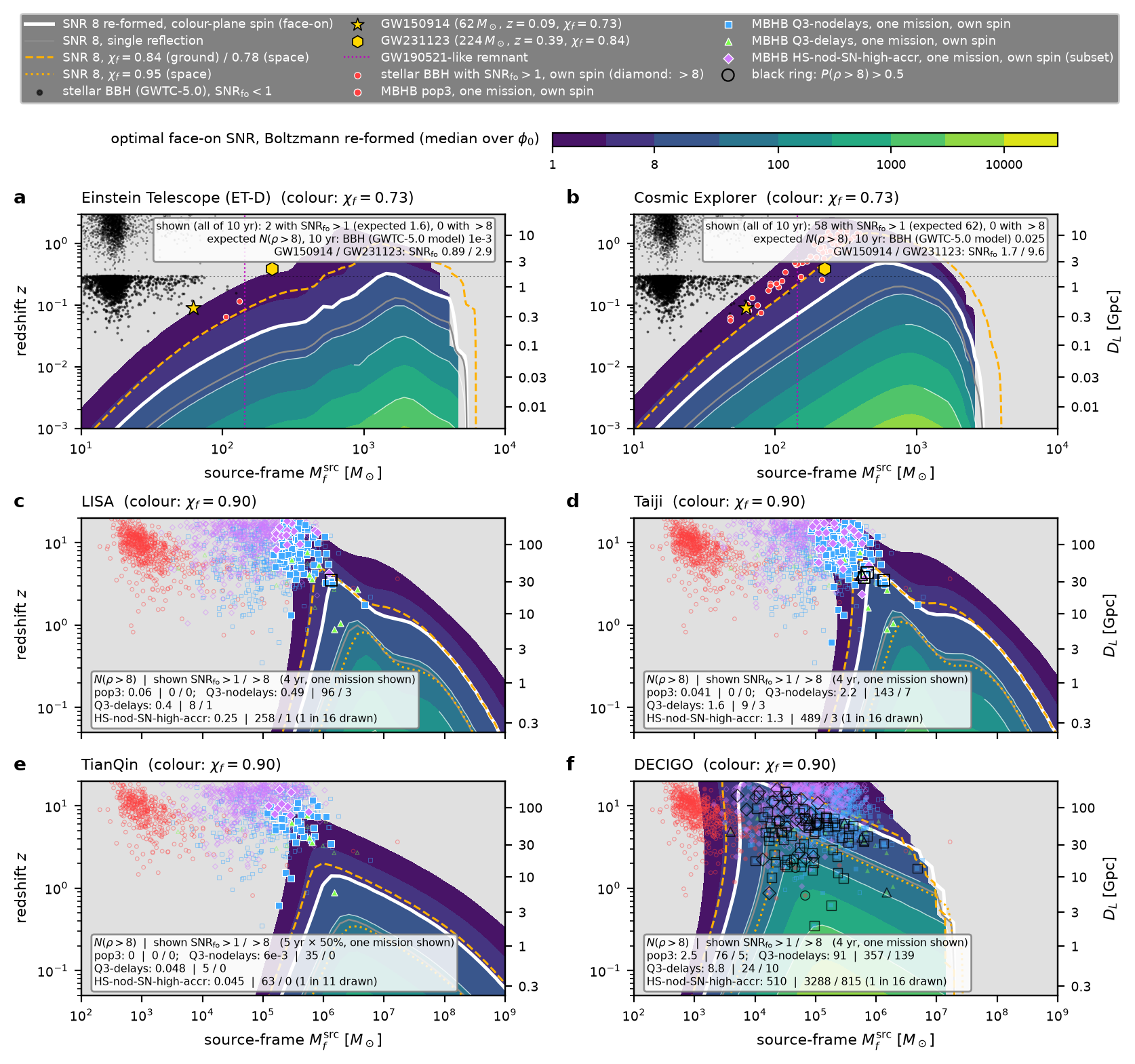}
\caption{Waterfall plots with astrophysical populations, for ET-D, CE (a, b) and LISA, Taiji, TianQin, DECIGO (c--f). Colour: optimal face-on SNR for the re-formed Boltzmann surface at the median $\phi_0$ (white lines SNR 30, 100, 1000) in the plane of source-frame remnant mass and redshift, for $\chi_f=0.7268$ (a, b) and $\chi_f=0.90$ (c--f); thick white: SNR 8 for this spin, grey: SNR 8 for a single reflection; orange: SNR 8 for $\chi_f=0.84$ (a, b) or $0.78$ (c--f, dashed) and $0.955$ (c--f, dotted); right axes give the Planck18 luminosity distance. Panels (a, b): every stellar-mass BBH merger (GWTC-5.0 Default model, posterior-predictive rate, spins from the Default spin model) with face-on, overhead SNR$_{\rm fo}>1$ in 10~yr, evaluated with its own remnant spin, is drawn as a red point without sub-sampling (diamonds: SNR$_{\rm fo}>8$); black dots are the remaining mergers, all of 10~yr at $z<0.3$ (below the dotted line) and a random one-year sample above. Star: GW150914 ($M_f^{\rm src}=62\,\Msun$, $z=0.09$); hexagon: GW231123 ($M_f^{\rm src}=224\,\Msun$, $z=0.39$); magenta line: GW190521-like remnant. Panels (c--f): one complete mission of the pop3 (red circles), Q3-nodelays (blue squares) and Q3-delays (green triangles) massive-BH catalogues \cite{Klein2016}, i.e.\ all events of one 5-yr realisation with coalescence time inside $T_{\rm obs}$, and one Poisson realisation of the PTA-calibrated model HS-nod-SN-high-accr \cite{Barausse2023} (purple diamonds; one event in $n$ is drawn, with $n$ given in the box), each evaluated with its own remnant spin; filled symbols have SNR$_{\rm fo}>1$, open ones not, and black rings mark an orientation-averaged probability $>0.5$ of SNR $>8$. The boxes give the numbers of plotted sources with SNR$_{\rm fo}>1$ and $>8$, and the expected numbers with SNR $>8$ over the stated observing time, averaged over orientation and, for the catalogues of Ref.~\cite{Klein2016}, over the ten realisations, at the median over a common $\phi_0$ (Table~\ref{tab:rates} gives the range over $\phi_0$). Adopted backreaction filter, $\epsilon_E=1$; ground: single interferometer, overhead source for the colour scale; space: sky- and polarisation-averaged noise curves, face-on source.}
\label{fig:wfpop}
\end{figure*}

\paragraph*{Waterfall plots.} Figure~\ref{fig:wfpop} extends Table~\ref{tab:cosmo} to the whole plane of source-frame remnant mass and redshift for the third-generation and space-based detectors, in the style of the LISA ``waterfall'' plots and with the same model and conventions. Because the SNR depends on the remnant spin, each panel is drawn for one reference spin. The colour shows the re-formed SNR at the median $\phi_0$ for $\chi_f=0.7268$ (ground, typical of stellar-mass remnants) and for $\chi_f=0.90$ (space, typical of massive-BH remnants). The SNR-8 contours for other spins are overlaid. Table~\ref{tab:waterfall} lists, for all eleven detectors and three spins, the largest redshift at which SNR 8 is reached for any mass. For $\chi_f=0.7268$ the re-formed surface extends the reach of a single reflection by a factor 2.4--2.7 in redshift (median $\phi_0$). The reach does not grow monotonically with spin. For $\chi_f=0.84$ it is 3.0--3.9 times larger than for $\chi_f=0.7268$ (median $\phi_0$). For $\chi_f=0.90$ the hole before absorption is close to extremal ($\chi_i=0.989$), the layer is thick and the packet amplitude is only $0.42$ times that on the remnant (Table~\ref{tab:progenitor}). The re-formed gain then depends strongly on $\phi_0$, and the reach is between 0.3 and 2.0 times that of $\chi_f=0.7268$ (1.4--1.5 times at the median $\phi_0$).

The second-generation detectors reach SNR 8 only within $17$~Mpc (KAGRA) to $66$~Mpc (LIGO A+), i.e.\ $z\le0.0039$ to $0.015$ at the median $\phi_0$, for $\chi_f=0.7268$, around $M_f^{\rm src}\sim240$--$460\,\Msun$. For $\chi_f=0.84$ the LIGO A+ reach grows to $z\simeq$ 0.012--0.074. That volume contains essentially no remnants of such mass: their regions with face-on SNR $>1$ contain only $3\times10^{-3}$ (LIGO at design sensitivity) and $0.023$ (LIGO at A+ sensitivity) expected mergers of a 10-yr stellar-BBH population, so we do not plot them. ET-D and CE extend the reach to $z=$0.31--0.34 and 0.45--0.51 for remnants of ${\sim}1600$ and ${\sim}640\,\Msun$ (0.28--1.3 and 0.42--1.9 for $\chi_f=0.84$). LISA reaches $z=$2.2--2.6 near $1.4\times10^{6}\,\Msun$, Taiji $z=$3.0--3.4 near $8.0\times10^{5}\,\Msun$, TianQin 0.95--1.1, and DECIGO 13--15 for $M_f^{\rm src}\sim6700\,\Msun$ ($\chi_f=0.7268$). For $\chi_f=0.90$ LISA reaches $z=$0.81--4.6, Taiji 1.1--6.1 and TianQin 0.34--1.9, and DECIGO runs past the edge of the grid at $z=20$. DECIGO is the only detector in which the signal would be visible at cosmological distances over several decades in mass. The sharp edges at high mass are where $f_H/(1+z)$ leaves the frequency range of the noise curve.

\begin{table}
\caption{SNR-8 reach at fixed source-frame mass (Planck18, optimal face-on, $\epsilon_E=1$, adopted backreaction filter; LIGO rows: Hanford--Livingston network): largest redshift $z_8$ over all masses for the re-formed surface (range over $\phi_0$) and for a single reflection, and the source-frame mass at which it is reached (re-formed, median $\phi_0$), for the $\chi_f=0.7268$ signal shape; last two columns: re-formed reach for the $\chi_f=0.84$ (GW231123-like) and $\chi_f=0.90$ shapes. Space detectors: generic templates with these remnant spins. From the waterfall grids.}
\label{tab:waterfall}
\begin{ruledtabular}
\scriptsize\setlength{\tabcolsep}{0.2pt}
\begin{tabular}{@{}lccccc@{}}
& \multicolumn{3}{c}{$\chi_f=0.7268$} & $0.84$ & $0.90$\\
detector & re-formed & single & $M_f^{\rm src}/\Msun$ & re-formed & re-formed\\
\hline
LIGO O4 & 0.0081--0.0094 & 0.0032 & 416 & 0.0075--0.045 & 0.0024--0.019\\
LIGO design & 0.0085--0.0099 & 0.0034 & 416 & 0.0078--0.047 & 0.0026--0.020\\
LIGO A+ & 0.014--0.016 & 0.0054 & 464 & 0.012--0.074 & 0.0041--0.031\\
AdV & 0.0038--0.0044 & 0.0015 & 373 & 0.0034--0.021 & 0.0011--0.0088\\
KAGRA & 0.0036--0.0042 & 0.0014 & 240 & 0.0033--0.020 & 0.0011--0.0085\\
ET-D & 0.31--0.34 & 0.13 & 1551 & 0.28--1.3 & 0.10--0.63\\
CE & 0.45--0.51 & 0.20 & 645 & 0.42--1.9 & 0.16--0.91\\
LISA & 2.2--2.6 & 1.0 & $1.4\times10^{6}$ & 2.0--10 & 0.81--4.6\\
Taiji & 3.0--3.4 & 1.3 & $8.0\times10^{5}$ & 2.6--13 & 1.1--6.1\\
TianQin & 0.95--1.1 & 0.43 & $1.4\times10^{6}$ & 0.83--4.0 & 0.34--1.9\\
DECIGO & 13--15 & 5.8 & 6690 & 12--$\ge20$ & 4.6--$\ge20$\\
\end{tabular}
\end{ruledtabular}
\end{table}

\begin{table*}
\caption{Expected numbers of signals with SNR $>8$ (first block) and SNR $>5$ (second block) over the observing time $T_{\rm obs}$ for the re-formed Boltzmann surface ($N\to\infty$), smallest and largest value over a Boltzmann reflectivity phase $\phi_0$ common to all sources ($256$ phases), averaged over source orientation, with the remnant spin of every source. Third block: the loudest expected signal, i.e.\ the SNR $\rho_1$ for which one signal above $\rho_1$ is expected in $T_{\rm obs}$, for each population. Stellar BBH: GWTC-5.0 Default model, posterior-predictive expectation (for CE the 5--95\% interval of the population posterior rescales the numbers by $0.33$--$2.3$). GW231123-like: single-event cross-check at the median rate of $0.08$~Gpc$^{-3}$\,yr$^{-1}$ with $\chi_f=0.84$ (its interval rescales the numbers by $0.12$--$3.4$); it overlaps with the stellar-BBH column and is not to be added to it. Massive-BH binaries: pop3, Q3-nodelays and Q3-delays are the catalogues of Ref.~\cite{Klein2016} (mean over the ten 5-yr realisations); HS-nod (B+23) is the PTA-calibrated model HS-nod-SN-high-accr of Ref.~\cite{Barausse2023} (weighted catalogue). The four are alternative models and are not to be added. Last two columns: local comoving rate $R_1^{300}$, $R_1^{1000}$ (Gpc$^{-3}$\,yr$^{-1}$) of equal-mass binaries with remnant mass $300$ or $1000\,\Msun$, with remnant spins drawn from the spin model (median $\chi_f=0.70$), that would give one signal above the threshold of the block in $T_{\rm obs}$ (median $\phi_0$; ``--'': not detectable at any distance). LIGO design and LIGO A+: Hanford--Livingston network, 10~yr at that sensitivity. Rows ``$f_H>0.2$~Hz'' below DECIGO: only the signals whose line, at the observed frequency $f_H=m\Omega_H/2\pi$, lies above the white-dwarf foreground (see text). $T_{\rm obs}$ is time of data taking; only the TianQin entry includes a duty cycle. Single-reflection numbers and the other scenarios are in Table~\ref{tab:reformed}.}
\label{tab:rates}
\begin{ruledtabular}
\footnotesize\setlength{\tabcolsep}{1.2pt}
\begin{tabular}{llcccccccc}
detector & $T_{\rm obs}$ & stellar BBH & GW231123-like & pop3 & Q3-nodelays & Q3-delays & HS-nod (B+23) & $R_1^{300}$ & $R_1^{1000}$\\
\hline
\multicolumn{10}{c}{expected number with SNR $>8$}\\
LIGO design & 10 yr & $2\times10^{-6}$--$3\times10^{-6}$ & $2\times10^{-6}$--$5\times10^{-4}$ & $0$ & $0$ & $0$ & $0$ & $4000$ & $8500$\\
LIGO A+ & 10 yr & $1\times10^{-5}$--$2\times10^{-5}$ & $1\times10^{-5}$--$3\times10^{-3}$ & $0$ & $0$ & $0$ & $0$ & $1000$ & $2100$\\
ET-D & 10 yr & $7\times10^{-4}$--$1\times10^{-3}$ & $6\times10^{-4}$--$0.14$ & $0$ & $0$ & $0$ & $2\times10^{-3}$ & $22$ & $0.55$\\
CE & 10 yr & $0.015$--$0.031$ & $0.014$--$3.9$ & $0$--$7\times10^{-3}$ & $0$ & $0$ & $0$ & $0.34$ & $0.062$\\
LISA & 4 yr & $0$ & $0$ & $0$--$0.15$ & $0.084$--$0.79$ & $0.046$--$0.56$ & $0.077$--$0.37$ & -- & --\\
Taiji & 4 yr & $0$ & $0$ & $0$--$0.19$ & $0.53$--$3.0$ & $0.22$--$2.1$ & $0.45$--$2.0$ & -- & --\\
TianQin & 5 yr$\times$50\% & $0$ & $0$ & $0$--$0.028$ & $0$--$0.027$ & $0.013$--$0.080$ & $0.015$--$0.063$ & -- & --\\
DECIGO & 4 yr & $9\times10^{-9}$--$1\times10^{-8}$ & $2\times10^{-9}$--$6\times10^{-7}$ & $0.30$--$9.4$ & $43$--$100$ & $2.7$--$14$ & $300$--$570$ & $5\times10^{5}$ & $77$\\
\quad$f_H>0.2$~Hz & 4 yr & -- & -- & $0$--$2.7$ & $2.7$--$5.1$ & $0.16$--$0.82$ & $35$--$61$ & & \\
\hline\multicolumn{10}{c}{expected number with SNR $>5$}\\
LIGO design & 10 yr & $7\times10^{-6}$--$1\times10^{-5}$ & $9\times10^{-6}$--$2\times10^{-3}$ & $0$ & $0$ & $0$ & $0$ & $980$ & $2100$\\
LIGO A+ & 10 yr & $5\times10^{-5}$--$9\times10^{-5}$ & $4\times10^{-5}$--$0.010$ & $0$ & $0$ & $0$ & $0$ & $250$ & $540$\\
ET-D & 10 yr & $3\times10^{-3}$--$5\times10^{-3}$ & $3\times10^{-3}$--$0.53$ & $0$ & $0$ & $0$ & $3\times10^{-3}$ & $5.7$ & $0.12$\\
CE & 10 yr & $0.075$--$0.15$ & $0.057$--$18$ & $0$--$0.047$ & $0$ & $0$ & $0$ & $0.080$ & $0.028$\\
LISA & 4 yr & $0$ & $0$ & $0$--$0.20$ & $0.50$--$2.8$ & $0.19$--$1.7$ & $0.37$--$1.8$ & -- & --\\
Taiji & 4 yr & $0$ & $0$ & $2\times10^{-3}$--$0.34$ & $1.8$--$9.0$ & $0.62$--$4.9$ & $2.1$--$9.4$ & -- & --\\
TianQin & 5 yr$\times$50\% & $0$ & $0$ & $0$--$0.074$ & $0.041$--$0.34$ & $0.058$--$0.42$ & $0.045$--$0.21$ & -- & --\\
DECIGO & 4 yr & $4\times10^{-8}$--$6\times10^{-8}$ & $1\times10^{-8}$--$2\times10^{-6}$ & $0.49$--$14$ & $67$--$150$ & $4.6$--$19$ & $540$--$900$ & $1\times10^{5}$ & $18$\\
\quad$f_H>0.2$~Hz & 4 yr & -- & -- & $1\times10^{-3}$--$5.1$ & $3.8$--$7.0$ & $0.25$--$1.0$ & $60$--$94$ & & \\
\hline\multicolumn{10}{c}{loudest expected signal: SNR $\rho_1$ exceeded by one source in $T_{\rm obs}$}\\
LIGO design & 10 yr & $0.11$--$0.13$ & $0.10$--$0.63$ & -- & -- & -- & -- & & \\
LIGO A+ & 10 yr & $0.20$--$0.25$ & $0.16$--$1.0$ & -- & -- & -- & -- & & \\
ET-D & 10 yr & $0.79$--$0.97$ & $0.64$--$4.0$ & -- & -- & -- & -- & & \\
CE & 10 yr & $2.3$--$2.9$ & $2.0$--$12$ & -- & -- & -- & -- & & \\
LISA & 4 yr & -- & -- & $0.11$--$1.6$ & $3.8$--$7.4$ & $2.3$--$6.4$ & $3.8$--$5.9$ & & \\
Taiji & 4 yr & -- & -- & $0.19$--$2.5$ & $6.2$--$13$ & $3.8$--$12$ & $6.2$--$10$ & & \\
TianQin & 5 yr$\times$50\% & -- & -- & $<0.04$--$0.56$ & $1.9$--$3.5$ & $0.97$--$3.4$ & $1.9$--$3.0$ & & \\
DECIGO & 4 yr & -- & -- & $3.2$--$47$ & $88$--$190$ & $20$--$61$ & $140$--$270$ & & \\
\end{tabular}
\end{ruledtabular}
\end{table*}

\paragraph*{Expected numbers from astrophysical populations.} To turn the reach into expected numbers we fold the face-on SNR grids with source orientations and population models (\code{POP}~\cite{EchoRepo}). Because the signal depends strongly on the remnant spin (Fig.~\ref{fig:reformed}, Table~\ref{tab:waterfall}), we compute the full chain (packet, barrier data, backreaction filter, re-formed gain) for a grid of templates, $\chi_f={}$$0.4$, $0.55$, $0.65$, $0.727$, $0.78$, $0.795$, $0.81$, $0.825$, $0.84$, $0.9$, $0.955$. The templates at $\chi_f=0.7268$ and $0.84$ are the two reference sources. These are stellar-mass binaries, and we use them only for the stellar-mass population. For the massive-BH catalogues we replace them by generic templates with the same remnant spins, built like the rest of the grid (mass ratio $0.8$, equal aligned spins); this changes the massive-BH counts by at most $3\%$ ($0.3\%$ in DECIGO). Every source is assigned the SNR interpolated in $\chi_f$ (linearly in $\ln\rho$) between the two neighbouring templates. The grid is densest between $0.78$ and $0.84$, where the hole before absorption changes branch (Sec.~\ref{sec:backreaction}) and the SNR is not monotonic in the spin: at $\chi_f=0.81$ the single-reflection SNR is $1.6$--$1.8$ times that at $0.78$ and $1.1$--$1.3$ times that at $0.84$. Leaving out one template of this interval and interpolating across the gap changes the median SNR at its spin by at most $16\%$. With no template between $0.78$ and $0.84$, as in the earlier version of the grid, the interpolated SNR at $0.81$ would be a factor $1.36$ too low. Each template is rescaled to the remnant mass $M_f^{\rm src}=M_{\rm tot}[1-0.048(4\eta)^2]$ and to the symmetric mass ratio $\eta$ of the source ($\rho\propto\eta$, as for the dominant ringdown mode). Remnant spins follow from the component spins with the precessing fit of Hofmann, Barausse and Rezzolla \cite{HBR2016}. Sources outside the template range are assigned the nearest template, and we report the share of such sources. We set $\epsilon_E=1$, except for the $\chi_f=0.955$ template, for which the sub-extremality cap of Sec.~\ref{sec:backreaction} gives $\epsilon_E=0.45$. A source counts as detected when its SNR, for a randomly drawn orientation, exceeds the threshold (8 or 5). The Boltzmann reflectivity phase is taken to be the same for all sources. We compute every count for each of $256$ equally spaced values of $\phi_0$, giving each source the gain $G(M_f,\chi_f,\phi_0)$ of its own mass and spin at that phase, and quote the smallest and the largest count as the range and the median over $\phi_0$ where a single number is given. The mean over $\phi_0$ is $0.79$--$1.52$ times the median for the entries of Table~\ref{tab:rates} above $0.01$. It is the expectation for a uniform prior on a common phase, and also for phases drawn independently for each source. For ET-D and CE we draw sky position, polarisation and inclination for a single L-shaped interferometer \cite{FinnChernoff1993}, and for LIGO we draw them for the Hanford--Livingston network. For the space detectors, whose noise curves are already sky- and polarisation-averaged, we draw only the inclination. We use the following populations. (i) Stellar-mass BBHs from the Default model of the GWTC-5.0 population analysis \cite{GWTC5pop,GWTC5popData}, fitted to $259$ binary black holes; it is the model introduced with GWTC-4.0 \cite{GWTC4pop}. The primary mass follows a broken power law with two Gaussian peaks, which steepens from $m_1^{-1.5}$ to $m_1^{-5.4}$ at $37\,\Msun$ and extends to $300\,\Msun$. The mass ratio follows a power law. The comoving rate is $R(z=0.2)=30.4$ ($25.0$--$36.9$)~Gpc$^{-3}$\,yr$^{-1}$ and grows as $(1+z)^\kappa$ with $\kappa=2.5$ ($1.8$--$3.2$). We use this growth up to $z=1.5$, the range probed by the catalogue, and hold the comoving rate constant beyond, which gives $9.8\times10^4$ mergers per year out to $z=4$. Masses, rate and redshift evolution are marginalised over $400$ samples of the population posterior, so every stellar-mass count is a posterior-predictive expectation. The spins follow the Default spin model at the posterior medians of its parameters: magnitudes from a Gaussian with $\mu_\chi=0.06$ and $\sigma_\chi=0.37$ truncated to $[0,1]$, and tilts from a mixture of a Gaussian in $\cos\theta$ (mean $0.32$, width $1.1$, weight $0.67$) and an isotropic part. This gives $\chi_f=0.28$--$0.76$ (5--95\%, median $0.64$). Unlike the GWTC-3 Power Law + Peak model \cite{GWTC3pop}, whose mass function ended at $87\,\Msun$, this population contains binaries like GW190521 \cite{GW190521} and GW231123. (ii) As a cross-check on the heaviest binaries, a GW231123-like class ($137+103\,\Msun$, $\chi_f=0.84$) with the single-event rate of that merger, $0.08^{+0.19}_{-0.07}$~Gpc$^{-3}$\,yr$^{-1}$, constant in comoving volume \cite{GW231123}. It overlaps with population (i) and is not added to it. (iii) Massive black-hole binaries from the ten 5-yr realisations of the pop3 (light seeds), Q3-nodelays and Q3-delays (heavy seeds) catalogues of Klein \emph{et al.} \cite{Klein2016,Barausse2012}, as distributed with GWToolbox \cite{GWToolbox,GWToolboxData}, with $175$, $122$ and $8.18$ mergers per year. The catalogues give the spin vectors of both holes, from which we compute $\chi_f$ source by source. Their remnants spin rapidly: the median $\chi_f$ is $0.95$ (pop3), $0.78$ (Q3-nodelays) and $0.93$ (Q3-delays), and $27\%$, $14\%$ and $18\%$ of the sources lie outside the template range. (iv) The most recent public catalogues of the same semi-analytic model \cite{Barausse2023}. The model was updated in Ref.~\cite{Barausse2020} and then compared with the gravitational-wave background reported by the pulsar-timing arrays (PTAs). We use the model that reproduces the amplitude of that background, HS-nod-SN-high-accr: heavy seeds, no delays between the galaxy and the black-hole mergers, supernova feedback and enhanced accretion. The recent comparison of merger-rate predictions by the LISA Astrophysics Working Group adopts the same version \cite{IzquierdoVillalba2026}. It has $2350$ mergers per year, most of them at $z\simeq4.2$--$18$ with total masses of $8\times10^{3}$--$4\times10^{5}\,\Msun$ (5--95\%). Its remnants spin more slowly than those of (iii): the median $\chi_f$ is $0.61$, and $21\%$ of the sources lie outside the template range. These are slowly spinning remnants below the grid, and they do not contribute to the detections. This release is the complete weighted catalogue of the model and not a set of realisations, so the expected numbers carry no realisation scatter. It also lists the remnant mass of every merger, which we use. For $six$ of its entries, with $0.002\%$ of the merger rate, the spin fit returns a value slightly above one (at most $1.0005$); they are assigned the edge template like all spins above the grid. The other models of the release, among them updated versions of the three catalogues of (iii), are compared in Appendix~\ref{app:b23}. The observing times are 10~yr for ET-D, CE and the LIGO network, 4~yr for LISA, Taiji and DECIGO, and 5~yr at a 50\% duty cycle for TianQin. These are years of data. For Taiji we take the same four years as for LISA, the longest time for which Ref.~\cite{TaijiConfusion2023} gives the confusion noise. Apart from TianQin no duty cycle is applied, and the LIGO numbers assume that both detectors operate together for the whole time. A duty cycle $d$ multiplies every count by $d$ and lowers the loudest expected SNR only by about $d^{1/3}$. Figure~\ref{fig:wfpop} overlays the populations on the waterfall planes, Table~\ref{tab:rates} lists the expected numbers, and Table~\ref{tab:reformed} compares all scenarios.

Stellar-mass BBHs remain marginal even in the third-generation detectors. With the re-formed surface (median $\phi_0$) their remnants enter the coloured region (face-on SNR $>1$) only at $z<1.8$ in CE and $z<0.55$ in ET-D. In 10~yr we expect $62$ such mergers for CE and $1.6$ for ET-D, and $0.059$ and $3\times10^{-3}$ with face-on SNR above 8. After orientation averaging the population gives $0.075$--$0.15$ (CE) and $3\times10^{-3}$--$5\times10^{-3}$ (ET-D) expected signals with SNR $>5$ in 10~yr, and $0.015$--$0.031$ and $7\times10^{-4}$--$1\times10^{-3}$ with SNR $>8$ (ranges over $\phi_0$). The 5--95\% interval of the population posterior rescales the CE numbers by $0.33$--$2.3$. These counts use $200$ spin draws per mass cell. Three independent draws give $0.025$--$0.034$ for CE at the median $\phi_0$ and SNR $>8$, a sampling scatter of ${\sim}29\%$. A single reflection gives ${\sim}29$ times fewer in CE (Table~\ref{tab:reformed}). The second-generation detectors see nothing, even over 10~yr at design sensitivity: the expected numbers of stellar-mass signals with SNR $>8$ in the LIGO network are $2\times10^{-6}$--$3\times10^{-6}$ (design sensitivity) and $1\times10^{-5}$--$2\times10^{-5}$ (A+) for the re-formed surface and $1\times10^{-7}$ and $8\times10^{-7}$ for a single reflection.

The CE number comes mainly from binaries like GW231123. Of the CE signals with SNR $>5$, $69\%$ have primary masses above $100\,\Msun$ and $83\%$ above $60\,\Msun$ (median $\phi_0$; $69$--$69\%$ above $100\,\Msun$ over $\phi_0$). The detected remnants have $M_f^{\rm src}\simeq63$--$477\,\Msun$ (5--95\%, median $266\,\Msun$) and lie at $z\lesssim1.2$ (median $0.25$). In ET-D, whose reach is smaller, the few expected signals are nearby binaries of ordinary mass, and only $19\%$ have primaries above $100\,\Msun$. In the population fit this part of the mass function rests on one event. GW231123 is the only binary of the fit whose median primary mass lies above $100\,\Msun$ ($126\,\Msun$, 5--95\% interval $112$--$140\,\Msun$, under the population prior); the next heaviest, GW231028, has $81\,\Msun$. The model assigns to primaries above $100\,\Msun$ a rate of $0.013$ ($0.004$--$0.029$)~Gpc$^{-3}$\,yr$^{-1}$ at $z=0.2$, to be compared with the single-event estimate of $0.08$ ($0.01$--$0.27$). Moreover, $30\%$ of the CE signals come from primaries above $200\,\Msun$, heavier than any observed binary component, where the model is the continuation of the power law to its upper limit of $300\,\Msun$. The stellar-mass count in CE should therefore be read mainly as a count of GW231123-like binaries, with the uncertainty of a rate inferred from one event.

Two cross-checks support this reading. First, among the confident LVK events evaluated one by one for a CE interferometer (see below), GW231123 is the only one whose echo would exceed SNR 5. Second, the single-event class (ii) gives $0.057$--$18$ echoes with SNR $>5$ in 10~yr of CE and $0.014$--$3.9$ with SNR $>8$, and $3\times10^{-3}$--$0.53$ and $6\times10^{-4}$--$0.14$ in ET-D (ranges over $\phi_0$; the rate interval rescales them by $0.12$--$3.4$). These numbers lie well above the population ones, for two reasons. The population model gives primaries above $100\,\Msun$ a rate about $6$ times below the median single-event estimate. And class (ii) gives every binary the remnant spin of GW231123, $\chi_f=0.84$, whereas the Default spin model does not depend on mass and gives the heaviest binaries $\chi_f\simeq0.64$ like all others. If every remnant of population (i) had $\chi_f=0.84$, the CE count would be $0.016$--$9.4$ with SNR $>5$ and $0.0037$--$1.8$ with SNR $>8$. At the median $\phi_0$ this is $3.3$, about $26$ times the default. The range over $\phi_0$ is wide because the re-formed gain of this template varies between $0.68$ and $4.61$, and at the least favourable phase the count falls below the default one.

The spin also reconciles these counts with the GW231123-like reference source of Table~\ref{tab:survivor}. The GW231123-like echo has SNR $9.4$ in CE because its remnant has $\chi_f=0.84$. The Default spin model gives a binary of the same mass ratio $\chi_f=0.69$ ($0.61$--$0.77$, 5th to 95th percentile). With the same masses, distance and orientation, the echo then has SNR $2.1$ in CE ($0.78$--$5.6$ over that spin interval), a factor $4.5$ lower. For $\chi_f=0.65$, $0.75$ and $0.80$ it has $1.3$, $4.5$ and $8.9$ (median $\phi_0$). In the population model, binaries as heavy as GW231123 are therefore both rare and more slowly spinning, and a face-on echo at the median spin passes SNR 5 only within about $0.4$ of the distance of that event. One GW231123-like event with an echo of SNR $9.4$ and $0.075$--$0.15$ expected echoes from the population are thus two statements about different spins, not about the same sources. A minimal variant keeps the Default model and gives the remnant spin of GW231123 only to the binaries with $m_1>100\,\Msun$, the range populated by that event alone. CE then expects $0.035$--$7.2$ echoes with SNR $>5$ in 10~yr and $0.0079$--$1.3$ with SNR $>8$ (range over $\phi_0$), with $2.6$ above SNR 5 at the median $\phi_0$, $20$ times the default, and $99\%$ of them from $m_1>100\,\Msun$. Under the same assumption ET-D expects $0.0026$--$0.064$ echoes with SNR $>5$ in 10~yr, still fewer than one, and its loudest echo has $\rho_1=0.75$--$2.2$ instead of $0.79$--$0.97$; for the LIGO network $\rho_1=0.10$--$0.32$ at design sensitivity and $0.19$--$0.52$ at A+. The GWTC-5.0 analysis finds evidence for a rapidly spinning subpopulation at primary masses above ${\sim}45\,\Msun$, consistent with hierarchical mergers \cite{GWTC5pop}, which the Default model does not describe. The default count is therefore probably conservative for the heaviest binaries, and the two variants show how far it could rise.

The massive-BH catalogues contribute little to ET-D and CE, because their lightest members merge at $z>0.3$. Only pop3 gives $0$--$7\times10^{-3}$ in CE. A robust third-generation detection would therefore need either fast-spinning binaries at the top of the stellar mass range or a population of intermediate-mass binaries within the reach of Table~\ref{tab:waterfall}. One CE detection in 10~yr requires a local rate of equal-mass binaries of $0.34$, $0.057$ and $0.062$~Gpc$^{-3}$\,yr$^{-1}$ at $M_f^{\rm src}=300$, $500$ and $1000\,\Msun$. ET-D requires $22$, $0.55$ and $0.15$~Gpc$^{-3}$\,yr$^{-1}$ at $M_f^{\rm src}=300$, $1000$ and $2000\,\Msun$. These rates are for the median $\phi_0$, with the remnant spins of the equal-mass binaries drawn from the spin model (median $\chi_f=0.70$). In LISA the re-formed surface gives $0$--$0.15$ (pop3), $0.084$--$0.79$ (Q3-nodelays) and $0.046$--$0.56$ (Q3-delays) signals with SNR $>8$ in 4~yr, and TianQin $0$--$0.028$, $0$--$0.027$ and $0.013$--$0.080$. With SNR $>5$ the LISA numbers are $0$--$0.20$, $0.50$--$2.8$ and $0.19$--$1.7$, and the TianQin ones $0$--$0.074$, $0.041$--$0.34$ and $0.058$--$0.42$. The PTA-calibrated model gives $0.077$--$0.37$ in LISA and $0.015$--$0.063$ in TianQin with SNR $>8$, and $0.37$--$1.8$ and $0.045$--$0.21$ with SNR $>5$. Taiji gives $0$--$0.19$ (pop3), $0.53$--$3.0$ (Q3-nodelays), $0.22$--$2.1$ (Q3-delays) and $0.45$--$2.0$ (PTA-calibrated model) signals with SNR $>8$ in 4~yr, and $2\times10^{-3}$--$0.34$, $1.8$--$9.0$, $0.62$--$4.9$ and $2.1$--$9.4$ with SNR $>5$. For the last three models these are $3$--$5$ times the LISA numbers at the median $\phi_0$, because above about 5~mHz, where the lines of the lighter remnants lie, the Taiji noise amplitude is about half that of LISA.

DECIGO is the one detector with large expected numbers. In a 4-yr mission, $76$ pop3, $357$ Q3-nodelays and $24$ Q3-delays mergers fall in the coloured region (Fig.~\ref{fig:wfpop}f). We expect $0.30$--$9.4$ (pop3), $43$--$100$ (Q3-nodelays) and $2.7$--$14$ (Q3-delays) signals with SNR $>8$ in 4~yr, and $0.49$--$14$, $67$--$150$ and $4.6$--$19$ with SNR $>5$, where the ranges are over $\phi_0$. Of the pop3 detections, 0.27--3.3 come from remnants inside the template spin range (see below). Over the ten catalogue realisations the counts are 0--13, 39--120 and 0.55--20. These numbers are 3.5--5.0 times the single-reflection ones. The detections come from remnants of $M_f^{\rm src}\simeq1\times10^{4}$--$1\times10^{6}\,\Msun$ (5--95\% of the detection-weighted distribution) at $z\simeq0.4$--$12$ (medians $1.2$--$5.3$), with $\chi_f\simeq0.73$--$0.994$. The PTA-calibrated model gives far more: $300$--$570$ signals with SNR $>8$ and $540$--$900$ with SNR $>5$ ($100$ with SNR $>8$ for a single reflection). It has $19$ times as many mergers as Q3-nodelays, and most of them lie in the DECIGO band: its detections come from remnants of $M_f^{\rm src}\simeq8\times10^{3}$--$2\times10^{5}\,\Msun$ at $z\simeq1.9$--$16$ (median $7.3$) with $\chi_f\simeq0.69$--$0.87$. The spread between the four models, more than two orders of magnitude, is that of the massive-BH merger rate itself (Appendix~\ref{app:b23}).

\paragraph*{DECIGO above 0.2~Hz.} The DECIGO curve used here \cite{DECIGO2011} is the instrumental noise and contains no astrophysical foreground. Below about 0.2~Hz the gravitational waves of white-dwarf binaries may form a confusion foreground, and Ref.~\cite{DECIGO2011} starts its signal integrals at 0.2~Hz for this reason. The echo is a narrow line at the observed frequency $f_H=m\Omega_H/2\pi$ (detector-frame mass), so a signal lies either above this frequency or below it. If we count only the catalogue signals with $f_H\ge0.2$~Hz and leave their SNRs unchanged, the numbers with SNR $>5$ in 4~yr become $1\times10^{-3}$--$5.1$ (pop3), $3.8$--$7.0$ (Q3-nodelays), $0.25$--$1.0$ (Q3-delays) and $60$--$94$ (PTA-calibrated model), and those with SNR $>8$ become $0$--$2.7$, $2.7$--$5.1$, $0.16$--$0.82$ and $35$--$61$ (rows ``$f_H>0.2$~Hz'' of Table~\ref{tab:rates}). At the median $\phi_0$ this is $5$--$22\%$ of the full counts with SNR $>5$, and $10\%$ for the PTA-calibrated model. Most of the DECIGO forecast therefore comes from the band in which the foreground would first have to be removed. This is a sensitivity scenario, not a property of the detector: a full treatment would add the residual foreground to the noise and repeat the SNR integral.

\begin{table*}
\caption{Expected numbers of signals with SNR $>8$ in all scenarios, over the observing times of Table~\ref{tab:rates} (orientation-averaged, optimal SNR, $\epsilon_E=1$ except for the capped $\chi_f=0.955$ template, Boltzmann surface). Single: the surface does not re-form. Re-formed ($N\to\infty$, range over a common $\phi_0$): with the remnant spin of every source (default); with the width of the backreaction filter, $k_c\to0.5k_c$ and $2k_c$ (median $\phi_0$); with every SNR multiplied by $0.82$, the wide-band normalisation of Appendix~\ref{app:setup} (median $\phi_0$); without the $\chi_f=0.90$ template in the spin grid (median $\phi_0$); and with every source given $\chi_f=0.7268$ (spin-blind, median $\phi_0$). Last column: largest redshift at which SNR 8 is reached for an optimally oriented $\chi_f=0.7268$ source (any mass).}
\label{tab:reformed}
\begin{ruledtabular}
\footnotesize\setlength{\tabcolsep}{2.5pt}
\begin{tabular}{llcccccc}
detector & scenario & stellar BBH & pop3 & Q3-nodelays & Q3-delays & HS-nod (B+23) & $z_8$\\
\hline
ET-D & single & $5\times10^{-5}$ & $0$ & $0$ & $0$ & $9\times10^{-4}$ & 0.13\\
 & re-formed & $7\times10^{-4}$--$1\times10^{-3}$ & $0$ & $0$ & $0$ & $2\times10^{-3}$ & 0.30--0.35\\
 & $k_c\to0.5k_c$ & $5\times10^{-5}$ & $0$ & $0$ & $0$ & $9\times10^{-4}$ & 0.13\\
 & $k_c\to2k_c$ & $0.023$ & $0$ & $0$ & $0$ & $4\times10^{-3}$ & 0.74\\
 & wide-band norm. & $6\times10^{-4}$ & $0$ & $0$ & $0$ & $2\times10^{-3}$ & 0.27\\
 & no $\chi_f=0.90$ template & $1\times10^{-3}$ & $0$ & $0$ & $0$ & $2\times10^{-3}$ & 0.33\\
 & all $\chi_f=0.73$ & $9\times10^{-4}$ & $0$ & $0$ & $0$ & $3\times10^{-4}$ & 0.33\\
\hline
CE & single & $9\times10^{-4}$ & $0$ & $0$ & $0$ & $0$ & 0.20\\
 & re-formed & $0.015$--$0.031$ & $0$--$7\times10^{-3}$ & $0$ & $0$ & $0$ & 0.45--0.51\\
 & $k_c\to0.5k_c$ & $9\times10^{-4}$ & $0$ & $0$ & $0$ & $0$ & 0.20\\
 & $k_c\to2k_c$ & $0.75$ & $0.020$ & $0$ & $0$ & $0$ & 1.1\\
 & wide-band norm. & $0.013$ & $0$ & $0$ & $0$ & $0$ & 0.41\\
 & no $\chi_f=0.90$ template & $0.025$ & $0$ & $0$ & $0$ & $0$ & 0.48\\
 & all $\chi_f=0.73$ & $0.018$ & $0$ & $0$ & $0$ & $0$ & 0.48\\
\hline
LISA & single & $0$ & $0$ & $0$ & $0$ & $5\times10^{-3}$ & 1.0\\
 & re-formed & $0$ & $0$--$0.15$ & $0.084$--$0.79$ & $0.046$--$0.56$ & $0.077$--$0.37$ & 2.2--2.5\\
 & $k_c\to0.5k_c$ & $0$ & $0$ & $0$ & $0$ & $8\times10^{-3}$ & 1.0\\
 & $k_c\to2k_c$ & $0$ & $0.21$ & $7.3$ & $3.2$ & $7.5$ & 5.5\\
 & wide-band norm. & $0$ & $0.036$ & $0.20$ & $0.18$ & $0.13$ & 2.0\\
 & no $\chi_f=0.90$ template & $0$ & $0.046$ & $0.42$ & $0.31$ & $0.24$ & 2.4\\
 & all $\chi_f=0.73$ & $0$ & $0.16$ & $0.020$ & $0.35$ & $0.051$ & 2.4\\
\hline
Taiji & single & $0$ & $0$ & $9\times10^{-3}$ & $0.019$ & $0.035$ & 1.3\\
 & re-formed & $0$ & $0$--$0.19$ & $0.53$--$3.0$ & $0.22$--$2.1$ & $0.45$--$2.0$ & 3.0--3.2\\
 & $k_c\to0.5k_c$ & $0$ & $0$ & $0.035$ & $0.052$ & $0.053$ & 1.4\\
 & $k_c\to2k_c$ & $0$ & $0.27$ & $20$ & $5.8$ & $31$ & 7.2\\
 & wide-band norm. & $0$ & $0.022$ & $1.4$ & $1.1$ & $0.74$ & 2.7\\
 & no $\chi_f=0.90$ template & $0$ & $0.032$ & $2.0$ & $1.2$ & $1.3$ & 3.1\\
 & all $\chi_f=0.73$ & $0$ & $0.25$ & $0.18$ & $1.5$ & $0.34$ & 3.1\\
\hline
TianQin & single & $0$ & $0$ & $0$ & $0$ & $1\times10^{-3}$ & 0.43\\
 & re-formed & $0$ & $0$--$0.028$ & $0$--$0.027$ & $0.013$--$0.080$ & $0.015$--$0.063$ & 0.95--1.1\\
 & $k_c\to0.5k_c$ & $0$ & $0$ & $0$ & $0$ & $2\times10^{-3}$ & 0.44\\
 & $k_c\to2k_c$ & $0$ & $0.034$ & $1.2$ & $0.93$ & $0.75$ & 2.3\\
 & wide-band norm. & $0$ & $0$ & $0$ & $0.013$ & $0.025$ & 0.87\\
 & no $\chi_f=0.90$ template & $0$ & $0$ & $5\times10^{-3}$ & $0.030$ & $0.038$ & 1.0\\
 & all $\chi_f=0.73$ & $0$ & $0.045$ & $0$ & $0.015$ & $0.014$ & 1.0\\
\hline
DECIGO & single & $6\times10^{-10}$ & $0.49$ & $25$ & $2.5$ & $100$ & 5.8\\
 & re-formed & $9\times10^{-9}$--$1\times10^{-8}$ & $0.30$--$9.4$ & $43$--$100$ & $2.7$--$14$ & $300$--$570$ & 13--15\\
 & $k_c\to0.5k_c$ & $6\times10^{-10}$ & $0.71$ & $32$ & $3.8$ & $130$ & 5.9\\
 & $k_c\to2k_c$ & $3\times10^{-7}$ & $8.9$ & $180$ & $16$ & $1300$ & $\ge20$\\
 & wide-band norm. & $7\times10^{-9}$ & $1.9$ & $76$ & $7.8$ & $410$ & 12\\
 & no $\chi_f=0.90$ template & $1\times10^{-8}$ & $2.4$ & $81$ & $7.2$ & $480$ & 14\\
 & all $\chi_f=0.73$ & $1\times10^{-8}$ & $14$ & $58$ & $8.7$ & $530$ & 14\\
\end{tabular}
\end{ruledtabular}
\end{table*}

\paragraph*{Loudest expected signal.} Where fewer than one signal above the threshold is expected, a more useful number is the loudest echo that the observing time is expected to contain. We define it as the SNR $\rho_1$ for which the expected number of signals above $\rho_1$ is one, for each population (last block of Table~\ref{tab:rates}). For the stellar-mass population it is $0.11$--$0.13$ in the LIGO network at design sensitivity and $0.20$--$0.25$ at A+ sensitivity in 10~yr, $0.79$--$0.97$ in ET-D and $2.3$--$2.9$ in CE. The single-event GW231123-like class alone would give $2.0$--$12$ in CE. For TianQin $\rho_1$ is $<0.04$--$0.56$ (pop3), $1.9$--$3.5$ (Q3-nodelays) and $0.97$--$3.4$ (Q3-delays), for LISA $0.11$--$1.6$, $3.8$--$7.4$ and $2.3$--$6.4$, for Taiji $0.19$--$2.5$, $6.2$--$13$ and $3.8$--$12$, and for DECIGO $3.2$--$47$, $88$--$190$ and $20$--$61$. For the PTA-calibrated model it is $1.9$--$3.0$ in TianQin, $3.8$--$5.9$ in LISA, $6.2$--$10$ in Taiji and $140$--$270$ in DECIGO. The ranges are again over a common $\phi_0$.

\begin{figure*}
\includegraphics[width=\textwidth]{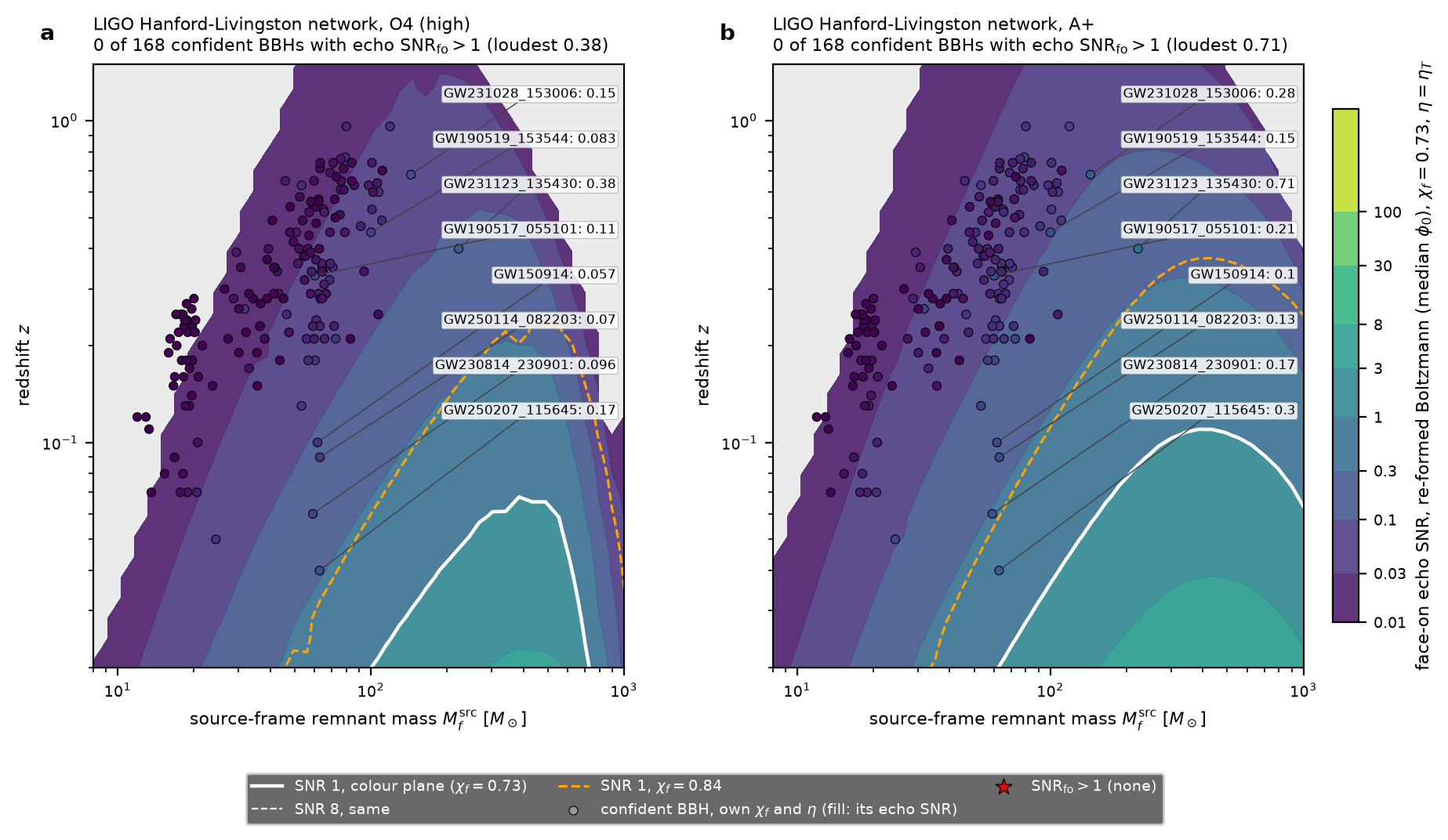}
\caption{Confident LVK binary black holes (false-alarm rate $<10^{-3}$~yr$^{-1}$, $168$ events, median catalogue parameters) on the plane of source-frame remnant mass and redshift for the LIGO Hanford--Livingston network at O4 (a) and A+ (b) sensitivity. Colour: optimal face-on network SNR of the re-formed Boltzmann echo (median $\phi_0$) for the $\chi_f=0.7268$ signal shape and $\eta=\eta_T$, with its SNR-1 contour (white); orange dashed: SNR 1 for $\chi_f=0.84$. Circles: the detected events, filled with their own echo SNR (own $\chi_f$ and $\eta$); the eight loudest are labelled with their SNR. No event exceeds SNR 1 (red stars would mark such events). $\epsilon_E=1$, adopted backreaction filter.}
\label{fig:gwtc}
\end{figure*}

\paragraph*{Effect of the remnant spin.} Assigning every source the remnant spin of the GW150914-like source, $\chi_f=0.7268$ (for the massive-BH catalogues with the generic template of that spin), as a spin-blind estimate would do, changes these numbers (last row of each block of Table~\ref{tab:reformed}; we compare counts at the median $\phi_0$). For stellar-mass BBHs, whose remnant spins cluster around $0.64$, the spin-blind CE count is $0.7$ times the spin-aware one. For the massive-BH catalogues, whose remnants spin faster, the spin-blind counts are $0.6$--$5.6$ times the spin-aware ones in DECIGO and $0.04$--$2.6$ times in LISA. For the PTA-calibrated model, whose remnants spin more slowly, the spin-blind DECIGO count is $1.0$ times the spin-aware one. The spin-aware counts of the catalogues dominated by fast-spinning remnants also vary much more with $\phi_0$ than the spin-blind ones (LISA pop3: $0$--$0.15$ against $0.16$), because the re-formed gain of the fastest templates does. The fastest remnants are not the loudest: the holes that absorbed their packets were close to extremal, their layers are thick and their packet energy is capped (Table~\ref{tab:progenitor}), so the spin-blind estimate is not simply a lower bound. The source-dependent spin is therefore an essential ingredient of the expected numbers.

\paragraph*{Confident LVK detections.} The populations above are models. As a direct check we place the $168$ binary black holes detected by the LVK network with a false-alarm rate below $10^{-3}$~yr$^{-1}$ on the second-generation planes (Fig.~\ref{fig:gwtc}). We use the median source-frame masses, effective spin, redshift and final mass of each event from the most recent catalogue that contains it (GWTC-2.1, GWTC-3, GWTC-4.0 and GWTC-5.0, as released by GWOSC \cite{GWTC1,GWTC21,GWTC3,GWTC4cat,GWTC5,GWOSC2023}), and give each event its own remnant spin and symmetric mass ratio; the calculation uses the detector-frame mass $(1+z)M_f^{\rm src}$ of each event. The released parameters contain no final spin, so we estimate it from the aligned-spin fit of Ref.~\cite{HBR2016} with both component spins equal to $\chi_{\rm eff}$. For GW231123, whose large in-plane spins this misses, we use the published $\chi_f=0.84$ \cite{GW231123}. No event reaches a face-on echo SNR of 1 in the LIGO Hanford--Livingston network at O4 or at A+ sensitivity. The loudest is GW231123, with $0.38$ in O4 and $0.71$ in A+. Next come GW250207\_115645 and GW231028\_153006 ($0.30$ and $0.28$ in A+). In A+, only one event (GW231123) exceeds $0.3$ and 10 events exceed $0.1$ (1 and 4 in O4). The loudest IMR signal, GW250114, gives $0.13$, because its remnant spin is only $0.68$. The same events can be evaluated for a single third-generation interferometer, as if it had observed them. In CE, GW231123 is the only event whose echo would exceed SNR 5: it reaches $9.2$ ($2.0$--$12$ over $\phi_0$). The next loudest, GW231028\_153006 and GW250207\_115645, reach $3.3$ and $2.3$, and $8$ events exceed SNR 1. In ET-D the loudest is again GW231123, with $2.8$, and $3$ events exceed SNR 1. A detection of an echo from the current catalogue would therefore require the third-generation detectors, and even for them it rests on one event.

These numbers should be read with care. The counts grow steeply with the amplitude, as $\rho^3$ for a Euclidean volume and more slowly at high redshift. The order-unity amplitude uncertainties of Appendix~\ref{app:robust} (shape and width of the backreaction filter, static background, $\epsilon_E$) therefore translate into uncertainties of a few in the DECIGO numbers. For example, $k_c\to0.5k_c$ and $2k_c$ (with the re-formed gain held at its nominal value) change the median DECIGO counts from $2.5$, $91$ and $8.8$ to $0.71$--$8.9$, $32$--$180$ and $3.8$--$16$ (pop3, Q3-nodelays, Q3-delays), a factor $2$--$4$. The CE counts of stellar-mass BBHs change by a factor ${\approx}28$ (the Euclidean $\rho^3$ scaling would give $20$). The spin interpolation between templates adds a smaller uncertainty, but the clamping at the upper end of the template range ($\chi_f=0.955$) does not. The massive-BH remnants spin fast, and the share of the expected detections that comes from remnants above $0.955$, evaluated at the edge template, is 0\% (pop3), 0\% (Q3-nodelays) and 0\% (Q3-delays) for LISA, 0--23\%, 0\% and 0--8\% for Taiji, 0\%, 0\% and 0\% for TianQin, and 8--65\%, 0--3\% and 0--11\% for DECIGO (ranges over $\phi_0$). The detection-weighted median spin of the DECIGO pop3 sources is $0.956$, and the 95th percentile is $0.994$. The ground-based detectors are not affected: pop3 gives no signal with SNR $>8$ in ET-D and at most $0.0067$ in CE, none of it from remnants above $0.955$. Within the validated spin range the DECIGO pop3 count is 0.27--3.3 and the LISA one 0--0.15 (Taiji: 0--0.14). The rest is an extrapolation whose sign is not guaranteed, since the SNR need not keep growing with spin once the threshold alignment, the temperature and the detector band all change. Even inside the grid the interpolation matters. Dropping the $\chi_f=0.90$ template and interpolating across the gap changes the median Q3-nodelays counts from $0.49$ to $0.42$ in LISA and from $0.0060$ to $0.0046$ in TianQin (Table~\ref{tab:reformed}). The space-detector counts use sky- and polarisation-averaged noise curves with the inclination drawn per source. Averaging the response before the threshold is an approximation for threshold counts, and a short echo does not average over the orbit. The ranges over $\phi_0$ are the extremes of the counts over a phase common to all sources. The median over $\phi_0$ is not a phase-marginalised expectation; the mean is (see above), and it differs from the median by up to a factor $1.5$. The scaling of each source to the templates ($\rho\propto\eta$, a non-spinning mass-loss fit, interpolation in $\chi_f$) is a population surrogate that does not capture precession or large mass ratios. Finally, the counts are optimal matched-filter estimates with nominal thresholds. The threshold of 5 assumes a search triggered by the merger with few independent templates. A search that must scan wide ranges of the echo time, the Boltzmann reflectivity phase and the in-band phase needs a higher threshold, and the SNR $>8$ numbers are then the safer reference. Both sets of counts are therefore optimistic.

\paragraph*{Relation to current echo searches.} Boltzmann-echo searches quote an amplitude $A$ relative to the canonical unfiltered model, and the combined GWTC-3 analysis gives $A<0.4$ \cite{Abedi2023,AbediGW190521}. Measured by SNR in the same detector, the survivor, the echo that survives the effective horizon backreaction, carries only $0.2$--$1.7\%$ of the unfiltered Boltzmann train of our pipeline, and the re-formed signal roughly $0.5$--$5\%$, one to two orders of magnitude below that bound. This is a ratio of optimal SNRs, not a recast of the likelihood bound: constraining $A$ for the surviving waveform would need an injection study with that template. The waveform is also different: a narrow line at $f_H$ with a central dip, or a few cavity lines within $\pm k_c$ of it, rather than a train of broadband pulses. Current non-detections therefore do not test this scenario, and templates built on the unfiltered train would not be matched to it. Model-agnostic searches for long-lived QNMs \cite{WuGaoRenAfshordi2023,WuZhangHuangRen2026} and for comb-like spectra \cite{ConklinHoldomRen2018,RenWu2021} are closer to the re-formed signal, provided they extend down to the narrow band around $f_H$.

\section{Other multipoles}\label{sec:modes}
So far we have used only the dominant $\ell=m=2$ mode, whose escape band sits at $k=\omega-2\Omega_H$. Every other $(\ell,m)$ mode of the ingoing ringdown has its own superradiant threshold, $\omega=m\Omega_H$, and a backreaction filter centred there picks out a different part of its spectrum. For $m=0$ the threshold is at $\omega=0$. We therefore repeat the calculation mode by mode. The source modes come from the NR surrogate NRHybSur3dq8 \cite{Varma2019NRHybSur}, which provides the modes with $\ell\le4$ except $(4,1)$ and $(4,0)$, and the $(5,5)$ mode, for the parameters of our source. Its $m=0$ modes are therefore $(2,0)$ and $(3,0)$. These are the least accurate modes of the surrogate and do not contain the non-oscillatory memory. Its $(2,2)$ mode has the same peak strain as IMRPhenomT, and its $(3,3)$ and $(4,4)$ amplitudes agree with IMRPhenomTHM \cite{IMRPhenomTHM} to $1\%$. (For this aligned-spin source the precessing models IMRPhenomTPHM and IMRPhenomXPHM reduce to their aligned-spin versions, which have no $m=0$ mode.)

For each of the $11$ modes with $m\ge0$ we run the full chain (the $-m$ partners are mirror images). The chain consists of a time window centred on the peak of that mode, the $1{+}1$ Teukolsky evolution with the $(\ell,m)$ spheroidal eigenvalue, exact barrier data for that $(\ell,m)$ and the gross-energy normalisation $\epsilon_E=1$ for that mode. It is followed by the backreaction filter $S(k)$ with $k=\omega-m\Omega_H$ and the common width $k_c$, the Boltzmann reflectivity $e^{-|k|/2T_H}$, and the re-formed surface (\code{RMD} and \code{MSV}~\cite{EchoRepo}). A common $k_c$ is justified because the subdominant modes add only ${\approx}2\%$ to the absorbed energy. For $m=0$ the band $|\omega|<1.3k_c$ requires barrier data down to $M_f\omega=5\times10^{-4}$, which we obtain by moving the outer boundary to $40/\omega$; energy conservation, $|\RBH|^2+\Gamma=1$, holds there to $2\times10^{-4}$. This mode-by-mode calculation is done on the remnant, with the amplitude of each mode fixed by its gross flux onto the hole before absorption of the $(2,2)$ calculation (Sec.~\ref{sec:backreaction}), and we use it for the ratios between modes. As a check, the NRHybSur3dq8 $(2,2)$ mode reproduces the IMRPhenomT survivor: SNR $6.23$ against $6.21$ (CE, $300\,\Msun$) and $151$ against $150$ (DECIGO, $10^5\,\Msun$).

\begin{figure*}
\includegraphics[width=\textwidth]{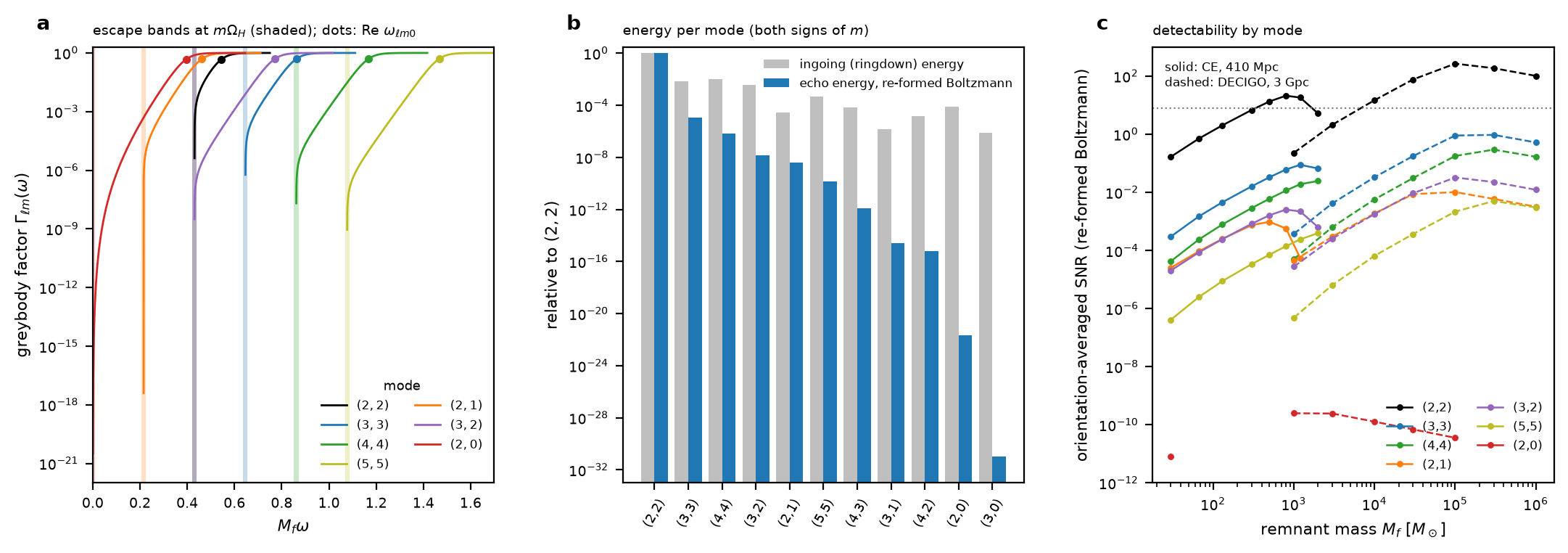}
\caption{Other multipoles (NRHybSur3dq8 source). (a) Greybody factors $\Gamma_{\ell m}(\omega)$ of the Kerr remnant ($\chi_f=0.7268$) with the escape band $|\omega-m\Omega_H|<1.3k_c$ of each mode (shaded; for $(2,0)$ at $\omega\approx0$, too narrow to see) and the real part of the $(\ell,m,0)$ QNM frequency (dots); $\Gamma<0$ (superradiant amplification) below $m\Omega_H$ is not shown. (b) Ringdown energy (grey) and echo energy of the re-formed Boltzmann signal (blue) of each mode relative to $(2,2)$. (c) Orientation-averaged SNR of the re-formed Boltzmann signal versus remnant mass in CE (410~Mpc, solid) and DECIGO (3~Gpc, dashed) for the main modes; the $(2,0)$ mode enters the CE band only at the lowest mass.}
\label{fig:modes}
\end{figure*}

The face-on harmonic $_{-2}Y_{\ell m}(0)$ vanishes for $m\neq2$, so we compare the modes through the orientation-averaged SNR, to which each mode group $(\ell,|m|)$ contributes in quadrature (Table~\ref{tab:modes}, Fig.~\ref{fig:modes}). The $(2,2)$ mode carries all but at most $1.8\times10^{-4}$ of the averaged SNR$^2$, both of the survivor and of the re-formed signal, in all eleven detectors and for remnant masses from $30$ to $2000\,\Msun$ (ground) and $10^3$ to $3\times10^7\,\Msun$ (space). The other modes do best in CE at $2000\,\Msun$, where $f_H$ of the $(2,2)$ mode drops to the low-frequency edge of the detector. The next mode, $(3,3)$, contributes $5.5\times10^{-6}$ (CE, $300\,\Msun$) and $1.1\times10^{-5}$ (DECIGO, $10^5\,\Msun$), and $(4,4)$ contributes $1.8\times10^{-7}$ and $4.4\times10^{-7}$. Face-on, where only $m=2$ radiates, adding $(3,2)$ and $(4,2)$ coherently to $(2,2)$ changes the SNR by less than $1.1\times10^{-4}$. At any inclination the other modes carry at most $4.1\times10^{-4}$ of SNR$^2$ (edge-on, CE, $2000\,\Msun$), so they would change the expected numbers of Sec.~\ref{sec:detect} by less than $6.2\times10^{-4}$; the detectability figures and tables therefore use the $(2,2)$ mode alone.

This is far below the share of the higher modes in the IMR signal and in the ringdown energy [Fig.~\ref{fig:modes}(b)]. The $(3,3)$ and $(4,4)$ modes carry $7.1\times10^{-3}$ and $1.1\times10^{-2}$ of the $(2,2)$ ringdown energy, but only $1.2\times10^{-5}$ and $6.7\times10^{-7}$ of its echo energy. The barrier is responsible. For every mode the escape band lies below the peak of its own potential, at a distance that grows with $\ell$ in units of the damping rate: $(\omega_R-m\Omega_H)/|\omega_I|=1.43$, $2.64$ and $3.69$ for $(2,2)$, $(3,3)$ and $(4,4)$. Transmission through the barrier is a tunnelling factor in this parameter, and $\Gamma$ at the band edge is smaller than for $(2,2)$ by $2.1\times10^{-2}$ and $7.1\times10^{-4}$ [Fig.~\ref{fig:modes}(a)]. Modes with $m<\ell$ have their thresholds even further below the barrier and are suppressed more.

The $m=0$ modes are suppressed most of all. Their escape band is centred on $\omega=0$, where there is no superradiance, the greybody factor takes its low-frequency form $\Gamma_{\ell0}\propto(M_f\omega)^{2\ell+2}$ ($\Gamma_{20}=2.0\times10^{-13}$ at the band edge), and the ingoing $\psi_4$ spectrum of the ringdown is itself small. The $(2,0)$ echo carries $2.2\times10^{-22}$ of the $(2,2)$ echo energy, and $(3,0)$ carries $1.1\times10^{-31}$. On top of this, the band lies below $1.3f_c=4.7$~Hz $(67.21\,\Msun/M_f)$, at or below the low-frequency edge of the ground-based detectors even for the lightest remnants. The largest $(2,0)$ SNR in any of the eleven detectors is $3.9\times10^{-10}$ (DECIGO, $1000\,\Msun$, where the $(2,2)$ signal has SNR $0.36$). The nonlinear memory is not included. Its power lies at frequencies where $\Gamma_{20}$ vanishes as $\omega^6$, so we expect it to be suppressed as well, but a bound would need a separate calculation. Within the model the echo that survives the effective horizon backreaction is therefore an $\ell=m=2$ signal to within $1.8\times10^{-4}$ in SNR$^2$, for this near-equal-mass, moderately spinning source and the detector and mass rescalings tested. This share is built from the per-mode median SNRs, and taking the share at each common Boltzmann reflectivity phase before the maximum gives $2.1\times10^{-4}$. Each spherical-harmonic mode of the surrogate is propagated in the radial channel of the same $(\ell,m)$. The spherical--spheroidal mixing \cite{BertiKlein2014}, of order $0.06$ in amplitude between the $(2,2)$ spheroidal and $(3,2)$ spherical states at $a\omega\simeq0.3$, is not included and would redistribute a comparable fraction of the amplitude between same-$m$ channels. For unequal-mass, rapidly spinning or precessing sources, whose higher modes are stronger and whose corotating thresholds move with the spin, the mode study would have to be repeated.

\begin{table*}
\caption{Echo contribution of the individual $(\ell,m)$ modes of the ingoing ringdown (NRHybSur3dq8 source, evaluated on the remnant, $\epsilon_E=1$ for each mode, backreaction filter centred on $m\Omega_H$, common $k_c$). $f_{\rm thr}=m\Omega_H/2\pi$ of the remnant, $M_f=67.21\,\Msun$ and $\Omega_H/2\pi=103.6$~Hz; $(\omega_R-m\Omega_H)/|\omega_I|$: distance of the threshold below the $(\ell,m,0)$ QNM frequency in units of its damping rate; $E_{\rm RD}$: ringdown energy of the mode (both signs of $m$) relative to $(2,2)$; $\Gamma_{\rm edge}$: greybody factor at the edge of the escape band, $|k|=1.3k_c$ ($\omega=1.3k_c$ for $m=0$); $E_{\rm echo}$: energy of the survivor (single reflection) and of the re-formed Boltzmann signal (median over $\phi_0$) relative to $(2,2)$; last column: share of the orientation-averaged SNR$^2$ of the re-formed Boltzmann signal in DECIGO ($10^5\,\Msun$, 3~Gpc).}
\label{tab:modes}
\begin{ruledtabular}
\footnotesize
\begin{tabular}{lccccccc}
mode & $f_{\rm thr}$ [Hz] & $(\omega_R-m\Omega_H)/|\omega_I|$ & $E_{\rm RD}/E_{\rm RD}^{22}$ & $\Gamma_{\rm edge}$ & $E_{\rm echo}/E_{\rm echo}^{22}$ (survivor) & $E_{\rm echo}/E_{\rm echo}^{22}$ (re-formed) & share of $\langle\rho^2\rangle$\\
\hline
$(2,2)$ & 207 & 1.43 & $1.0$ & $3.5\times10^{-3}$ & $1.0$ & $1.0$ & $1.0$\\
$(3,3)$ & 311 & 2.64 & $7.1\times10^{-3}$ & $7.3\times10^{-5}$ & $1.2\times10^{-5}$ & $1.2\times10^{-5}$ & $1.1\times10^{-5}$\\
$(4,4)$ & 414 & 3.69 & $1.1\times10^{-2}$ & $2.5\times10^{-6}$ & $6.6\times10^{-7}$ & $6.7\times10^{-7}$ & $4.4\times10^{-7}$\\
$(3,2)$ & 207 & 4.09 & $3.6\times10^{-3}$ & $4.1\times10^{-7}$ & $1.5\times10^{-8}$ & $1.5\times10^{-8}$ & $1.5\times10^{-8}$\\
$(2,1)$ & 104 & 3.03 & $2.8\times10^{-5}$ & $9.6\times10^{-6}$ & $3.9\times10^{-9}$ & $4.0\times10^{-9}$ & $1.5\times10^{-9}$\\
$(5,5)$ & 518 & 4.65 & $4.7\times10^{-4}$ & $1.1\times10^{-7}$ & $1.4\times10^{-10}$ & $1.4\times10^{-10}$ & $6.5\times10^{-11}$\\
$(4,3)$ & 311 & 5.05 & $6.6\times10^{-5}$ & $2.0\times10^{-8}$ & $1.2\times10^{-12}$ & $1.2\times10^{-12}$ & $1.2\times10^{-12}$\\
$(3,1)$ & 104 & 5.60 & $1.5\times10^{-6}$ & $4.1\times10^{-10}$ & $2.5\times10^{-15}$ & $2.5\times10^{-15}$ & $9.3\times10^{-16}$\\
$(4,2)$ & 207 & 6.47 & $1.5\times10^{-5}$ & $6.3\times10^{-11}$ & $6.2\times10^{-16}$ & $6.2\times10^{-16}$ & $6.2\times10^{-16}$\\
$(2,0)$ & 0 & 4.72 & $7.6\times10^{-5}$ & $2.0\times10^{-13}$ & $4.1\times10^{-22}$ & $2.2\times10^{-22}$ & $1.8\times10^{-26}$\\
$(3,0)$ & 0 & 7.17 & $7.4\times10^{-7}$ & $2.0\times10^{-20}$ & $2.2\times10^{-31}$ & $1.1\times10^{-31}$ & $7.6\times10^{-36}$\\
\end{tabular}
\end{ruledtabular}
\end{table*}

\section{Conclusions}\label{sec:conclusions}
(1) We have formulated a phenomenological model in which the backreaction of the absorbed post-merger energy filters the first reflection from a near-horizon Boltzmann surface down to a band $|\omega-m\Omega_H|\lesssim k_c=\pi/L_*$, with $L_*$ obtained from an equal-area mapping of the backreacted horizon. For a GW150914-like remnant $L_*=418.5M_f$, and it grows logarithmically with mass. The model does not answer the causal objection of Guo and Mathur. It says what would be observed if a band near the superradiant threshold did escape. In that case the surviving first reflection is a narrow packet at the horizon frequency, with half-width at half-maximum $f_c=3.61\,{\rm Hz}\,(67.21\,\Msun/M_f)$ (backreaction-filter support $\pm1.3f_c$). With the packet normalised to a gross absorbed energy equal to the ringdown energy, its amplitude is $1.21\times10^{-4}$ of the merger peak.

(2) Inside the escape band the Boltzmann reflectivity is $\ge0.861$, so the survivor is set by the barrier transmission and by the part of the ingoing packet near threshold. The amplitude therefore depends on how close the ringdown frequency is to $m\Omega_H$. For a GW231123-like remnant ($\chi_f=0.84$, $M_f\omega_{220}=0.615$ against $mM_f\Omega_H=0.545$) $6.4\%$ of the field power of the packet lies in the band, against $5.4\%$ for GW150914. In gross horizon energy, which vanishes at the threshold, the shares are $0.23\%$ and $0.14\%$. The survivor is $1.8$ times stronger relative to the merger. Most of this factor comes from the barrier: at the higher spin the transmission rises $3.8$ times more steeply from its zero [Fig.~\ref{fig:dip}(a)]. The packet is $1.9$ times stronger at the threshold, because the ringdown carries more energy and lies closer to the threshold, and the narrower band and lower frequency take back a factor $0.25$. The remnant mass and spin already contain the packet, so the hole that absorbs it is the progenitor of the remnant. For the GW150914-like source it is lighter by $1.77\%$ and has $\chi_i=0.689$. For the GW231123-like source it spins faster than the remnant ($\chi_i=0.919$). The packet lies below its threshold, extracts energy and angular momentum (net flux $-0.986$ of the gross) and spins it down to the observed remnant, while its area grows. For $\chi_f\gtrsim0.95$ a sub-extremal progenitor exists only for a packet energy below the ringdown energy ($\epsilon_E\le0.45$ at $\chi_f=0.955$).

(3) Because the transmission vanishes at the superradiant threshold, the survivor is the time derivative of a smooth packet. With an even, zero-phase filter its envelope has a central dip at the geometric echo time, with hump separation $2x_1L_*/\pi=530M_f$ ($0.175$~s; $L_*=418.5M_f$) for the adopted backreaction filter. A minimum-phase (causal-response) filter of the same width delays the dip by $384M_f$ and makes the humps unequal, and a quadratic in-band phase of order $\pi$ fills it in. The dip belongs to the first echo and survives the re-formation of the surface, whose cavity ringing starts one round trip later. It is a prediction of the backreaction-filter model: its existence is tied to the Kerr transmission zero, its timing and shape to the unknown escape response. A double-peaked echo with the same origin, the zero of the transfer function at $\omega=m\Omega_H$, was found for unfiltered Boltzmann echoes in Ref.~\cite{LongoMicchi2021}.

(4) With the re-formed surface, our default, stellar-mass remnants at the distances of the observed events stay below threshold in the second-generation ground-based detectors, and only the heaviest of them reach it in CE. None of the $168$ confident LVK binary black holes would give an echo with SNR above 1 even at A+ sensitivity. The loudest, GW231123, reaches $0.71$ (Fig.~\ref{fig:gwtc}). At 410~Mpc the optimal face-on SNR of the GW150914-like signal is $0.11$--$0.13$ in the LIGO Hanford--Livingston network (design and O4 sensitivity) and $1.8$ in CE ($0.66$ for a single reflection). The GW231123-like source at $2.2$~Gpc gives $0.40$ in the LIGO network at O4 sensitivity and $9.4$ in CE. Scaled to the observed network SNR, an echo of GW231123 would have SNR ${\approx}0.097$ in the O4 network. (We use 20.7, the value of the published version of Ref.~\cite{GW231123}.) Normalising the packet to its net rather than its gross absorbed energy is consistent with the measured remnant only for a net energy up to $0.34\%$ of the ringdown energy, and lowers this to ${\approx}0.03$ (Sec.~\ref{sec:detect}). Over all masses (Fig.~\ref{fig:wfpop}, Table~\ref{tab:waterfall}) the reach is $z\simeq$0.31--0.34 for ET-D, 0.45--0.51 for CE, 2.2--2.6 for LISA, 3.0--3.4 for Taiji and 13--15 for DECIGO for $\chi_f=0.7268$ (generic templates for the space detectors). It is 3.0--3.9 times larger for $\chi_f=0.84$ and only 1.4--1.5 times larger for $\chi_f=0.90$ (median $\phi_0$), for which the hole before absorption is nearly extremal (Tables~\ref{tab:waterfall} and \ref{tab:progenitor}). Folded with current population models, with the remnant spin of every source (Table~\ref{tab:rates}), this gives $0.075$--$0.15$ expected signals with SNR $>5$ from stellar-mass BBHs in 10~yr of CE ($0.015$--$0.031$ with SNR $>8$). Of these, $69\%$ come from binaries like GW231123, with primary masses above $100\,\Msun$, a part of the mass function that rests on that one event. The population model gives a binary of this mass ratio $\chi_f\simeq0.69$, for which the GW231123-like echo would have SNR $2.1$ in CE instead of $9.4$. If the binaries above $100\,\Msun$ had the spin of GW231123, the number would be $0.035$--$7.2$ ($0.0079$--$1.3$), with $2.6$ at the median $\phi_0$, and the loudest echo in ET-D would have SNR $0.75$--$2.2$ instead of $0.79$--$0.97$. If all remnants had that spin, the CE number would be $0.016$--$9.4$ ($0.0037$--$1.8$). For massive-BH binaries and SNR $>5$ it gives $0$--$0.20$ (pop3), $0.50$--$2.8$ (Q3-nodelays) and $0.19$--$1.7$ (Q3-delays) in 4~yr of LISA, $2\times10^{-3}$--$0.34$, $1.8$--$9.0$ and $0.62$--$4.9$ in 4~yr of Taiji, and $0.49$--$14$, $67$--$150$ and $4.6$--$19$ in 4~yr of DECIGO ($0.30$--$9.4$, $43$--$100$ and $2.7$--$14$ with SNR $>8$). The most recent catalogue of the same galaxy-formation model, calibrated on the pulsar-timing signal, gives $0.37$--$1.8$ in LISA, $2.1$--$9.4$ in Taiji and $540$--$900$ in DECIGO ($300$--$570$ with SNR $>8$). Only $5$--$22\%$ of the DECIGO signals with SNR $>5$ have their line above 0.2~Hz, where no white-dwarf foreground is expected ($60$--$94$ signals for the last model). Where less than one signal is expected, the loudest echo of the observing time has SNR $0.20$--$0.25$ in the LIGO network at A+ sensitivity, $0.79$--$0.97$ in ET-D, $2.3$--$2.9$ in CE and $1.9$--$3.5$ (Q3-nodelays) in TianQin (Table~\ref{tab:rates}). The DECIGO detections come from binaries of $10^4$--$10^6\,\Msun$ at $z\simeq0.4$--$12$. Ignoring the spin distribution and using $\chi_f=0.7268$ for every source would change the massive-BH counts by factors $0.2$--$25$. The fastest remnants are not the loudest, because the holes that absorbed their packets were nearly extremal and their packet energy is capped (Table~\ref{tab:progenitor}). A third-generation detection would need fast-spinning binaries at the top of the stellar mass range, or intermediate-mass binaries with local rates ${\gtrsim}0.062$~Gpc$^{-3}$\,yr$^{-1}$ at $M_f^{\rm src}\sim10^3\,\Msun$ (CE). If the surface does not re-form, all numbers drop to the single-reflection values of Table~\ref{tab:reformed}, for example $0.49$--$25$ in DECIGO. All of these are optimal matched-filter SNRs with nominal thresholds, not search sensitivities. The threshold of 5 presumes a search triggered by the merger. A search that scans wide ranges of the echo time and phases needs a higher one, so the counts are optimistic. The echo that survives the effective horizon backreaction also lies one to two orders of magnitude below the current bound on Boltzmann echoes \cite{Abedi2023}, so present non-detections do not test it.

(5) The absolute amplitude is uncertain by factors of order unity. There are four main sources. The first is the escape prescription, i.e.\ the shape and width of the backreaction filter. The second is the Kerr background of the first reflection, which we take to be the remnant. Relative to the hole before absorption this changes the amplitude by $+47\%$ (GW150914-like) and $+27\%$ (GW231123-like), because the threshold shifts by $3.2k_c$ ($-21k_c$) during an absorption that is sudden on the scale $1/k_c$. Taking the Boltzmann factor and the surface from the hole at the arrival of each slice would multiply it by $0.80$ ($0.48$). The third is the packet energy, which depends on $\epsilon_E$, on its normalisation, on the band on which it is imposed and on the start of the time window. Counting the out-of-band flux of the computed packet would change the amplitude by a factor $0.82$ (Appendix~\ref{app:setup}), and the start of the time window by a factor $0.34$--$1$ (Appendix~\ref{app:window}). The fourth is the Boltzmann reflectivity phase $\phi_0$, which moves the gain over 2.47--3.15 (0.68--4.61 for the GW231123-like source). The tested changes of the discretisation of the time-domain evolution, the barrier grid and the frequency step change the survivor SNR of the GW150914-like source by less than $0.05\%$ (Appendix~\ref{app:checks}). This budget does not include the finite extraction radius and frozen eigenvalue of the evolution (per cent level in the greybody factor, while the free ringing is overdamped by $22\%$; Appendix~\ref{app:td}), the band of the incident packet (Appendix~\ref{app:setup}), the source model or the escape assumption.

(6) The cavity (ergoregion) modes of the object, the poles of the echo transfer function, are absent from the single-reflection survivor. If the surface re-forms after the absorption, the surviving wave builds up the two or three in-band modes. This multiplies the SNR by 2.47--3.15 (0.68--4.61 for the GW231123-like source, where the Boltzmann reflectivity phase matters more), and the strongest mode alone is louder than the survivor. The modes are damped: the stability index $2T_H\ln|\RBH|/|k|$ stays below $0.22$ for all remnant spins up to $\chi_f=0.955$, so the re-formed Boltzmann surface is free of the ergoregion instability that excludes a perfect reflector. This margin has to be computed with the exact barrier data. The modes below the threshold lie at $M_f\omega\simeq0.42$ ($0.53$ for the GW231123-like remnant), where the low-frequency expansion ($M_f\omega\ll1$) of Ref.~\cite{Maggio2019} breaks down, as that work points out. The exact superradiant amplification there is $Z=|\RBH|^2-1=0.16$--$0.34\%$ ($0.78$--$1.7\%$), and the growth rate $\omega_I$ that the modes would have with a perfect surface is $67$--$81$ ($318$--$383$) times that of the low-frequency formula. With the Boltzmann surface the same modes decay $131$--$187$ ($26$--$36$) times faster than they would grow. After the escape prescription itself, whether the surface re-forms is the largest single uncertainty in the expected signal.

(7) For the near-equal-mass source studied, the other $(\ell,m)$ modes of the ringdown, each passed through the backreaction filter at its own threshold $m\Omega_H$, do not change these results. With all modes of NRHybSur3dq8 [$\ell\le4$ except $(4,0)$ and $(4,1)$, and $(5,5)$], the $(2,2)$ mode carries all but $1.8\times10^{-4}$ of the orientation-averaged SNR$^2$ (per-mode median SNRs; $2.1\times10^{-4}$ at a common Boltzmann reflectivity phase). The $(3,3)$ and $(4,4)$ modes carry ${\sim}1\%$ of the ringdown energy but only $1.2\times10^{-5}$ and $6.7\times10^{-7}$ of the echo energy, because their thresholds lie deeper below their potential barriers. The $m=0$ modes, whose backreaction filter is centred on $\omega=0$ where $\Gamma_{\ell0}\propto(M_f\omega)^{2\ell+2}$, give ${\sim}10^{-22}$ of the $(2,2)$ echo energy.

The decisive open problem is a causal test of the escape prescription: evolving the reflected wave on a dynamically accreting background, with the reflecting boundary condition imposed at a fixed proper distance from the evolving horizon. Numerical relativity with reflecting boundaries \cite{Dailey2023,Dailey2025} and the reconstruction of the near-horizon spacetime from merger waveforms \cite{Ma2022} provide the tools. A nonrotating Vaidya-type model is the simplest controlled starting point. It lacks the superradiant threshold that shapes the survivor here, so the band structure and the dip can only be tested in the rotating case. A second open question is what geometry the absorbed energy actually produces. A configuration close to marginal trapping develops a long throat, whose echo delay is set by the redshift at its bottom rather than by a logarithm, and whose barrier imposes a greybody cutoff of Boltzmann form \cite{AbediThroat2026}. If a backreacted remnant ended up in this regime, both $L_*$ and the reflectivity used here would change. Future detectors will in any case measure the reflectivity of remnants far more precisely than current ones \cite{MaggioTesta2019,Branchesi2023,Deppe2025}. One natural extension is the near-extremal limit $T_H\to0$, $m\Omega_H\to1/M_f$ for $m=2$, where accurate barrier data become difficult and our template grid ends. Others are a full Bayesian treatment of the spin distributions and searches for narrow-band signals at $f_H$ using the waveforms released with this paper.

\appendix
\section{Source parameters and frequency-domain scattering data}\label{app:setup}
\subsection{Source, remnant and ingoing packet}
The source is the $\ell=|m|=2$ mode of the aligned-spin IMRPhenomT waveform \cite{IMRPhenomT} with source-frame masses $m_1=36$, $m_2=29\,\Msun$, the values of Ref.~\cite{GW150914}, at $z=0.09$ ($D_L=410$~Mpc), with detector-frame masses $(1+z)m_i=39.24$ and $31.61\,\Msun$, $\chi_{1z}=0.3$, $\chi_{2z}=-0.1$, face-on, $f_{\rm low}=20$~Hz. Its ringdown decays smoothly to numerical zero (Appendix~\ref{app:window}). Final-state fits give $M_f=67.21\,\Msun$ in the detector frame ($M_f^{\rm src}=61.66\,\Msun$ in the source frame), $\chi_f=0.7268$, hence $r_+=1.687M_f$, $m\Omega_H=0.4309/M_f$ ($f_H=207$~Hz), $T_H=0.0324/M_f$ and the prograde light ring $r_{\rm ph}=1.962M_f$. This is the hole after the absorption (Sec.~\ref{sec:backreaction}). The time-domain packet is evolved on it and the first reflection is evaluated on it (run \texttt{V0}, and run \texttt{GW231123} for the GW231123-like source). The hole before the absorption has $M_i=66.02\,\Msun$, $\chi_i=0.6892$, $m\Omega_H^{(i)}=0.4068/M_f$ ($196$~Hz) and $T_H^{(i)}=0.0340/M_f$. Table~\ref{tab:progenitor} lists the holes before absorption of all templates.

\begin{table*}
\caption{Holes before absorption of the reference sources and spin templates [Eq.~\eqref{eq:progenitor}; code \code{KEP}~\cite{EchoRepo}]. $E_{\rm RD}/M_f$: ringdown energy of the source; $\epsilon_E$: gross flux onto the absorbing hole in units of $E_{\rm RD}$ (below one where the cap $\chi_i\le0.998$ applies); $\chi_i$: spin of the hole before absorption; $\Delta M/M_f=(M_f-M_i)/M_f$: net absorbed energy (negative when the packet extracts energy); net/gross: net over gross flux onto that hole; shift: $(m\Omega_H-m\Omega_H^{(i)})/k_c$; $L_*$: thickness of the layer; last column: amplitude of the packet relative to one that delivers the gross flux $E_{\rm RD}$ to the remnant itself. $^\ast$: three holes satisfy Eq.~\eqref{eq:progenitor} ($\chi_i=0.799, 0.827\ {\rm and}\ 0.875$); the one nearest to $\chi_f$ is listed and used.}
\label{tab:progenitor}
\begin{ruledtabular}
\footnotesize
\begin{tabular}{lccccccccc}
template & $\chi_f$ & $E_{\rm RD}/M_f$ (\%) & $\epsilon_E$ & $\chi_i$ & $\Delta M/M_f$ (\%) & net/gross & shift ($k_c$) & $L_*/M_f$ & amplitude\\
\hline
$\chi_f=0.40$ template & 0.400 & 1.07 & 1.00 & 0.365 & $+1.07$ & $+1.000$ & $+2.1$ & 367 & 0.95\\
$\chi_f=0.55$ template & 0.550 & 1.29 & 1.00 & 0.513 & $+1.29$ & $+0.998$ & $+2.4$ & 382 & 0.92\\
$\chi_f=0.65$ template & 0.650 & 1.54 & 1.00 & 0.612 & $+1.53$ & $+0.991$ & $+2.8$ & 399 & 0.89\\
GW150914-like & 0.727 & 1.86 & 1.00 & 0.689 & $+1.77$ & $+0.955$ & $+3.2$ & 419 & 0.85\\
$\chi_f=0.78$ template & 0.780 & 2.17 & 1.00 & 0.748 & $+1.86$ & $+0.855$ & $+3.1$ & 440 & 0.86\\
$\chi_f=0.795$ template & 0.795 & 2.29 & 1.00 & 0.768 & $+1.78$ & $+0.777$ & $+2.6$ & 450 & 0.88\\
$\chi_f=0.81$ template$^\ast$ & 0.810 & 2.42 & 1.00 & 0.799 & $+1.31$ & $+0.540$ & $+0.7$ & 471 & 0.98\\
$\chi_f=0.825$ template & 0.825 & 2.57 & 1.00 & 0.901 & $-2.43$ & $-0.945$ & $-16.7$ & 608 & 1.00\\
GW231123-like & 0.840 & 2.70 & 1.00 & 0.919 & $-2.66$ & $-0.986$ & $-20.5$ & 665 & 0.86\\
$\chi_f=0.90$ template & 0.900 & 3.75 & 1.00 & 0.989 & $-3.75$ & $-1.000$ & $-93.8$ & 1453 & 0.42\\
$\chi_f=0.955$ template & 0.955 & 4.74 & 0.45 & 0.998 & $-2.15$ & $-1.000$ & $-181.9$ & 3132 & 0.29\\
\end{tabular}
\end{ruledtabular}
\end{table*}
 The Kerr fundamental mode is $M_f\omega_{220}=0.5451-0.0797i$ \cite{qnmpkg}, and a free single-mode fit to the IMRPhenomT ringdown $30$--$120M_f$ after the peak gives $0.5434-0.0792i$. The time-windowed post-merger signal (ramp length $8M_f$) radiates $E_{\rm RD}=1.25\,\Msun=1.86\%$ of $M_f$. The ingoing packet is normalised so that its gross horizon energy $\int|dE_H/d\omega|d\omega$ onto the hole before absorption equals $\epsilon_EE_{\rm RD}$ (net: $0.955\,\epsilon_EE_{\rm RD}$). It is injected $5M_f$ inside the light ring, where the spurious outgoing energy is $1.3\times10^{-7}$ of the horizon flux. Its $|\tilde\Phi_{\rm in}|^2$-weighted mean frequency is $\bar\omega=0.494/M_f$ (peak $0.521/M_f$).

The GW231123-like source has source-frame masses $m_1=137$, $m_2=103\,\Msun$, the median values of Ref.~\cite{GW231123}, at $z=0.39$ ($D_L=2.2$~Gpc), face-on, $f_{\rm low}=6$~Hz, with detector-frame masses $(1+z)m_i$. The published spin magnitudes $\chi_1=0.90$, $\chi_2=0.80$ are projected on the orbital angular momentum with a common factor, $\chi_{1z}=0.568$, $\chi_{2z}=0.505$, chosen so that the aligned-spin remnant has the published $\chi_f=0.84$. Final-state fits give $M_f=$$311\,\Msun$ in the detector frame ($M_f^{\rm src}=$$224\,\Msun$ in the source frame), $\chi_f=0.84$, $m\Omega_H=0.545/M_f$ ($f_H=56.6$~Hz), $T_H=0.0280/M_f$ and $M_f\omega_{220}=0.615-0.072i$. The time-windowed post-merger signal radiates $E_{\rm RD}=2.70\%$ of $M_f$. The hole before absorption has $M_i=1.0266\,M_f$ and $\chi_i=0.919$, with $m\Omega_H^{(i)}=0.641/M_f$ ($66.6$~Hz), which is $4\%$ above the ringdown frequency. Of the gross horizon flux onto it $99.3\%$ is superradiant, and the net flux is $-0.986$ of the gross; on the remnant itself the same packet would have a net flux of $+0.097$ of the gross. The surface of the remnant lies at $\Delta t_{\rm echo}=1059M_f=1.62$~s, with the detector-frame mass in the logarithm. The spin templates of Sec.~\ref{sec:detect} use a $65\,\Msun$ binary at 410~Mpc with mass ratio $q=0.8$ ($q=0.5$ for $\chi_f=0.955$) and equal aligned spins chosen to give $\chi_f={}$$0.4$, $0.55$, $0.65$, $0.727$, $0.78$, $0.795$, $0.81$, $0.825$, $0.84$, $0.9$, $0.955$.

\subsection{Barrier scattering data}\label{sec:barrier}
Outgoing waves near the horizon behave as $\Phi\propto\Delta^{-2}e^{ikr_*}$ and grow like $e^{-4\kappa r_*}$ ($\kappa=0.2036/M_f$) toward the horizon, by a factor ${\sim}10^{159}$ at the surface $r_*\simeq-450M_f$. This is representable in double precision, but the outgoing and ingoing parts of $\Phi$ then differ in magnitude by this factor. The reflected wave cannot be resolved against the ingoing one (cancellation), and the corresponding fluxes ($\propto|\Phi|^2$, ${\sim}10^{318}$) overflow. A grid that resolves $e^{-4\kappa r_*}$ over $450M_f$ would also be required. We therefore use the time-domain code only to obtain the ingoing $\psi_4$ spectrum $\tilde\Phi_{\rm in}(\omega)$ at the reference point $r_*^p=-20M_f$, where the Teukolsky operator is constant to $10^{-4}$, and treat the near-horizon cavity analytically in the frequency domain. The frequency dependence of the reflection of gravitational waves by Kerr black holes, and a nonthermal frequency scale that enters it, are discussed in Ref.~\cite{Abedi:2026nonthermal}.

The barrier seen from inside is characterised by the Teukolsky ``up'' solution (pure outgoing at infinity, $\Phi\to e^{i\omega r_*}$). Integrated inward, it is decomposed at $r_*=-6M_f$ against the two Frobenius solutions about $r_+$, $\Phi^H_{\rm in,out}=y^{\nu_{\rm in,out}}\sum_n b_n y^n$, $y=r-r_+$, $\nu_{\rm in}=-ik\,2M_fr_+/(r_+-r_-)$, $\nu_{\rm out}=-\nu_{\rm in}-2$ (condition number ${\sim}300$). Each basis solution is converted separately to the Sasaki--Nakamura (SN) variable \cite{SasakiNakamura1982,Hughes2000}
\begin{equation}
X=\sqrt{r^2+a^2}\,r^2\,J_-J_-\!\left(R/r^2\right),\qquad J_-=\partial_r-\frac{iK}{\Delta},
\end{equation}
which is a pure plane wave $e^{\mp ikr_*}$ near the horizon. This avoids the $\Delta^{-2}$ cancellations of a direct conversion. The outputs are the reflection coefficient from inside $\RBH(\omega)$, the transmission $\Tr(\omega)$ from a unit outgoing SN amplitude at the horizon to $\psi_4$ at infinity, and $A_{\rm trans}(\omega)$, the SN amplitude of a unit ingoing $\Phi$ wave. The grid has 101 frequencies on $M_f\omega\in[0.25,0.75]$. It is widened to $[m M_f\Omega_H-0.10,\,\max(m M_f\Omega_H,M_f\omega_{220})+0.15]$ when the threshold or the ringdown frequency lies near an edge, e.g.\ $[0.25,0.760]$ for the GW231123-like remnant. It is supplemented by 41 points with $\Delta k=10^{-3}/M_f$ around $k=0$ to resolve the zero of $\Tr$ (Sec.~\ref{sec:dip}). Its convergence is shown in Appendix~\ref{app:checks}.

\subsection{Energy normalisation and the reciprocity test}\label{sec:energy}
The SN equation has a complex potential in Kerr, so $|X|^2$ is not a flux. We use SN plane waves only as the phase convention and fix energy weights from physics: the ingoing horizon flux is $dE_H/d\omega=\alpha|Z_H|^2/4\pi\omega^2$ with the Teukolsky--Press factor \cite{TeukolskyPress1974}
\begin{equation}
\begin{aligned}
\alpha&=\frac{256(2M_fr_+)^5\,k\,(k^2+4\epsilon^2)(k^2+16\epsilon^2)\,\omega^3}{|C|^2},\\
\epsilon&=\frac{\sqrt{M_f^2-a^2}}{4M_fr_+},
\end{aligned}
\end{equation}
where $|C|^2=[(\lambda+2)^2+4ma\omega-4a^2\omega^2](\lambda^2+36ma\omega-36a^2\omega^2)+(2\lambda+3)(96a^2\omega^2-48ma\omega)+144\omega^2(M_f^2-a^2)$ is the Teukolsky--Starobinsky constant. The flux at infinity is $|Z_\infty|^2/4\pi\omega^2$. Here $Z_H$ and $Z_\infty$ are the amplitudes at the horizon and at infinity. Reference~\cite{Hughes2000} labels them $Z^\infty$ and $Z^H$, after the boundary condition of the solution that produces each one. Finally, the outgoing-at-horizon weight follows from energy conservation of the up solution. The factors $f_{\rm in}$ and $f_{\rm out}$ in Eq.~\eqref{eq:h1} are these horizon fluxes per unit squared SN amplitude. The greybody factor obtained from the up solution agrees with the one from the independent in-solution to within $1.1\times10^{-3}$. This maximum occurs at the barrier top, and the difference is $\le1.0\times10^{-3}$ for $M_f\omega\ge0.5$ and $\le1.2\times10^{-4}$ for $M_f\omega<0.5$, which contains the escape band. The energy-weighted $|\RBH|^2$ equals $1-\Gamma$ and exceeds unity for $\omega<m\Omega_H$ (superradiance), and the reflection phase is $\simeq\pi$ near the barrier top. The remaining convention freedom---SN versus a Hartle--Hawking $\psi_0/\psi_4$ definition of the reflected phase \cite{Xin2021}---is smooth across the narrow escape band and is absorbed into the constant Boltzmann reflectivity phase $\phi_0$.

\paragraph*{Band of the incident packet.} All spectral quantities of the packet---its echoes, $E_{\rm in}$, $\Delta J$ and the area change---are evaluated on the stored band $0.25\le M_f\omega\le0.75$, which contains the ringdown line and the threshold, and the normalisation $\int|dE_H/d\omega|\,d\omega=E_{\rm RD}$ is imposed on this band. The target $E_{\rm RD}$ is the spectral energy $\int|\tilde F|^2/4\pi\omega^2\,d\omega/2\pi$ of the time-windowed $\psi_4$ profile $F(u)$ for $u\ge0$, over the band padded by $0.05/M_f$ at each end. The profile starts at the middle of the $8M_f$ ramp, and the injected packet also carries the first half of the ramp, $-4<u<0\,M_f$. The strict band holds $95.6\%$ ($96.9\%$) of this target and including the first half of the ramp raises it by $+2.0\%$ ($+1.6\%$). Both together change it by $-2.2\%$ ($-1.2\%$), i.e.\ $-1.1\%$ ($-0.6\%$) in amplitude. The target also agrees with the time-domain energy of the time-windowed strain to $0.6\%$. The incident packet of the model is therefore a band-limited packet by definition. The time-domain packet at the horizon-side probe carries additional gross horizon flux above the band. Integrating $|dE_H/d\omega|$ over $0.05<M_f\omega<3$ gives $1.47$ times the in-band value for the GW150914-like source and $1.53$ for the GW231123-like one, with $30\%$ and $29\%$ of the total above $M_f\omega=0.75$. The integral has not converged at $M_f\omega=3$. This content lies far from the ringdown line, where the frozen-eigenvalue evolution and the WKB injection are least accurate, and we do not count it as part of the ringdown packet. Had the whole probe spectrum been normalised to $E_{\rm RD}$, the in-band amplitude would be multiplied by $0.82$ ($0.81$), $k_c$ would change by only $-0.3\%$ ($-0.3\%$), and the survivor SNR by $0.82$ ($0.81$). Re-evaluating every source of the populations with the uniform factor $0.82$ (Table~\ref{tab:reformed}, row ``wide-band norm.'') retains $52\%$ of the CE stellar count, $59\%$, $41\%$ and $44\%$ of the LISA counts and $77\%$, $84\%$ and $89\%$ of the DECIGO counts (pop3, Q3-nodelays, Q3-delays). The Euclidean $\rho^3$ scaling therefore holds only for the local stellar population, while the massive-BH counts are set by sources far above threshold and by the catalogue itself. This factor belongs with the $\epsilon_E$ uncertainty of Sec.~\ref{sec:detect}.

\subsection{Surface location and first reflection}
For the remnant of this paper the convention of Sec.~\ref{sec:wall}, $r_*^{\rm wall}=-\Delta t_{\rm echo}/2$, places the surface at $r_*^{\rm wall}=-450.0M_f$, i.e.\ $\Delta t_{\rm echo}=900M_f=0.298$~s (hole before absorption: $-428.3M_f$, $0.284$~s). The largest round-trip gain $|\RBH\Rwall|$ of the Boltzmann surface on the frequency grid is $0.9998$, reached at threshold, where both factors tend to unity. It stays below one at every $k\neq0$ (Sec.~\ref{sec:reformed}).

\section{Step-by-step derivation of the backreaction-filter model}\label{app:derivation}
This appendix derives the relations of Sec.~\ref{sec:backreaction} in the order in which they are used, for the reference remnant ($M_f=67.21\,\Msun$, $\chi_f=0.7268$, $\ell=m=2$, $\epsilon_E=1$). Units are $G=c=1$, and $M_f$ is the (detector-frame) mass of the remnant, i.e.\ of the hole \emph{after} the absorption; quantities of the hole before absorption carry the index $i$. Every number below is computed with \code{EC} and inserted into the text by \code{FIL}~\cite{EchoRepo}.
Table~\ref{tab:derivstatus} at the end separates exact relations from prescriptions.

\paragraph*{Step 0: Kerr quantities and the horizon-frame frequency.}
\begin{equation}
\begin{aligned}
r_\pm&=M_f\pm\sqrt{M_f^2-a^2},\qquad \Omega_H=\frac{a}{2M_fr_+},\\
\kappa&=\frac{r_+-r_-}{2(r_+^2+a^2)},\qquad T_H=\frac{\kappa}{2\pi},
\end{aligned}
\end{equation}
give, for the remnant, $r_+=1.6868M_f$, $r_-=0.3132M_f$, $m\Omega_H=0.43089/M_f$ ($f_H=207.2$~Hz), $\kappa=0.2036/M_f$, $T_H=0.0324/M_f$ and $A=4\pi(r_+^2+a^2)=42.394M_f^2$. The tortoise coordinate
\begin{equation}
r_*=r+\frac{2M_fr_+}{r_+-r_-}\ln\frac{r-r_+}{2M_f}-\frac{2M_fr_-}{r_+-r_-}\ln\frac{r-r_-}{2M_f}
\label{eq:Drstar}
\end{equation}
behaves near the horizon as
\begin{equation}
r_*\simeq\frac{1}{2\kappa}\ln\frac{r-r_+}{2M_f}+\text{const},
\label{eq:DrstarNH}
\end{equation}
since $2M_fr_+/(r_+-r_-)=1/2\kappa=2.456M_f$: $r_*$ is logarithmic in $r-r_+$, and this controls everything below. Near the horizon the SN equation reduces to $\partial_{r_*}^2X+k^2X=0$ with $k=\omega-m\Omega_H$. It is translation invariant in $r_*$, and its solutions $e^{\pm ikr_*}$ have the fixed tortoise wavelength $\lambda_*=2\pi/|k|$. ``Low frequency'' in Sec.~\ref{sec:backreaction} therefore means $\omega$ close to $m\Omega_H$, not small $\omega$.

\paragraph*{Step 1: the reflecting surface.} On the axis, $g_{rr}=\Sigma/\Delta=(r^2+a^2)/[(r-r_+)(r-r_-)]$, and Eq.~\eqref{eq:planck} becomes
\begin{equation}
2\sqrt{\frac{r_+^2+a^2}{r_+-r_-}}\sqrt{\Delta r}=l_p\;\Rightarrow\;\Delta r=\frac{(r_+-r_-)\,l_p^2}{4(r_+^2+a^2)},
\end{equation}
which is Eq.~\eqref{eq:planck} after $r_+^2+a^2=2M_fr_+$ and $r_+-r_-=2M_f\sqrt{1-\chi^2}$. Inserting $\Delta r$ into Eq.~\eqref{eq:DrstarNH}, with $\ln\Delta r\simeq2\ln l_p$ and $1/\kappa=2M_f(1+1/\sqrt{1-\chi^2})=4.912M_f$,
\begin{equation}
r_*^{\rm wall}\simeq-\frac{1}{\kappa}\ln\frac{M_f}{l_p}=-2M_f\Big(1+\frac{1}{\sqrt{1-\chi^2}}\Big)\ln\frac{M_f}{l_p},
\end{equation}
and $\Delta t_{\rm echo}=2|r_*^{\rm wall}|$. With $\ln(M_f/l_p)=91.62$: $r_*^{\rm wall}=-450.0M_f$ and $\Delta t_{\rm echo}=900M_f=0.298$~s for the remnant. For the hole before absorption (Step~6) the same expressions, with $(M_i,\chi_i)$ in place of $(M_f,\chi_f)$, give $r_*^{{\rm wall},i}=-428.3M_f$ and $\Delta t_{\rm echo}=0.284$~s.

\paragraph*{Step 2: the unfiltered first reflection} is Eq.~\eqref{eq:h1}. The barrier transmission $\Tr$ maps a unit outgoing amplitude at the horizon to $\psi_4$ at infinity and vanishes linearly at $k=0$, which produces the central dip (Sec.~\ref{sec:dip}). The factor $-2/\omega^2$ converts $\psi_4$ to strain, and $\sqrt{f_{\rm in}/f_{\rm out}}$ maps an ingoing onto an outgoing energy-normalised amplitude at the surface.

\paragraph*{Step 3: the hypothesis and the confinement heuristic.} Equation~\eqref{eq:filter} is an assumption. Its width comes from a standing-wave count. In a layer $r_*^{\rm wall}<r_*<r_*^{\rm wall}+L_*$ with nodes at both ends, the modes are
\begin{equation}
X_n\propto\sin\frac{n\pi\,(r_*-r_*^{\rm wall})}{L_*},\qquad|k_n|=\frac{n\pi}{L_*},\quad n=1,2,\dots
\end{equation}
The lowest one has $|k_1|=\pi/L_*$, i.e.\ half a wavelength fits, $\lambda_*/2=L_*$. No standing wave exists for $|k|<\pi/L_*$ [Fig.~\ref{fig:schematic}(b)], which motivates the definition $k_c\equiv\pi/L_*$. The boundary condition at the new trapped surface is not actually Dirichlet, so this sets a scale, not a theorem. Steps 4--7 compute $L_*$.

\paragraph*{Step 4: energy and angular momentum absorbed.} The packet is normalised by its gross flux onto the horizon of the hole that absorbs it, $(M_i,a_i)$ of Step~6: $\int|dE_H/d\omega|\,d\omega=\epsilon_EE_{\rm RD}$ with $E_{\rm RD}=1.86\times10^{-2}M_f$, and the net absorbed energy is $E_{\rm in}=\int(dE_H/d\omega)\,d\omega=0.955\,\epsilon_EE_{\rm RD}$. For a mode $e^{-i\omega t+im\phi}$ the fluxes obey $dJ=(m/\omega)\,dE$. The horizon energy spectrum is $dE_H/d\omega=\alpha_i|Z_H|^2/4\pi\omega^2$ with the Teukolsky--Press factor of $(M_i,a_i)$, $\alpha_i\propto k_i=\omega-m\Omega_H^{(i)}$ (Appendix~\ref{app:setup}), so Eq.~\eqref{eq:dJ} gives $\Delta J=3.490\,M_f\,E_{\rm in}$. The mean-frequency shortcut $m/\bar\omega=4.048M_f$ overestimates it by $16\%$. Modes with $k_i<0$ have $\alpha_i<0$: they are superradiant and carry $2.3\%$ of the gross flux. A normalisation of the net energy would be ill-conditioned for a packet that straddles the threshold of the absorbing hole, where positive and negative contributions nearly cancel, and for the GW231123-like packet the net flux is negative (net/gross $=-0.986$).

\paragraph*{Step 5: first law and area increase.} From $dM=(\kappa/8\pi)\,dA+\Omega_H\,dJ$ for the absorbing hole,
\begin{equation}
\frac{\kappa_i}{8\pi}\,dA=dE_H-\Omega_H^{(i)}\frac{m}{\omega}\,dE_H=\frac{k_i}{\omega}\,dE_H\;\propto\;\frac{k_i^2}{\omega}\;\ge0,
\end{equation}
because $dE_H\propto\alpha_i\propto k_i$. Every mode increases the area, the superradiant ones included. Integrated over the packet, $(\kappa_i/8\pi)\Delta A=0.2901\,E_{\rm in}$ at linear order. The exact Kerr areas of Step 6 give $0.2755\,E_{\rm in}$, i.e.\ $0.950$ times the linear value.

\paragraph*{Step 6: the hole before absorption.} The remnant $(M_f,a)$ is the hole after the absorption (Sec.~\ref{sec:backreaction}), so
\begin{equation}
\begin{aligned}
M_i&=M_f-E_{\rm in},\quad J_i=aM_f-\Delta J,\quad\chi_i=J_i/M_i^2,\\
r_\pm^{(i)}&=M_i\pm\sqrt{M_i^2-a_i^2},\quad A_i=8\pi M_ir_+^{(i)} .
\end{aligned}
\end{equation}
Because $E_{\rm in}$ and $\Delta J$ depend on $(M_i,a_i)$ through $\alpha_i$, these relations are solved numerically for $\chi_i$; the root is unique for this source. They give $E_{\rm in}=0.01772M_f$, $\Delta J=0.06184M_f^2$, $M_i=0.98228M_f$, $J_i=0.66499M_f^2$, $\chi_i=0.68920$, $r_+^{(i)}=1.69401M_f$, $m\Omega_H^{(i)}=0.40684/M_f$ ($195.6$~Hz), $\kappa_i=0.2139/M_f$, $A_i=41.821M_f^2$ and $\Delta A/A_i=1.372\times10^{-2}$. Comparing $r_+^{(i)}$ with $r_+$ is meaningless: they belong to two different Kerr metrics, which is why Step 7 uses the area.

\paragraph*{Step 7: equal-area mapping, $L_*$ and $k_c$.} The induced metric on $t,r={\rm const}$ of the geometry $(M_i,a_i)$ has $g_{\theta\theta}=\Sigma$ and $g_{\phi\phi}=[(r^2+a_i^2)^2-\Delta_i a_i^2\sin^2\theta]\sin^2\theta/\Sigma$. Hence $\sqrt{g_{\theta\theta}g_{\phi\phi}}=\sin\theta\,[(r^2+a_i^2)^2-\Delta_i a_i^2\sin^2\theta]^{1/2}$, which gives Eq.~\eqref{eq:area}; we solve $A(r_{\rm eq})=A$, the area of the remnant horizon. Differentiating at $r_+^{(i)}$, where $\Delta_i=0$ and $\Delta_i'=r_+^{(i)}-r_-^{(i)}$,
\begin{equation}
\frac{dA}{dr}\Big|_{r_+^{(i)}}=8\pi r_+^{(i)}-\frac{4\pi}{3}\frac{(r_+^{(i)}-r_-^{(i)})\,a_i^2}{r_+^{(i)2}+a_i^2}=41.754M_f,
\end{equation}
so $r_{\rm eq}-r_+^{(i)}\simeq\Delta A/(dA/dr)=0.57369M_f^2/41.754M_f=1.374\times10^{-2}M_f$. The exact root is $1.3682\times10^{-2}M_f$; dropping the $\Delta_i$ term [i.e.\ $A=4\pi(r^2+a_i^2)$] gives $1.3422\times10^{-2}M_f$. Equation~\eqref{eq:Drstar}, written for $(M_i,a_i)$, then gives $r_*(r_{\rm eq})=-9.788M_f$. With the surface of that hole at $r_*^{{\rm wall},i}=-428.3M_f$ (Step~1), $L_*=418.51M_f$ and $k_c=\pi/L_*=7.507\times10^{-3}/M_f$. With $M_f=$$3.310\times10^{-4}$~s, $f_c=k_c/2\pi$$=3.61$~Hz, and the support of the backreaction filter, $|k|<1.3k_c$, is $\pm4.69$~Hz around the threshold of the remnant, $f_H=207.2$~Hz.

The details of the mapping barely matter. Near the horizon
\begin{equation}
L_*\simeq\frac{1}{2\kappa_i}\ln\frac{r_{\rm eq}-r_+^{(i)}}{\Delta r_i},
\end{equation}
so a fractional error $\delta$ in $r_{\rm eq}-r_+^{(i)}$ shifts $L_*$ by only $\delta/2\kappa_i$; the approximate area moves $L_*$ by $0.05M_f$. Moreover $L_*$ equals $|r_*^{{\rm wall},i}|$ to within $2.3\%$, and the surface of the remnant lies at a similar distance, so $k_c$ is close to $\pi/|r_*^{\rm wall}|=6.98\times10^{-3}/M_f$, the cavity mode spacing, and the band $|k|<1.3k_c$ spans $2.80$ mode spacings. The proper thickness of the layer is $0.358M_f$ on the axis [$\simeq2((r_+^2+a^2)(r_{\rm eq}-r_+)/(r_+-r_-))^{1/2}$ with the quantities of $(M_i,a_i)$] and $0.333M_f$ in the equatorial plane ($\sqrt{g_{rr}}=r/\sqrt{\Delta_i}$).

\paragraph*{Step 8: the shape of the backreaction filter.} For Eq.~\eqref{eq:survivor}, $S(k_c)=\tfrac12$, because the cosine argument is $\pi/2$ at $x=1$. The roll-off is antisymmetric about $\tfrac12$ around $x=1$, so
\begin{equation}
\begin{aligned}
\int S\,dk&=2k_c\Big[0.7+\!\int_{0.7}^{1.3}\!\Big(\tfrac12+\tfrac12\cos\frac{\pi(x-0.7)}{0.6}\Big)dx\Big]\\
&=2k_c(0.7+0.3)=2k_c,
\end{aligned}
\end{equation}
and $k_c$ is both the half-width at half-maximum and the equivalent rectangular half-width. $S$ is real and even in $k$ (zero phase), so the surviving pulse stays centred on the echo time; the minimum-phase alternative is treated in Appendix~\ref{app:robust}.

\paragraph*{Step 9: the reflectivity inside the band.} $L_*$ depends only on $E_{\rm in}$ and $\Delta J$, not on the reflectivity. Inside the band the Boltzmann reflectivity is $e^{-|k|/2T_H}\ge e^{-1.3k_c/2T_H}=0.861$, since $k_c/2T_H=0.116$, and it is $0.911$ at the lobe centres $|k|\simeq0.8k_c$. The backreaction filter therefore keeps the band in which the surface reflects almost perfectly, and the survivor is controlled by $\Tr$ and by the packet near threshold.

\paragraph*{Step 10: shift of the horizon frequency and the sudden limit.} The threshold of the remnant, $m\Omega_H=0.43089/M_f$, lies $\Delta(m\Omega_H)=0.0240/M_f=3.20\,k_c$ above that of the hole before absorption, a shift larger than the bandwidth. With $\Delta t_{\rm abs}=22M_f$ for 80\% of the ingoing energy, the adiabaticity parameter is
\begin{equation}
\begin{aligned}
\frac{d(m\Omega_H)/dt}{k_c^2}&\simeq\frac{\Delta(m\Omega_H)/k_c}{k_c\,\Delta t_{\rm abs}}\\
&=\frac{3.20}{7.51\times10^{-3}\times22}\approx19\gg1 .
\end{aligned}
\end{equation}
The sweep is sudden compared with the coherence time $1/k_c=133M_f$ and with $\Delta t_{\rm echo}$. This is why Sec.~\ref{sec:backreaction} evaluates the first reflection (Steps 1, 2, 8 and 9: surface, barrier, backreaction filter and reflectivity) on the remnant, keeping $L_*$ and $k_c$ from the mapping of Step 7. Repeating everything on the hole before absorption instead gives survivor SNRs lower by a factor $1.49$ for this source and $1.16$ for the GW231123-like source. If only the reflection (Steps 1 and 9: surface and reflectivity) is evaluated slice by slice on the hole found at arrival, the survivor SNR changes by a factor $0.80$ ($0.48$) (\code{ARC}~\cite{EchoRepo}).

\begin{table}
\caption{Status of the steps in Appendix~\ref{app:derivation}.}
\label{tab:derivstatus}
\begin{ruledtabular}
\footnotesize
\begin{tabular}{p{0.5\columnwidth}p{0.42\columnwidth}}
item & status\\
\hline
Kerr quantities, $r_*$, surface location, $\Delta t_{\rm echo}$ & exact, given the Planck-distance surface\\
$\Delta J=\int(m/\omega)\,dE_H$; $(\kappa/8\pi)\,dA=(k/\omega)\,dE_H\ge0$ & exact flux relations and first law\\
$M_f$ is the hole after absorption, $M_i=M_f-E_{\rm in}$ ($E_{\rm in}$ net) & follows from $M_f=M_{\rm ADM}-E_{\rm rad}$\\
fluxes from the packet evolved on the remnant, with the flux weights of the absorbing hole & approximation, checked by evolving the packet on the hole before absorption\\
new horizon at the equal-area sphere & prescription, not a computed marginally trapped surface\\
$k_c=\pi/L_*$ & definition motivated by the standing-wave count\\
$\tilde h_{\rm surv}=S(k)\,\tilde h_1$, raised-cosine $S$ & phenomenological hypothesis (Appendix~\ref{app:robust})\\
gross absorbed energy $=\epsilon_EE_{\rm RD}$, $\epsilon_E=1$ & normalisation choice; amplitude $\propto\sqrt{\epsilon_E}$; $\epsilon_E<1$ where $\chi_i$ would exceed $0.998$\\
static Kerr background of the first reflection: the remnant & default, motivated by the sudden absorption (Sec.~\ref{sec:backreaction}); compared with the evaluation on the hole before absorption and with the arrival variant\\
\end{tabular}
\end{ruledtabular}
\end{table}

\section{Post-merger time window}\label{app:window}
\begin{figure*}
\includegraphics[width=\textwidth]{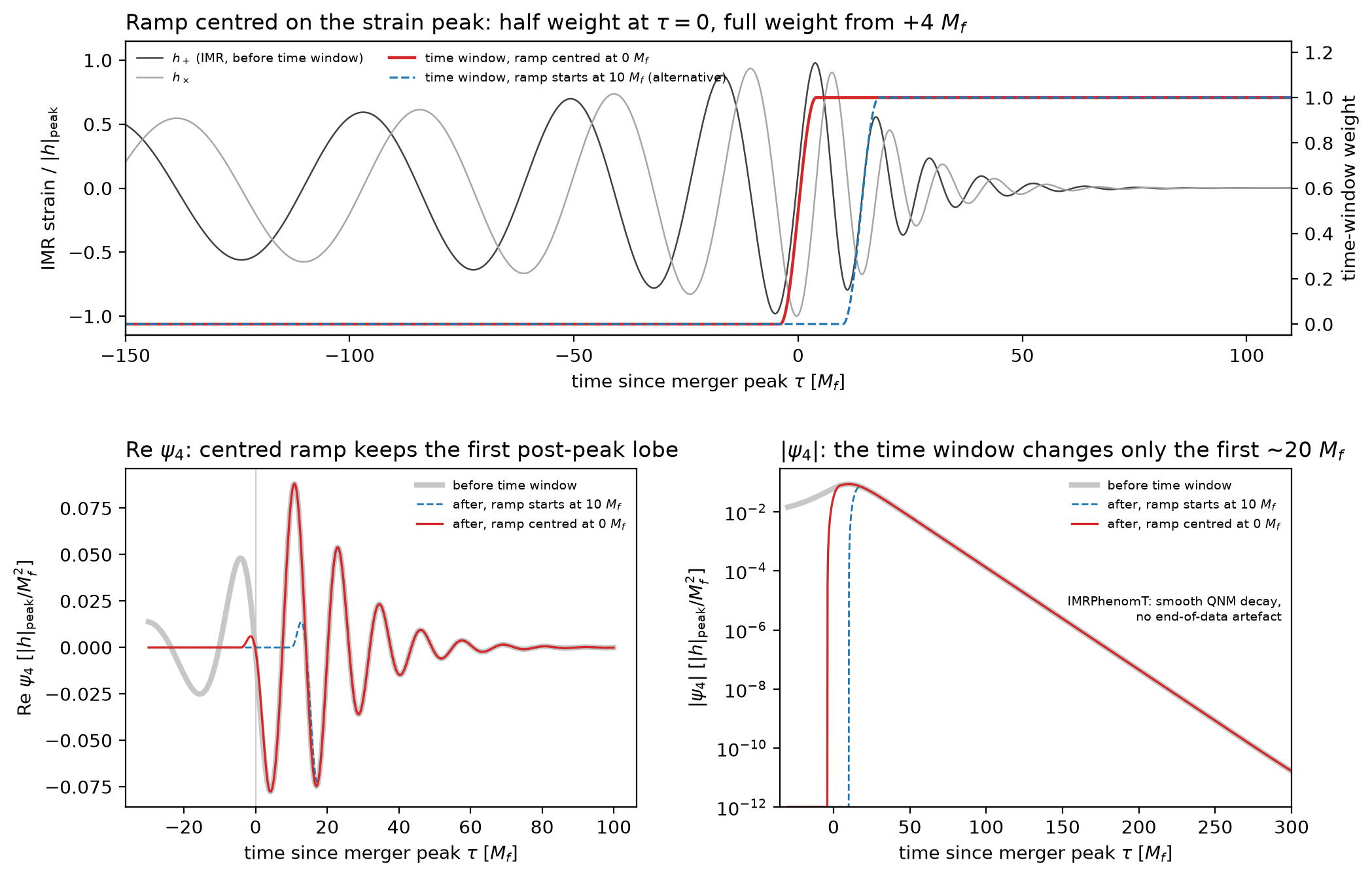}
\caption{Post-merger time window for the IMRPhenomT source. (a) IMR strain $h_+,h_\times$ (peak-normalised) with the ramp centred on the strain peak (red) and a ramp starting $10M_f$ after the peak (blue dashed). (b) ${\rm Re}\,\psi_4$ and (c) $|\psi_4|$ before (grey) and after the two time windows; the IMRPhenomT ringdown decays smoothly to numerical zero.}
\label{fig:window}
\end{figure*}

The ingoing packet is built from $\psi_4=\tfrac12\ddot h_{22}$ (cubic-spline second derivative of the peak-normalised strain) multiplied by a raised-cosine ramp $w(\tau)=\tfrac12[1-\cos(\pi x)]$, $x={\rm clip}[(\tau-\tau_0)/\ell,0,1]$, with $\ell=8M_f$ and $\tau_0=-4M_f$, i.e.\ centred on the strain peak. The IMRPhenomT data extend to $527M_f$ after the strain peak, where $|h|/|h|_{\rm peak}=1.4\times10^{-18}$, and follow the Kerr 220 exponential to $2\%$. A cosine roll-off over the last $20M_f$ of the data acts at $|h|/|h|_{\rm peak}\sim10^{-18}$ and has no effect. Figure~\ref{fig:window} shows the IMR strain with the time window and $\psi_4$ before and after windowing, compared with a ramp starting $10M_f$ after the peak. $\psi_4$ peaks $10M_f$ after the strain peak, so the centred ramp keeps the full $\psi_4$ peak while the $10M_f$ ramp cuts it to $82\%$. The instantaneous frequency at the start, centre and end of the centred ramp is $M_f\omega=0.31$, $0.35$, $0.39$ (against $0.545$ for the 220 mode), so the packet contains the end of the merger burst as well as the ringdown. The energy after the time window is $E_{\rm RD}=1.25\,\Msun$ for the centred ramp, $0.99\,\Msun$ for a ramp starting at the peak and $0.38\,\Msun$ for a ramp starting at $10M_f$. The survivor amplitude scales as $\sqrt{\epsilon_E E_{\rm RD}}$ times a spectral factor near $m\Omega_H$. Hence $\epsilon_E$ and the start of the time window should be read together as the modelling uncertainty on the overall echo scale. They affect $L_*$ only logarithmically and leave $k_c$ and the dip structure essentially unchanged.

\begin{figure}
\includegraphics[width=\columnwidth]{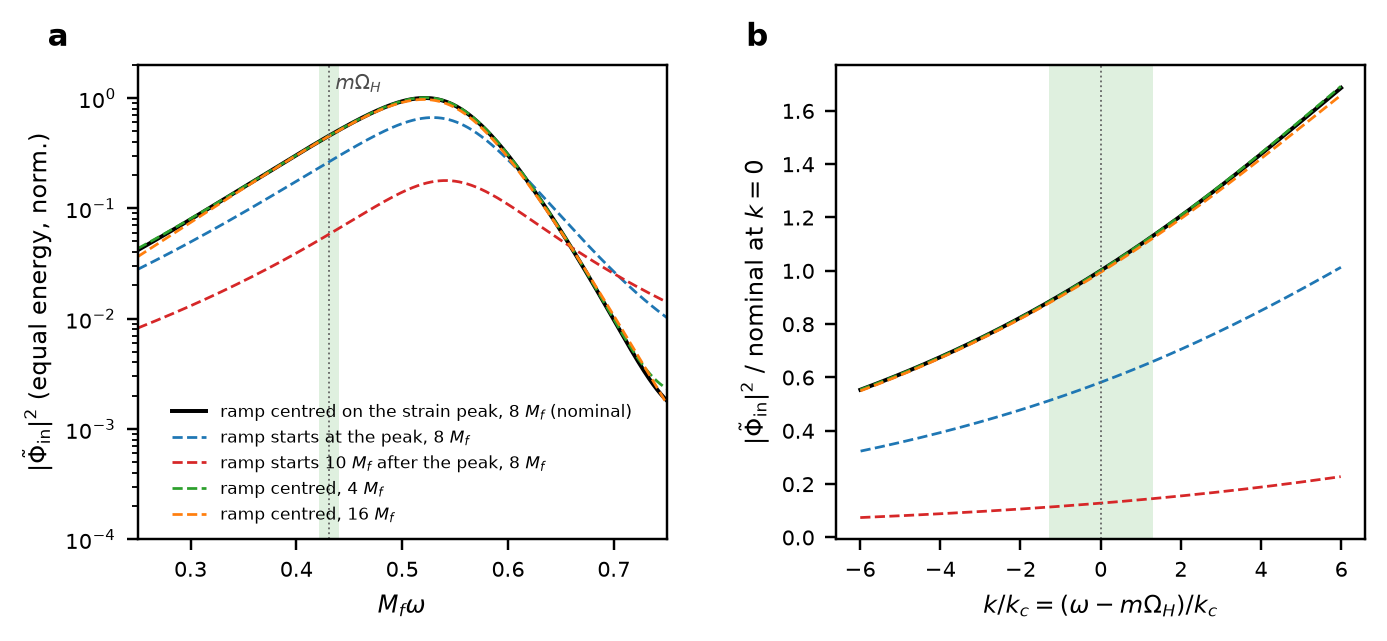}
\caption{Ingoing packet spectrum $|\tilde\Phi_{\rm in}|^2$ for the nominal time window (black) and four alternatives (gross-energy normalisation, $\epsilon_E=1$). (a) Full band, normalised to the nominal peak; (b) threshold region in units of $k_c$, normalised to the nominal value at $k=0$. Green: escape band $|k|<1.3k_c$.}
\label{fig:threshold}
\end{figure}

\paragraph*{Threshold-region spectrum and time-window dependence.} The survivor is built from the part of the ingoing packet within $|k|<1.3k_c$, a band below the spectral peak of the GW150914-like packet ($M_f\omega\simeq0.52$). Figure~\ref{fig:threshold} shows $|\tilde\Phi_{\rm in}|^2$ for the nominal time window and four alternatives, each a full rerun of the pipeline. The spectrum is smooth across the threshold and rises by ${\approx}30\%$ across the band. The band contains $5.4\%$ of $\int|\tilde\Phi_{\rm in}|^2d\omega$ for the nominal time window ($3.3$--$5.5\%$ for the alternatives). Table~\ref{tab:window} lists the survivor for each time window. The ramp length (4--16$M_f$) changes the survivor amplitude by a factor $0.992$--$1.003$. The ramp start is the dominant choice, with factors $0.75$ (ramp starting at the strain peak) and $0.34$ (starting $10M_f$ later). A later start removes the end of the merger burst, which carries most of the low-frequency content near $m\Omega_H$. The equal-$\psi_4$ normalisation (linear propagation of the time-windowed $\psi_4$ without energy rescaling, equivalent to $\epsilon_E=0.0229$) multiplies the survivor by $0.070$--$0.152$. All these factors enter the overall echo scale, which we treat as the dominant modelling uncertainty together with $\epsilon_E$.

\begin{table*}
\caption{Dependence of the survivor on the post-merger time window (each row a full pipeline run, with the hole before absorption solved for its own packet). Ramp start $\tau_0$ and length $\ell$ relative to the strain peak; ingoing energy $E_{\rm RD}$; share of $\int|\tilde\Phi_{\rm in}|^2d\omega$ inside the escape band $|k|<1.3k_c$; $L_*$; Boltzmann survivor peak relative to the nominal time window and face-on SNRs (CE, $300\,\Msun$, 410~Mpc; DECIGO, $10^5\,\Msun$, 3~Gpc) for the gross-energy normalisation ($\epsilon_E=1$); survivor peak relative to the nominal for the equal-$\psi_4$ normalisation.}
\label{tab:window}
\begin{ruledtabular}
\footnotesize\setlength{\tabcolsep}{3pt}
\begin{tabular}{lccccccccc}
time window & $\tau_0$ [$M_f$] & $\ell$ [$M_f$] & $E_{\rm RD}$ [$\Msun$] & band [\%] & $L_*$ [$M_f$] & peak & SNR CE & SNR DEC. & peak, eq.\ $\psi_4$\\
\hline
nominal & $-4$ & $8$ & 1.25 & 5.4 & 418.5 & 1.000 & 6.21 & 150 & 0.151\\
short ramp & $-2$ & $4$ & 1.28 & 5.4 & 418.1 & 1.003 & 6.22 & 151 & 0.152\\
long ramp & $-8$ & $16$ & 1.18 & 5.5 & 419.5 & 0.992 & 6.16 & 149 & 0.150\\
start at peak & $+0$ & $8$ & 0.99 & 4.5 & 422.9 & 0.747 & 4.66 & 113 & 0.126\\
start $+10M_f$ & $+10$ & $8$ & 0.38 & 3.3 & 431.2 & 0.338 & 2.13 & 52 & 0.070\\
\end{tabular}
\end{ruledtabular}
\end{table*}

\section{Detector noise curves and SNR conventions}\label{app:psd}
\begin{figure*}
\includegraphics[width=\textwidth]{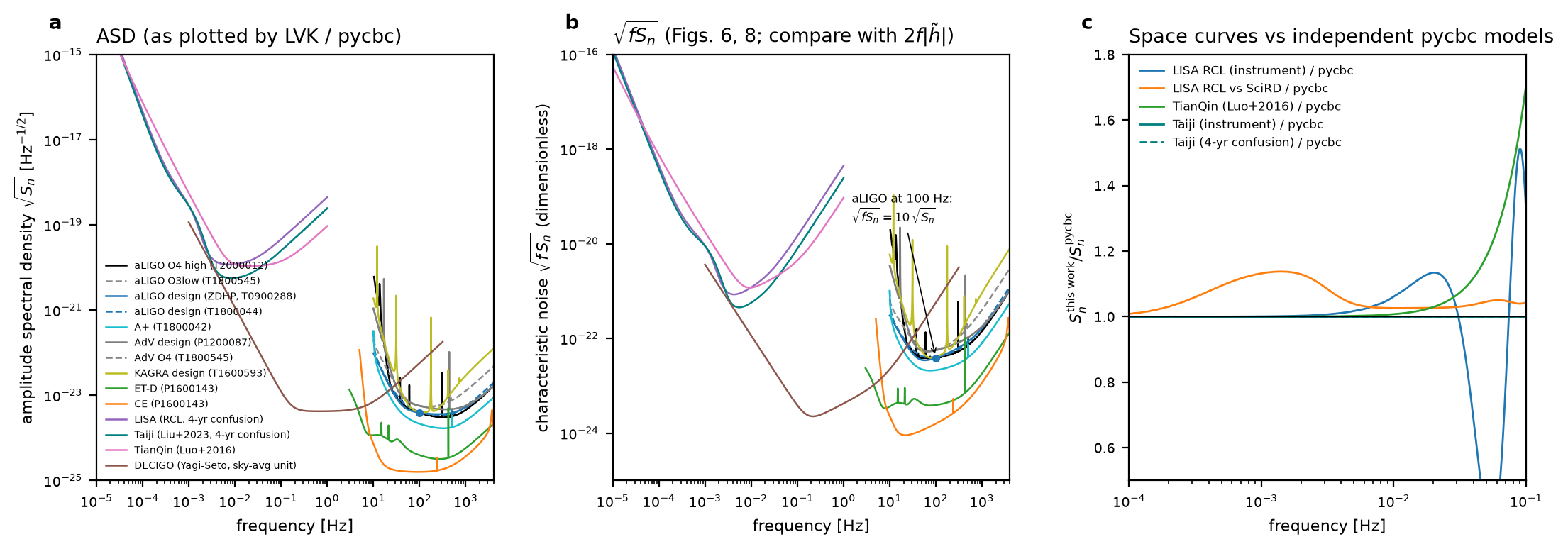}
\caption{Detector noise curves used in this work. (a) ASD $\sqrt{S_n}$ (the quantity plotted by LVK and \texttt{pycbc}). (b) Characteristic noise amplitude $\sqrt{fS_n}$ of one interferometer; in Fig.~\ref{fig:det}(a) the LIGO curves are lower by $W$, the effective noise of the Hanford--Livingston network. The dot marks the LIGO design curve (ZDHP) at 100~Hz in both panels. (c) Ratio of our analytic LISA (RCL, instrument part), Taiji (without and with the confusion noise) and TianQin curves to the independent \texttt{pycbc.psd.analytical\_space} models.}
\label{fig:psd}
\end{figure*}

\paragraph*{Plotted quantity.} Figures~\ref{fig:det}(a) and \ref{fig:psd}(b) show the dimensionless characteristic noise amplitude $\sqrt{fS_n(f)}$, to be compared with $2f|\tilde h(f)|$ \cite{Moore2015,Robson2019}, whereas LVK documents and \texttt{pycbc.psd} plot the amplitude spectral density (ASD) $\sqrt{S_n}$ in ${\rm Hz}^{-1/2}$ [Fig.~\ref{fig:psd}(a)]. The two differ by $\sqrt f$, a factor 10 at 100~Hz; the underlying one-sided PSDs are the same.

\paragraph*{Ground curves.} Every analytic ground PSD is taken from \texttt{pycbc.psd} and agrees bitwise with the corresponding direct LALSimulation call (Table~\ref{tab:psd}). As an external check, the Newtonian $1.4{+}1.4\,\Msun$ inspiral range (SNR 8, integrated to $f_{\rm ISCO}$, horizon distance$/2.2648$) reproduces the published BNS ranges to within $\lesssim8\%$ of the nearest end of the quoted ranges \cite{ObsScenario}. The LIGO O4 curve is the LVK observing-scenario ``O4 high'' projection (LIGO-T2000012, tabulated ASD), whose Newtonian BNS range we recover as 189~Mpc against the quoted 190~Mpc. The LALSimulation function \texttt{aLIGOAdVO4T1800545} is the Advanced Virgo O4 curve and is not used for LIGO. The LIGO curves and BNS ranges of this appendix are those of one interferometer; the SNRs of the paper combine Hanford and Livingston as described below.

\paragraph*{Space curves.} Our analytic RCL LISA curve (instrument part) agrees with the independent \texttt{pycbc} semi-analytical model to $<1\%$ for $0.1$--$3$~mHz, $6\%$ at 10~mHz and $13\%$ at 20~mHz, with larger deviations above the transfer frequency $f_*=19$~mHz where RCL's analytic response fit is approximate; it agrees with the LISA SciRD curve to within $13\%$ everywhere. Our Taiji curve agrees with the \texttt{pycbc} analytical Taiji model, without and with the four-year confusion noise, to within $0.2\%$. Our TianQin curve agrees with the \texttt{pycbc} analytical TianQin model to $<1\%$ below 10~mHz and $3\%$ at 20~mHz. The survivor bands relevant for LISA and TianQin ($f_H=4.6$~mHz at $3\times10^6\,\Msun$, 1.4~mHz at $10^7\,\Msun$) lie where the curves agree to $\lesssim1\%$. For DECIGO we use $S_n=\tfrac52S_h^{\rm YS}$ (sky-averaged, one unit of two interferometers), minimum ASD $4.2\times10^{-24}\,{\rm Hz}^{-1/2}$ at 0.66~Hz.

\paragraph*{Sources of the curves.} The ground PSDs are evaluated with \texttt{pycbc.psd} \cite{PyCBC} (v2.11.0), which wraps LALSimulation \cite{LALSuite} (v6.2.1); the function names are given in parentheses. LIGO O4: the ``O4 high'' (190~Mpc) tabulated ASD \texttt{aligo\_O4high.txt} of LIGO-T2000012 \cite{T2000012}, used in the observing-scenarios update \cite{ObsScenario} (an identical copy is distributed with \texttt{bilby} \cite{bilby}). LIGO design: the Advanced LIGO zero-detuned high-power curve LIGO-T0900288 \cite{Shoemaker2010,aLIGO2015} (\texttt{aLIGOZeroDetHighPower}). LIGO A+: LIGO-T1800042 \cite{T1800042} (\texttt{aLIGOAPlusDesignSensitivityT1800042}). Advanced Virgo: design curve of LIGO-P1200087 \cite{ObsScen2018,AdV2015} (\texttt{AdVDesignSensitivityP1200087}). KAGRA: design scenario of LIGO-T1600593 \cite{T1600593,KAGRA2013} (\texttt{KAGRADesignSensitivityT1600593}). ET-D: the ET-D model of Hild \emph{et al.} \cite{Hild2011}, a single interferometer, as tabulated in LIGO-P1600143 \cite{P1600143} (\texttt{EinsteinTelescopeP1600143}). CE: LIGO-P1600143 \cite{P1600143} (\texttt{CosmicExplorerP1600143}). The two comparison curves of Table~\ref{tab:psd} are the updated aLIGO design curve LIGO-T1800044 \cite{T1800044} and the O3/O4 projections of LIGO-T1800545 \cite{T1800545}. LISA: the analytic sky-averaged curve of Robson, Cornish and Liu \cite{Robson2019}, including the four-year Galactic confusion noise, for the mission of Ref.~\cite{LISA2017}; it is compared with the \texttt{pycbc} semi-analytical model and the SciRD curve \cite{LISASciRD}. Taiji: the analytic sky-averaged curve of Liu, Ruan and Guo \cite{TaijiConfusion2023}, of the same form as the LISA curve, with arm length $3\times10^6$~km, position noise 8~pm/$\sqrt{\rm Hz}$, acceleration noise 3~fm\,s$^{-2}$/$\sqrt{\rm Hz}$ and their fit of the Galactic confusion noise for four years of data (valid for 0.1--10~mHz), for the mission of Refs.~\cite{Taiji2017,Taiji2020,TaijiOverview2021}. TianQin: the curve of Luo \emph{et al.} \cite{TianQin2016} with the sky-averaging factor $10/3$, compared with the \texttt{pycbc} analytical TianQin model. DECIGO \cite{Seto2001}: Eq.~(5) of Yagi and Seto \cite{DECIGO2011} in the form corrected by their erratum (arXiv:1101.3940v2), with $S_n=\tfrac52S_h^{\rm YS}$.

\paragraph*{SNR conventions.} (i) Ground, single interferometer with $F_+=1$, face-on (Advanced Virgo, KAGRA, ET-D, CE): $\rho^2=4\int|\tilde h_+|^2/S_n\,df$ with $|\tilde h_+|^2=|\tilde h|^2/4$; for the IMRPhenomT source at 410~Mpc this agrees with \texttt{pycbc.filter.sigma}($h_+$) to $0.2\%$. (ii) LIGO: the Hanford--Livingston network. With the response tensors $D^{ij}_d$ of the two detectors \cite{LALSuite}, $F_{+,\times}=D^{ij}_de^{+,\times}_{ij}$, a face-on source at sky position $\hat n$ gives $\rho^2=\rho_1^2\sum_d(F_+^2+F_\times^2)_d$, where $\rho_1$ is the single-interferometer value of (i) with the same noise curve in both detectors. We quote the maximum over $\hat n$, $\rho=W\rho_1$ with $W^2=1.89$ ($W=1.37$), reached midway between the sites, which are $27^\circ$ apart; each detector then contributes $0.94$. In the code this is an effective noise curve $S_n/W^2$ (\code{EC}~\cite{EchoRepo}; the antenna patterns agree with \texttt{pycbc.detector} to $10^{-6}$; \code{HLN}). For the population counts the network SNR of a source with random sky position, polarisation and inclination is $w\rho$, with $w^2=\sum_d[F_+^2(1+\cos^2\iota)^2/4+F_\times^2\cos^2\iota]_d/W^2$ and $\langle w^2\rangle=0.170$ ($4/25$ for one interferometer). For the noise measured at GW231123, which differs between the detectors, the maximum over $\hat n$ of $\sum_d(F_+^2+F_\times^2)_d\,\rho_{1,d}^2$ is used. (iii) Space, sky- and polarisation-averaged $S_n$ (RCL for LISA and Taiji, TianQin with the $10/3$ factor, DECIGO $\tfrac52S_h^{\rm YS}$): $\rho^2=4\int(|\tilde h_+|^2+|\tilde h_\times|^2)/S_n\,df$ \cite{Robson2019} with $|\tilde h_+|^2+|\tilde h_\times|^2=|\tilde h|^2/2$ face-on.

\begin{table*}
\caption{Ground noise curves (one interferometer each): deviation from direct LALSimulation evaluation (n/a for the tabulated O4 curve), minimum ASD, and Newtonian BNS range compared with published values \cite{ObsScenario,psds}. ET-D (single interferometer) and CE ranges are Euclidean and not comparable with cosmological quotes.}
\label{tab:psd}
\begin{ruledtabular}
\begin{tabular}{lcccc}
curve & lalsim dev. & min ASD [${\rm Hz}^{-1/2}$] & BNS range [Mpc] & published\\
\hline
aLIGO O4 high (T2000012) \cite{T2000012} & n/a & $3.03\times10^{-24}$ @ 329 Hz & 189 & 160--190\\
aLIGO O3low (T1800545) \cite{T1800545} & 0 & $5.41\times10^{-24}$ @ 149 Hz & 121 & 110--130\\
aLIGO design (ZDHP, T0900288) \cite{Shoemaker2010} & 0 & $3.61\times10^{-24}$ @ 253 Hz & 198 & 190--200\\
aLIGO design (T1800044) \cite{T1800044} & 0 & $3.35\times10^{-24}$ @ 282 Hz & 181 & 173\\
A+ (T1800042) \cite{T1800042} & 0 & $1.68\times10^{-24}$ @ 321 Hz & 346 & 320--330\\
AdV design (P1200087) \cite{ObsScen2018} & 0 & $4.68\times10^{-24}$ @ 375 Hz & 128 & 125\\
AdV O4 (T1800545) \cite{T1800545} & 0 & $4.68\times10^{-24}$ @ 376 Hz & 128 & 90--120\\
KAGRA design (T1600593) \cite{T1600593} & 0 & $3.99\times10^{-24}$ @ 124 Hz & 140 & 130+\\
ET-D (P1600143) \cite{Hild2011,P1600143} & 0 & $3.19\times10^{-25}$ @ 271 Hz & 2840 & --\\
CE (P1600143) \cite{P1600143} & 0 & $1.59\times10^{-25}$ @ 94 Hz & 8350 & --\\
\end{tabular}
\end{ruledtabular}
\end{table*}

\section{Time-domain Teukolsky construction of the ingoing packet}\label{app:td}
\subsection{A $1{+}1$ formulation regular at both ends}
For an $m$-mode of the $s=-2$ Teukolsky equation, with Teukolsky's separation constant $\lambda=A+a^2\omega^2-2am\omega$ frozen at its value for the real carrier frequency $\omega_R={\rm Re}\,\omega_{220}$, the radial equation with $\omega\to i\partial_t$ becomes a $1{+}1$ wave equation. What is held fixed is $\lambda$ as a whole, that is, the angular eigenvalue $A$ together with the terms $a^2\omega^2-2am\omega$, and not $A$ alone; ``frozen eigenvalue'' refers to this throughout. Freezing $\lambda$ is the only approximation relative to the full $2{+}1$ problem. It neglects the frequency dependence of $\lambda$ and the associated $\ell$-mixing, and it changes the free ringing of the equation. A Leaver continued-fraction solution of the radial equation with $\lambda$ held at its value for ${\rm Re}\,\omega_{220}$ has its fundamental pole at $M_f\omega=0.54578-0.09757i$ for the GW150914-like remnant and $0.61513-0.08890i$ for the GW231123-like one, against the Kerr values $0.54507-0.07970i$ and $0.61454-0.07238i$. The damping rate is therefore $22\%$ and $23\%$ too large. A Gaussian packet scattered freely in our evolution rings down at $M_f\omega\simeq0.5457-0.0974i$ for the GW150914-like remnant, unchanged between $\Delta r_*=0.1$ and $0.05M_f$ (\code{TDS}~\cite{EchoRepo}). The packet of this paper is prescribed rather than freely rung, and Appendix~\ref{app:checks} tests the sensitivity of the survivor to the freezing frequency, but this damping error is a limitation of the $1{+}1$ equation. The Teukolsky function $R$ behaves as $\Delta^2 e^{-ikr_*}$ for ingoing waves at the horizon and as $r^3e^{i\omega r_*}$ for outgoing waves at infinity. The variable
\begin{equation}
\Phi \equiv \frac{r\,R}{\Delta^2}
\end{equation}
is $O(1)$ in both limits and satisfies
\begin{equation}
\partial_t^2\Phi=\partial_{r_*}^2\Phi+A_s(r)\,\partial_{r_*}\Phi+A_t(r)\,\partial_t\Phi+A_0(r)\,\Phi ,
\label{eq:phi}
\end{equation}
with
\begin{align}
A_s&=\frac{2(2r^4+a^2r^2-2M_fr^3-a^4)}{r(r^2+a^2)^2},\\
A_t&=\frac{2(2r^3-6M_fr^2+2a^2r+2M_fa^2)-2iam(r^2+a^2)}{(r^2+a^2)^2},\\
A_0&=-\frac{\mathcal{N}(r)}{r^2(r^2+a^2)^2},\\
\mathcal{N}&=\lambda r^4-2M_f\lambda r^3+(a^2\lambda-a^2m^2-2a^2+4M_f^2)r^2\nonumber\\
&\quad+4iam\,r^3-4iM_fam\,r^2+2M_fa^2r-2a^4-2M_fr^3,
\end{align}
where $\lambda=A+a^2\omega_R^2-2am\omega_R$ is Teukolsky's constant ($A=2.8245$, $\lambda=1.3968$ here). All coefficients are regular at the horizon; at $r_+$, $A_s+A_t=-2im\Omega_H$, so the near-horizon equation is solved exactly by $\Phi=e^{-im\Omega_H t}g(t+r_*)$ for arbitrary $g$, while at large $r$, $A_s\simeq A_t\simeq 4/r$ makes outgoing waves $O(1)$ and ingoing ones $O(r^{-4})$, reproducing peeling. Both limits were verified symbolically. We discretise with fourth-order finite differences in $r_*\in[-150,350]M_f$ ($\Delta r_*=0.05M_f$), RK4 in time (CFL 0.5), Kreiss--Oliger dissipation, the exact co-rotating ingoing condition $\partial_t\Phi-\partial_{r_*}\Phi+im\Omega_H\Phi=0$ at the inner boundary and a Sommerfeld condition at the outer one.

\subsection{Ingoing packet and one-way injection}
The ingoing counterpart of $F(u)$ is the co-rotating WKB left-mover
\begin{equation}
\begin{aligned}
\Phi_{\rm inc}(t,r_*)&=B(r_*)\,F(t+r_*-r_*^{\rm inj})\,e^{i\phi},\\
(\partial_t-\partial_{r_*})B&=\tfrac12(A_s+A_t)B,
\end{aligned}
\end{equation}
which is exact where the coefficients are constant and encodes the horizon wavenumber shift $\omega\to k$. It is fed into the domain through a one-way (total-field/scattered-field) source at $r_*^{\rm inj}$: with $\mathcal{L}\Phi\equiv\partial_{r_*}^2\Phi+A_s\partial_{r_*}\Phi+A_t\partial_t\Phi+A_0\Phi-\partial_t^2\Phi$ and a smooth step $H(r_*)$ (unity inside $r_*^{\rm inj}$), the total field $\Phi=H\Phi_{\rm inc}+\Phi_{\rm sc}$ obeys $\mathcal{L}\Phi=0$ away from the interface if
\begin{equation}
\partial_t^2\Phi_{\rm sc}=\partial_{r_*}^2\Phi_{\rm sc}+A_s\partial_{r_*}\Phi_{\rm sc}+A_t\partial_t\Phi_{\rm sc}+A_0\Phi_{\rm sc}+H\,\mathcal{L}\Phi_{\rm inc} .
\end{equation}
The interface terms then radiate the incident wave, and the only source of outgoing radiation is the residual $\mathcal{L}\Phi_{\rm inc}$ of the ansatz inside the barrier. We verified numerically that the total field satisfies the homogeneous equation throughout the barrier region ($|\mathcal{L}\Phi|/|\partial_t^2\Phi|<10^{-4}$, limited by the finite-difference evaluation of the check). This is the precise sense in which an ``ingoing'' packet is defined in a region with a potential: no exactly ingoing solution exists there, and the residual of the best local ansatz is the physical leak. A global phase $\phi$ drops out of the linear evolution.

\subsection{Results: a clean fall to the horizon}
\begin{figure*}[t]
\includegraphics[width=\textwidth]{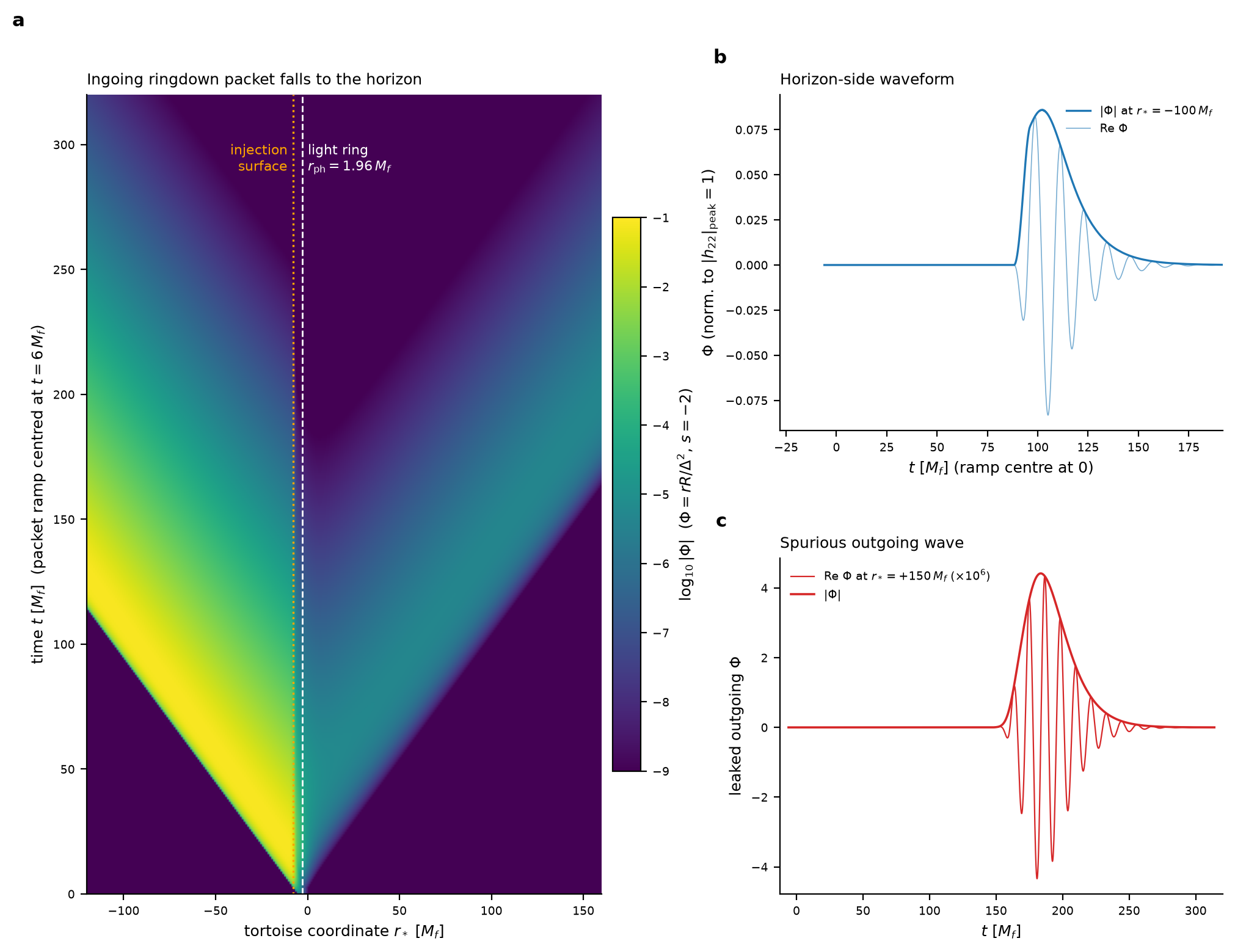}
\caption{Ingoing post-merger packet injected $5M_f$ (in $r_*$) inside the light ring of the $\chi_f=0.7268$ remnant. (a) Spacetime diagram of $\log_{10}|\Phi|$; dashed: light ring; dotted: injection surface. (b) Horizon-side waveform at $r_*=-100M_f$. (c) Leaked outgoing wave at $r_*=+150M_f$; $E_{\rm leak}/E_H=1.3\times10^{-7}$.}
\label{fig:packet}
\end{figure*}

\begin{figure*}[t]
\includegraphics[width=\textwidth]{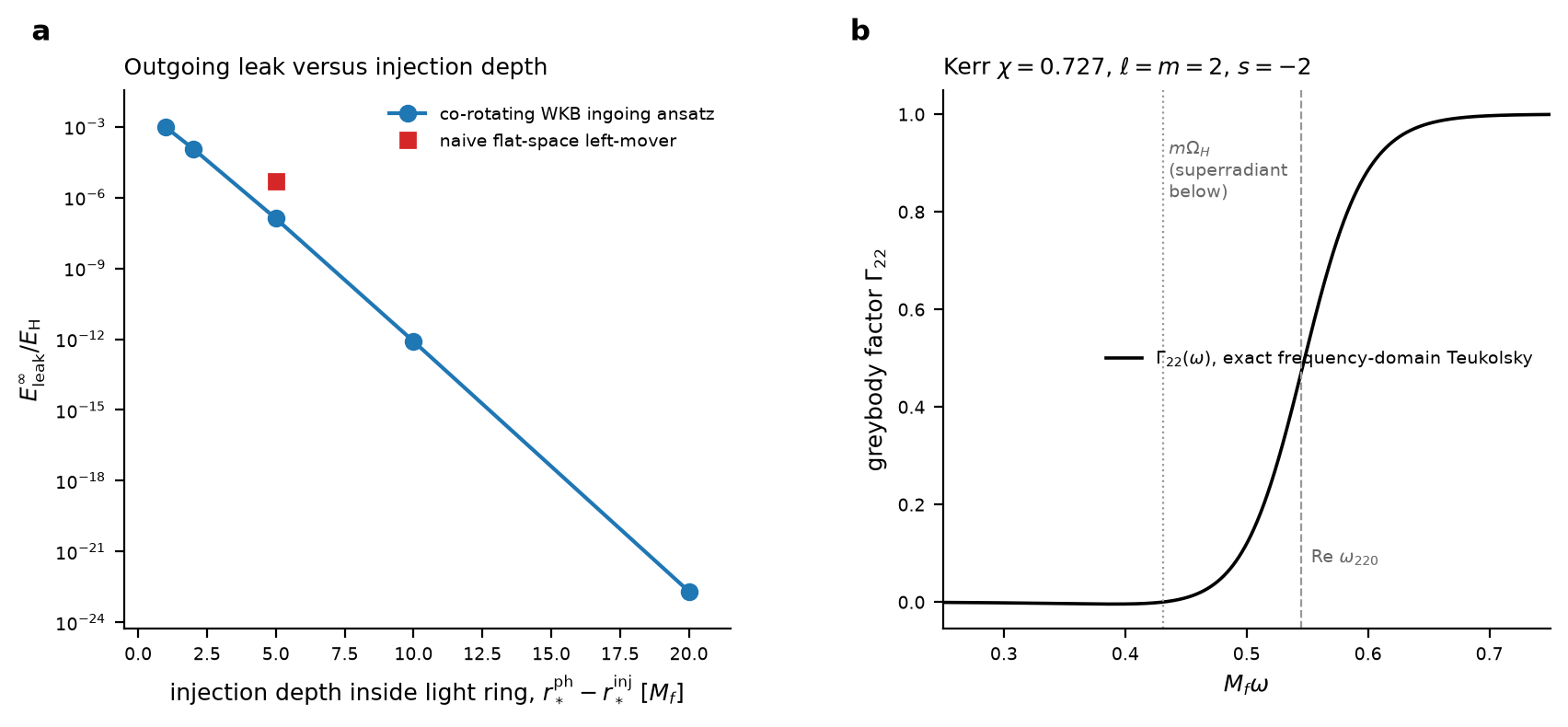}
\caption{(a) Leaked outgoing energy fraction versus injection depth inside the light ring for the co-rotating WKB ansatz (circles) and a naive flat-space left-mover (square). (b) Greybody factor $\Gamma_{22}(\omega)$ of the exact frequency-domain Teukolsky equation of Appendix~\ref{app:setup}; it is negative for $\omega<m\Omega_H$ (superradiance).}
\label{fig:scan}
\end{figure*}

Figure~\ref{fig:packet} shows the evolution for injection $5M_f$ inside the light ring. The packet reaches $r_*=-100M_f$ with its shape intact. The outgoing leak measured at $r_*=+150M_f$ is $4.4\times10^{-6}$ in $\Phi$ and carries $E_{\rm leak}/E_H=1.3\times10^{-7}$ of the horizon flux. In double precision the inversion $r(r_*)$ saturates at $r-r_+\simeq10^{-16}M_f$ for $r_*\lesssim-88M_f$, so the grid beyond that point uses the regular horizon limits of the coefficients. Moving the inner boundary from $-100$ to $-150M_f$ is therefore not a test of finite-radius corrections. Scanning the injection depth (Fig.~\ref{fig:scan}) gives $9.6\times10^{-4}$, $1.2\times10^{-4}$, $1.3\times10^{-7}$, $8.1\times10^{-13}$ and $2.1\times10^{-23}$ at 1, 2, 5, 10 and 20~$M_f$ inside the light ring, while a naive flat-space left-mover ansatz leaks $4.7\times10^{-6}$ at $5M_f$. The horizon-side wave keeps the frequency $\omega$ in $t$ but its radial wavenumber is $k=\omega-m\Omega_H$.


\section{Numerical checks and convergence}\label{app:checks}
\paragraph*{Frequency-domain solver.} Horizon limit of Eq.~\eqref{eq:phi}: the residual of $\Phi=e^{-im\Omega_Ht}e^{-ik(t+r_*)}$ decreases $\propto\Delta$; infinity limit: outgoing $O(1)$, ingoing $O(r^{-4})$. At $a=0$ the frequency-domain solver with the same coefficients gives $\Gamma_{\ell=2}(M_f\omega=0.3,0.4,0.5,0.6)=0.0548,\,0.6898,\,0.9845,\,0.9993$, identical to an independent Regge--Wheeler integration. Reciprocity $\Gamma_{\rm up}=\Gamma_{\rm in}$ holds to $\le1.1\times10^{-3}$ (maximum, barrier top) and to $\le1.2\times10^{-4}$ for $M_f\omega<0.5$, and the energy-weighted $|\RBH|^2=1-\Gamma$. The Sasaki--Nakamura conversion of the in-solution is a single plane wave to $10^{-8}$, that of the up-solution via the Frobenius basis to $10^{-4}$. The zero of $\Tr$ is resolved with 41 barrier points at $\Delta k=10^{-3}/M_f$; $\Tr$ is interpolated in (Re, Im).

\paragraph*{Time-domain versus frequency-domain greybody factor.} A Gaussian packet ($\sigma_t=4M_f$, carrier $M_f\omega=0.6$) is injected inward at $r_*=60M_f$ and scattered off the barrier. The greybody factor $\Gamma_{\rm TD}(\omega)=dE_H/(dE_H+dE_\infty)$ is computed from the horizon flux at $r_*=-100M_f$ and the reflected flux at an extraction radius $r_*^{\rm ex}$. At $a=0$ we find $\Gamma_{\rm TD}/\Gamma_{\rm FD}=0.682/0.687$, $0.984/0.984$ and $0.999/0.999$ at $M_f\omega=0.40$, $0.50$, $0.60$ ($r_*^{\rm ex}=300M_f$). For the Kerr remnant at $M_f\omega=0.545$, the frequency at which the eigenvalue is frozen (the Gaussian carrier is $0.6$), $\Gamma_{\rm TD}=0.455$ ($r_*^{\rm ex}=150M_f$) and $0.462$ ($300M_f$) against $\Gamma_{\rm FD}=0.467$. The residual difference decreases with the extraction radius and is consistent with the $O(1/\omega r)$ corrections to the outgoing wave at finite radius. At $a=0$ the frequency-domain in-solution itself changes by $2.6\%$ in amplitude between $r_*=150M_f$ and $1500M_f$. Away from the carrier, e.g.\ at $M_f\omega=0.50$ and $0.60$, $\Gamma_{\rm TD}=0.132$ and $0.860$ against $\Gamma_{\rm FD}=0.116$ and $0.887$ with the exact eigenvalue and $0.155$ and $0.844$ with the frozen one. The remaining differences reflect the frozen spheroidal eigenvalue of the $1{+}1$ equation. At the carrier frequency, halving the grid spacing gives $\Gamma_{\rm TD}=0.455$ and moving the horizon probe to $-50M_f$ gives $0.455$ (nominal $0.455$). Injecting at $100M_f$ and extracting at $250M_f$ gives $0.460$, following the extraction-radius trend. A packet with three times narrower bandwidth ($\sigma_t=12M_f$, carrier $0.545$) gives $0.465$ against $\Gamma_{\rm FD}=0.480$ at its grid point $M_f\omega=0.546$. The frozen eigenvalue enters only the ingoing packet. All barrier data used for the echoes come from the exact frequency-domain solver. The escape band lies at $\omega\simeq m\Omega_H$ rather than at $\omega_R$, so we repeated the full calculation with the eigenvalue frozen at $\omega=m\Omega_H$ ($\lambda=1.936$, $39\%$ above the nominal value). The survivor SNRs change by at most $0.01\%$ over all detectors and masses, because the packet is renormalised to $\epsilon_E=1$ and its near-horizon spectrum hardly changes shape across the narrow band.

\paragraph*{Convergence of the survivor.} Table~\ref{tab:conv} repeats the complete calculation (time-domain evolution, barrier grid, survivor, detectability) for the GW150914-like source and time window with each numerical parameter changed in turn. These tests use the default evaluation (run \texttt{V0}); in each of them the hole before absorption is solved again for the packet of that run. The survivor amplitude, its SNRs, the dip depth, the hump separation and the node time change by at most $0.04\%$ (amplitude), $0.05\%$ (SNRs) and $0.0015$ (dip depth, when the boundaries are moved), with no change in the hump separation or node time at the $1M_f$ sampling. These are the changes under the tested variations, not a bound on the numerical error. The three grid spacings of the time-domain evolution do not approach a limit monotonically at this level (the peak changes by $+0.04\%$ and then by $-0.006\%$ when the spacing is halved twice), so we quote no order of convergence. The frozen eigenvalue of the evolution (Appendix~\ref{app:td}; damping rate of the free ringing $22\%$ too large) and the sampling of the Boltzmann reflectivity phase (Sec.~\ref{sec:reformed}) are separate errors with their own scales.

\begin{table*}[t]
\caption{Convergence and boundary tests of the survivor (default evaluation, run \texttt{V0}; $\epsilon_E=1$, adopted backreaction filter, Boltzmann surface). $\delta$: change of the peak relative to the nominal run. SNRs: LIGO Hanford--Livingston network at O4 sensitivity at $67\,\Msun$, CE at $300\,\Msun$ (410~Mpc), LISA at $3\times10^6\,\Msun$ (3~Gpc). ``Boundaries moved'': time-domain domain $r_*\in[-200,450]M_f$, up-solution started at $r_*=2500M_f$, Frobenius decomposition at $-8M_f$. ``Barrier grid'': coarse spacing $0.0025$ ($0.01$) and threshold sampling $5\times10^{-4}$ ($2\times10^{-3}$).}
\label{tab:conv}
\begin{ruledtabular}
\footnotesize\setlength{\tabcolsep}{2.5pt}
\begin{tabular}{lccccccccc}
run & $L_*$ & peak & $\delta$ & \multicolumn{3}{c}{SNR} & dip & $\Delta u_{\rm hump}$ & node$-u_e$\\
 & [$M_f$] & & & O4 & CE & LISA & & [$M_f$] & [$M_f$]\\
\hline
nominal & 418.5 & $1.209\times10^{-4}$ & -- & 0.0474 & 6.207 & 13.57 & 0.113 & 533 & $-2$\\
$\Delta r_*=0.1M_f$ & 418.5 & $1.209\times10^{-4}$ & $-0.04\%$ & 0.0474 & 6.204 & 13.57 & 0.113 & 533 & $-2$\\
$\Delta r_*=0.025M_f$ & 418.5 & $1.209\times10^{-4}$ & $-0.01\%$ & 0.0474 & 6.206 & 13.57 & 0.113 & 533 & $-2$\\
barrier $\times2$ denser & 418.5 & $1.209\times10^{-4}$ & $+0.01\%$ & 0.0474 & 6.207 & 13.58 & 0.113 & 533 & $-2$\\
barrier $\times2$ coarser & 418.5 & $1.209\times10^{-4}$ & $-0.01\%$ & 0.0474 & 6.206 & 13.57 & 0.113 & 533 & $-2$\\
boundaries moved & 418.5 & $1.209\times10^{-4}$ & $-0.02\%$ & 0.0474 & 6.204 & 13.57 & 0.111 & 533 & $-2$\\
$\delta\omega=5\times10^{-5}/M_f$ & 418.5 & $1.209\times10^{-4}$ & $0.00\%$ & 0.0474 & 6.207 & 13.58 & 0.113 & 533 & $-2$\\
\end{tabular}
\end{ruledtabular}
\end{table*}

\paragraph*{Error budget.} The contributions to the absolute survivor amplitude are as follows. The tested numerical variations (Table~\ref{tab:conv}) change the survivor SNR of the GW150914-like source by at most $0.05\%$; the frozen eigenvalue and the finite extraction radius of the time-domain evolution are not part of this figure (Appendix~\ref{app:td}). The shape of the backreaction filter changes the SNR by $-11$ to $+23\%$, and its width ($\times0.5$--$2$) by a factor $0.37$--$2.7$. The background of the first reflection (remnant versus the hole before absorption) changes the SNR by $+49\%$. The packet energy enters as $\propto\sqrt{\epsilon_E E_{\rm RD}}$: $E_{\rm RD}$ falls from $1.25$ to $0.38\,\Msun$ for the $10M_f$ ramp, and $\epsilon_E=0.0229$ for the equal-$\psi_4$ normalisation. The net-energy normalisation is discussed in Sec.~\ref{sec:detect}. Finally, the SNR changes by factors of several between $\chi_f=0.4$ and $0.955$ (Table~\ref{tab:waterfall}), which the populations take into account source by source.

\section{Robustness to the backreaction filter}\label{app:robust}
\begin{figure*}
\includegraphics[width=\textwidth]{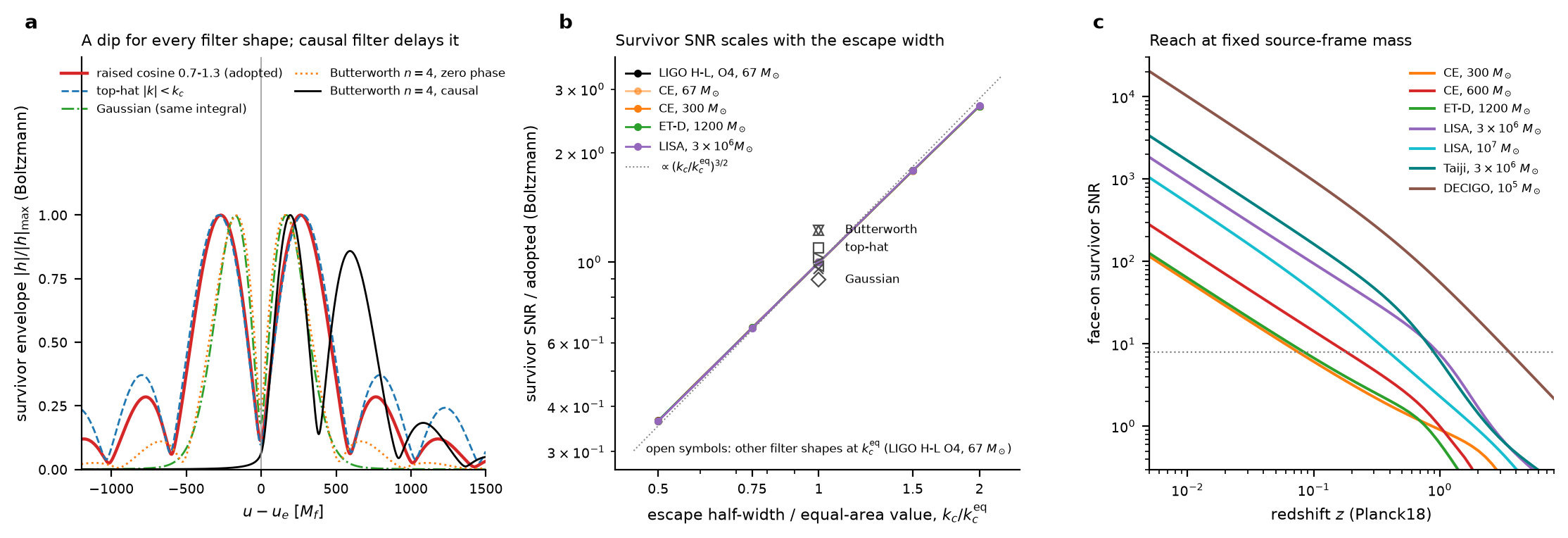}
\caption{Robustness of the survivor. (a) Envelope of the Boltzmann survivor for five shapes of the backreaction filter (normalised to their maxima) versus time from the geometric echo time $u_e$; every zero-phase filter gives a central dip at $u_e$, the causal (minimum-phase) Butterworth filter shifts it by $384M_f$ (its group delay is $348M_f$ at threshold and up to $521M_f$ in the lobes) and makes the humps unequal. (b) Survivor SNR relative to the adopted backreaction filter versus its width relative to the equal-area value, for five detector/mass cases (lines; dotted: $(k_c/k_c^{\rm eq})^{3/2}$), and for the other shapes of the backreaction filter at the nominal width (open symbols; LIGO O4 network, $67\,\Msun$). (c) Face-on survivor SNR versus redshift at fixed source-frame remnant mass (Planck18 cosmology; detector-frame mass $M_f=$$(1+z)M_f^{\rm src}$, surface location from the detector-frame mass); dotted: SNR 8.}
\label{fig:robust}
\end{figure*}

This appendix collects the tests of the backreaction filter summarised in Sec.~\ref{sec:backreaction}. Table~\ref{tab:robust} and Fig.~\ref{fig:robust} compare the adopted backreaction filter with several alternatives: raised cosines with different roll-off, a top hat, a Gaussian with the same $\int S\,dk$, a zero-phase fourth-order Butterworth response $|S|^2=1/[1+(k/k_c)^8]$, and the causal (minimum-phase) Butterworth response with the same $|S|$. The poles of the last one lie in the lower half $k$-plane, so the filtered signal cannot precede the unfiltered one. ``Causal'' refers here to the impulse response of the backreaction filter only. It does not describe causal transmission out of a trapped region and does not address the Guo--Mathur objection.

The optimal SNR depends only on $|S|$, so the zero-phase and minimum-phase Butterworth filters give identical SNRs (Table~\ref{tab:robust}) and differ only in the waveform. We keep the zero-phase raised cosine as the reference because its symmetry isolates the effect of the transmission zero. Zero-phase filters are acausal with respect to the arrival time of the unfiltered reflection (see the end of this appendix). Across the shapes the survivor peak changes by up to $57\%$ and the LIGO O4 SNR by $-11$ to $+23\%$. Changing the width, $k_c\to(0.5$--$2)\,k_c$, multiplies the survivor SNR by $0.37$--$2.7$ in every detector and mass case, i.e.\ $\rho\propto k_c^{1.44}$ (the in-band spectrum grows linearly in $|k|$, which alone would give $k_c^{3/2}$) [Fig.~\ref{fig:robust}(b)]. The prescription thus leaves the survivor amplitude uncertain by factors of order unity, while the narrow-band character of the signal and its central dip (for even filters) hold for every shape of the backreaction filter we tried.

\paragraph*{Phase of the reflectivity.} We also replace the constant phase by $\phi_0+c\,(k/k_c)^n$ inside the band, for $n=1,2,3$ and $c$ up to $2\pi$ (\code{R2C}~\cite{EchoRepo}; CE, $300\,\Msun$). The optimal SNR depends only on $|\tilde h|$ and does not change. The waveform, on the other hand, does change. A linear phase is just a time shift, $c/k_c$, which is $133M_f$ for $c=1$. A quadratic phase lowers the match to the constant-phase survivor to $0.990$, $0.961$ and $0.67$ for $c=0.5$, $1$ and $\pi$; the central dip survives for $c\le\pi/2$ (hump separation $547M_f$) but is filled in at $c=\pi$ (depth $1.00$). A cubic phase makes the two humps unequal, with a ratio of $0.49$ for $c=\pi$. The dip is thus a prediction for a constant or slowly varying phase, and a template search would need the in-band phase as a parameter. Conversely, reflectivity models whose phases differ within $\pm k_c$, or whose magnitude varies across the band, could in principle be told apart.

\paragraph*{Is a zero-phase response excluded?} With a zero-phase filter half of the survivor energy arrives before the geometric echo time $u_e$. This would be unphysical if the reflection took place at a sharply defined time, but in the model it does not. First, the retained waves are not localised. Inside the band the tortoise wavelength is $2\pi/|k|\ge2\pi/1.3k_c=644M_f$ ($1023M_f$ for the GW231123-like source), and it diverges at the threshold. This exceeds the depth of the whole cavity between the barrier and the surface, $450M_f$ ($530M_f$), so such a wave does not resolve where in the cavity it is turned back. Second, the layer that defines the backreaction filter is not thin. It is defined on the hole before absorption. There it reaches from the surface at $r_*=-428M_f$ ($-676M_f$ for the GW231123-like source) up to the equal-area sphere at $r_*=-9.8M_f$ ($-10.6M_f$), close to the inner foot of the angular-momentum barrier, so it covers almost the whole cavity of that hole. The first reflection is evaluated on the remnant, whose surface lies at $r_*=-450M_f$ ($-530M_f$), and the model does not say where the layer lies in that geometry. The packet enters the layer right after the merger, and a response generated anywhere inside it can reach infinity before $u_e$ without preceding its cause. The part of the zero-phase survivor that precedes the merger peak cannot come from any causal response; it holds $0.74\%$ ($3.2\%$) of the energy, measured here and below by $\int|\partial_uh|^2\,du$. For a response generated inside the layer the limit is later, but its value depends on where the layer is placed, so we give it only as an illustration. With the layer of the hole before absorption the earliest arrival is $-2r_*(r_{\rm eq})=20M_f$ ($21M_f$) after the merger peak, and $0.86\%$ ($3.2\%$) of the energy arrives earlier. If the thickness $L_*$ is instead laid off from the surface of the remnant, which is an additional prescription, the limit is $u_e-2L_*=63M_f$ for the GW150914-like source, with $1.2\%$ of the energy earlier, and lies before the merger peak for the GW231123-like one. These are necessary conditions. They show that this timing argument does not rule out the rest of the waveform, not that the rest is the output of a causal response. The zero-phase waveform is therefore not excluded on these grounds, but nothing in the model selects it. The zero-phase and minimum-phase Butterworth waveforms, which have the same $|S|$ and the same SNR, are the two cases we compare. They do not bound the possibilities: among causal responses with a given $|S|$ the minimum-phase one has the smallest group delay, and every other one adds the delay of an all-pass factor. As before, this concerns the response of the backreaction filter only, not the escape from a trapped region (\code{ZPP}~\cite{EchoRepo}).

\begin{table*}
\caption{Robustness of the survivor to the backreaction filter ($M_f=67.21\,\Msun$, $\epsilon_E=1$). Peak strain in units of the merger peak and face-on SNRs (LIGO Hanford--Livingston network at O4 sensitivity at $67\,\Msun$ and CE at $300\,\Msun$, both at 410~Mpc; LISA at $3\times10^6\,\Msun$, 3~Gpc); dip depth (envelope minimum over the lower hump), hump separation and node time relative to the geometric echo time $u_e$.}
\label{tab:robust}
\begin{ruledtabular}
\footnotesize\setlength{\tabcolsep}{3pt}
\begin{tabular}{lccccccc}
filter & peak & \multicolumn{3}{c}{SNR} & dip & $\Delta u_{\rm hump}$ & node$-u_e$\\
 & & O4 & CE & LISA & depth & [$M_f$] & [$M_f$]\\
\hline
raised cosine $[0.7,1.3]k_c$ (adopted) & $1.21\times10^{-4}$ & 0.047 & 6.2 & 13.6 & 0.11 & 533 & $-2$\\
raised cosine $[0.8,1.2]k_c$ & $1.21\times10^{-4}$ & 0.049 & 6.4 & 14.0 & 0.11 & 546 & $-2$\\
raised cosine $[0.6,1.4]k_c$ & $1.21\times10^{-4}$ & 0.046 & 6.1 & 13.2 & 0.12 & 516 & $-2$\\
raised cosine $[0.5,1.5]k_c$ & $1.21\times10^{-4}$ & 0.045 & 5.9 & 13.0 & 0.12 & 497 & $-2$\\
top hat $|k|<k_c$ & $1.21\times10^{-4}$ & 0.052 & 6.8 & 14.9 & 0.10 & 558 & $-2$\\
Gaussian (same $\int S\,dk$) & $1.30\times10^{-4}$ & 0.042 & 5.6 & 12.2 & 0.18 & 341 & $-2$\\
Butterworth $n=4$, zero phase & $1.70\times10^{-4}$ & 0.058 & 7.6 & 16.7 & 0.24 & 325 & $-3$\\
Butterworth $n=4$, causal & $1.90\times10^{-4}$ & 0.058 & 7.6 & 16.7 & 0.16 & 398 & $+384$\\
\hline
$k_c\to0.5\,k_c$ & $3.14\times10^{-5}$ & 0.017 & 2.3 & 4.94 & 0.06 & 1063 & $-2$\\
$k_c\to0.75\,k_c$ & $6.93\times10^{-5}$ & 0.031 & 4.1 & 8.94 & 0.09 & 709 & $-2$\\
$k_c\to1.5\,k_c$ & $2.63\times10^{-4}$ & 0.085 & 11 & 24.3 & 0.17 & 356 & $-2$\\
$k_c\to2\,k_c$ & $4.54\times10^{-4}$ & 0.13 & 17 & 36.7 & 0.22 & 268 & $-2$\\
\end{tabular}
\end{ruledtabular}
\end{table*}

\section{Other massive-black-hole catalogues}\label{app:b23}
The catalogues of Klein \emph{et al.} \cite{Klein2016} date from 2016. The same semi-analytic model \cite{Barausse2012} has since been updated, first with supernova feedback and with delays at kiloparsec separations \cite{Barausse2020}, and then compared with the gravitational-wave background reported by the pulsar-timing arrays \cite{Barausse2023}. The catalogues of Ref.~\cite{Barausse2023} are public. They contain the ten models for which that work gives LISA detection rates: seven without delays between the galaxy and the black-hole mergers, built on the earlier versions of the model (B12 \cite{Barausse2012}, K+16 \cite{Klein2016}) and on the updated one (B+20 \cite{Barausse2020}), with light (LS, popIII) or heavy (HS, Q3) seeds and with or without supernova feedback (SN, noSN); the model HS-nod-SN-high-accr with enhanced accretion, which reproduces the PTA amplitude; and the two models of Ref.~\cite{Klein2016} with delays. Each catalogue lists every merger of the model's merger trees with its comoving number density, at the finite resolution of the trees and, except for HS-nod-SN-high-accr, also extrapolated to infinite resolution; Ref.~\cite{Barausse2023} regards the extrapolation as an upper limit. Expected numbers are weighted sums over the entries, with the rate $4\pi c\,W D_M^2(z)$ per year for an entry of comoving density $W$ at comoving distance $D_M$. We take the remnant mass from the catalogue and compute the remnant spin from the spins and angles of the catalogue with the fit of Ref.~\cite{HBR2016}, as for all other sources; for the default model it agrees with the remnant spin listed in the catalogue to within $0.002$ for 95\% of the entries.

Table~\ref{tab:b23} lists the expected numbers of echoes for all models. Two points stand out. First, the models span a very wide range: with SNR $>5$ the DECIGO numbers run from $4\times10^{-3}$ to $2500$, and the LISA numbers from $2\times10^{-3}$ to $23$, and the Taiji numbers from $2\times10^{-3}$ to $190$. The heavy-seed models of Ref.~\cite{Barausse2020} have tens of thousands of mergers in four years, and DECIGO would detect the echoes of a thousand or more of them. The light-seed model with supernova feedback, which Ref.~\cite{Barausse2023} finds in marginal tension with the quasar luminosity function, gives only $4\times10^{-3}$--$0.013$. The model adopted in the main text lies between these. Second, the three models of Ref.~\cite{Klein2016} appear twice in this paper, as the Monte Carlo realisations distributed with GWToolbox (Table~\ref{tab:rates}) and as the weighted catalogues of the later release. The DECIGO numbers of the weighted catalogues are $0.81$ (popIII), $0.98$ (Q3-nodelays) and $1.04$ (Q3-delays) times those of the realisations at the median $\phi_0$.

\begin{table*}
\caption{Expected numbers of echoes for every model of the catalogue release of Ref.~\cite{Barausse2023} (re-formed Boltzmann surface, range over $\phi_0$, observing times of Table~\ref{tab:rates}). Model names as in Ref.~\cite{Barausse2023}; names without a label in brackets are models of Ref.~\cite{Barausse2020}; res.: ``finite'' is the resolution of the merger trees, ``extrap.'' the extrapolation to infinite resolution. $\dot N$: mergers per year; $\chi_f$: median remnant spin. The first row is the model used in the main text. Upper block: LISA and Taiji; lower block: TianQin and DECIGO.}
\label{tab:b23}
\begin{ruledtabular}
\scriptsize\setlength{\tabcolsep}{1.5pt}
\begin{tabular}{llcccccc}
model & res. & $\dot N$ [yr$^{-1}$] & $\chi_f$ & LISA, $>8$ & LISA, $>5$ & Taiji, $>8$ & Taiji, $>5$\\
\hline
HS-nod-SN-high-accr & finite & 2350 & 0.61 & $0.077$--$0.37$ & $0.37$--$1.8$ & $0.45$--$2.0$ & $2.1$--$9.4$\\
HS-nod-SN & finite & 4160 & 0.60 & $0.048$--$0.46$ & $0.35$--$2.6$ & $0.53$--$2.9$ & $2.6$--$13$\\
 & extrap. & 9200 & 0.61 & $0.28$--$2.1$ & $1.9$--$12$ & $2.1$--$11$ & $11$--$54$\\
HS-nod-noSN & finite & 4070 & 0.61 & $0.11$--$0.61$ & $0.49$--$2.6$ & $0.56$--$2.7$ & $2.6$--$13$\\
 & extrap. & 9840 & 0.61 & $0.76$--$4.9$ & $3.6$--$19$ & $3.3$--$17$ & $14$--$69$\\
LS-nod-SN & finite & 317 & 0.68 & $6\times10^{-4}$--$2\times10^{-3}$ & $2\times10^{-3}$--$4\times10^{-3}$ & $8\times10^{-4}$--$2\times10^{-3}$ & $2\times10^{-3}$--$5\times10^{-3}$\\
 & extrap. & 401 & 0.68 & $7\times10^{-4}$--$2\times10^{-3}$ & $2\times10^{-3}$--$5\times10^{-3}$ & $1\times10^{-3}$--$4\times10^{-3}$ & $3\times10^{-3}$--$7\times10^{-3}$\\
LS-nod-noSN & finite & 321 & 0.71 & $0.042$--$0.15$ & $0.098$--$0.31$ & $0.090$--$0.31$ & $0.20$--$0.57$\\
 & extrap. & 413 & 0.71 & $0.055$--$0.22$ & $0.14$--$0.45$ & $0.12$--$0.43$ & $0.27$--$0.82$\\
HS-nod (B12) & finite & 1560 & 0.80 & $0.090$--$2.2$ & $1.2$--$23$ & $2.2$--$34$ & $21$--$190$\\
 & extrap. & 1800 & 0.61 & $0.075$--$1.1$ & $0.74$--$10$ & $1.1$--$14$ & $10$--$83$\\
LS-nod (B12) & finite & 15600 & 0.77 & $0.026$--$0.10$ & $0.056$--$0.22$ & $0.051$--$0.21$ & $0.098$--$0.41$\\
 & extrap. & 16100 & 0.68 & $0.038$--$0.16$ & $0.088$--$0.39$ & $0.072$--$0.32$ & $0.15$--$0.71$\\
Q3-nod (K+16) & finite & 119 & 0.78 & $0.055$--$0.52$ & $0.26$--$2.3$ & $0.31$--$2.6$ & $1.4$--$8.6$\\
 & extrap. & 164 & 0.65 & $0.097$--$0.83$ & $0.46$--$3.3$ & $0.48$--$3.2$ & $1.8$--$9.8$\\
Q3-d (K+16) & finite & 8.07 & 0.92 & $0.12$--$0.92$ & $0.40$--$2.2$ & $0.46$--$2.4$ & $1.1$--$5.2$\\
 & extrap. & 18.4 & 0.93 & $0.14$--$1.2$ & $0.49$--$2.9$ & $0.49$--$2.9$ & $1.2$--$7.2$\\
popIII-d (K+16) & finite & 171 & 0.95 & $0.018$--$0.077$ & $0.032$--$0.15$ & $0.022$--$0.13$ & $0.039$--$0.28$\\
 & extrap. & 353 & 0.95 & $0.029$--$0.13$ & $0.055$--$0.31$ & $0.036$--$0.22$ & $0.070$--$0.54$\\
\end{tabular}
\end{ruledtabular}
\vspace{4pt}
\begin{ruledtabular}
\scriptsize\setlength{\tabcolsep}{1.5pt}
\begin{tabular}{llccc}
model & res. & TianQin, $>5$ & DECIGO, $>8$ & DECIGO, $>5$\\
\hline
HS-nod-SN-high-accr & finite & $0.045$--$0.21$ & $300$--$570$ & $540$--$900$\\
HS-nod-SN & finite & $0.028$--$0.18$ & $460$--$880$ & $860$--$1500$\\
 & extrap. & $0.12$--$0.71$ & $610$--$1300$ & $1200$--$2300$\\
HS-nod-noSN & finite & $0.050$--$0.29$ & $450$--$880$ & $860$--$1500$\\
 & extrap. & $0.22$--$1.5$ & $650$--$1400$ & $1300$--$2500$\\
LS-nod-SN & finite & $1\times10^{-4}$--$5\times10^{-4}$ & $2\times10^{-3}$--$6\times10^{-3}$ & $4\times10^{-3}$--$0.013$\\
 & extrap. & $2\times10^{-4}$--$7\times10^{-4}$ & $5\times10^{-3}$--$0.015$ & $9\times10^{-3}$--$0.028$\\
LS-nod-noSN & finite & $0.019$--$0.078$ & $12$--$26$ & $18$--$40$\\
 & extrap. & $0.022$--$0.098$ & $13$--$31$ & $22$--$48$\\
HS-nod (B12) & finite & $0.13$--$2.2$ & $360$--$1300$ & $720$--$2100$\\
 & extrap. & $0.11$--$0.90$ & $190$--$540$ & $380$--$950$\\
LS-nod (B12) & finite & $0.010$--$0.047$ & $2.4$--$14$ & $3.7$--$23$\\
 & extrap. & $0.014$--$0.071$ & $3.0$--$23$ & $4.6$--$41$\\
Q3-nod (K+16) & finite & $0.027$--$0.24$ & $43$--$100$ & $66$--$150$\\
 & extrap. & $0.042$--$0.31$ & $30$--$74$ & $49$--$110$\\
Q3-d (K+16) & finite & $0.072$--$0.49$ & $3.1$--$14$ & $5.3$--$20$\\
 & extrap. & $0.071$--$0.49$ & $3.2$--$19$ & $5.8$--$29$\\
popIII-d (K+16) & finite & $7\times10^{-3}$--$0.030$ & $0.23$--$8.0$ & $0.35$--$13$\\
 & extrap. & $0.011$--$0.046$ & $0.30$--$13$ & $0.45$--$24$\\
\end{tabular}
\end{ruledtabular}
\end{table*}

\section{Code and data}\label{app:code}
All codes, the run outputs and the copies of the population catalogues used here are available in the public repository \texttt{echo-backreaction}~\cite{EchoRepo}. A run is one complete calculation with \code{KEP} for one source and one background: runs \texttt{V0} and \texttt{GW231123} are the default evaluations of the GW150914-like and GW231123-like sources on the remnant, runs \texttt{I0} and \texttt{IG} the same sources evaluated on the hole before absorption (Sec.~\ref{sec:backreaction}), runs \texttt{T\_chi*} the generic spin templates, and runs \texttt{V1}--\texttt{V5}, \texttt{V7}, \texttt{V8\_lamthr}, \texttt{W\_*} and \texttt{modes} the convergence, time-window and multipole tests. Table~\ref{tab:code} lists the codes with the abbreviations used in the text. One script (\texttt{build\_v5.sh}) runs them in order and rebuilds every figure, table and number of the paper from the stored runs; the README of the repository gives the requirements, the commands to repeat the simulations and the sources of the external data.

\begin{table*}
\caption{Codes of the repository \texttt{echo-backreaction}~\cite{EchoRepo}, with the abbreviations used in the text. All are Python scripts in the folder \texttt{code}.}
\label{tab:code}
\begin{ruledtabular}
\scriptsize\setlength{\tabcolsep}{3pt}\renewcommand{\arraystretch}{0.95}
\begin{tabular}{@{}lll@{}}
name & file & what it computes (where it is used)\\
\hline
\multicolumn{3}{@{}l}{\emph{Pipeline and shared library}}\\
\code{KEP} & \texttt{kerr\_echo\_pipeline.py} & \parbox[t]{0.64\textwidth}{\raggedright time window and ingoing packet, $1{+}1$ Teukolsky evolution and leak scan, exact barrier data, first reflection and echo spectra; hole before absorption [Eq.~\eqref{eq:progenitor}] (Secs.~\ref{sec:model}, \ref{sec:backreaction}; Appendices~\ref{app:setup}, \ref{app:td}; Tables~\ref{tab:progenitor}, \ref{tab:conv})}\\
\code{EC} & \texttt{echo\_common.py} & \parbox[t]{0.64\textwidth}{\raggedright equal-area mapping, $k_c$, backreaction filters, hole before absorption, noise curves, Hanford--Livingston network response, SNRs (all sections)}\\
\code{SRM} & \texttt{settle\_remnant.py} & \parbox[t]{0.64\textwidth}{\raggedright applies the mapping to the hole before absorption to a finished run}\\
\code{BRD} & \texttt{backreaction\_detectability.py} & \parbox[t]{0.64\textwidth}{\raggedright survivor and its SNRs in all detectors against remnant mass (Sec.~\ref{sec:detect})}\\
\hline\multicolumn{3}{@{}l}{\emph{Survivor, dip and re-formed surface}}\\
\code{BDF} & \texttt{backreaction\_dip\_figures.py} & \parbox[t]{0.64\textwidth}{\raggedright Figs.~\ref{fig:walls}, \ref{fig:dip}, \ref{fig:det}; dip depth, hump separation, $x_1$ (Sec.~\ref{sec:dip})}\\
\code{SCF} & \texttt{schematic\_filter.py} & \parbox[t]{0.64\textwidth}{\raggedright Fig.~\ref{fig:schematic}}\\
\code{IEF} & \texttt{imr\_echo\_figure.py} & \parbox[t]{0.64\textwidth}{\raggedright Fig.~\ref{fig:imrecho}}\\
\code{MCF} & \texttt{metric\_choice\_figure.py} & \parbox[t]{0.64\textwidth}{\raggedright Fig.~\ref{fig:metric}}\\
\code{ARC} & \texttt{arrival\_reflection\_check.py} & \parbox[t]{0.64\textwidth}{\raggedright reflection on the hole found at arrival (Sec.~\ref{sec:backreaction}); uses \code{ABT} (\texttt{absorption\_trajectory.py})}\\
\code{RFC} & \texttt{reformed\_cavity.py} & \parbox[t]{0.64\textwidth}{\raggedright re-formed gain $G(M_f,\chi_f,\phi_0)$ and stability index (Sec.~\ref{sec:reformed}, Fig.~\ref{fig:reformed})}\\
\code{CMS} & \texttt{cavity\_modes.py} & \parbox[t]{0.64\textwidth}{\raggedright cavity (ergoregion) modes of the echo transfer function (Sec.~\ref{sec:reformed}); \code{MMQ} (\texttt{maggio\_modes\_quick.py}) compares them with Ref.~\cite{Maggio2019}}\\
\hline\multicolumn{3}{@{}l}{\emph{Detectability and populations}}\\
\code{SNG} & \texttt{snr\_gw231123.py} & \parbox[t]{0.64\textwidth}{\raggedright ground-detector SNRs of the two reference sources (Table~\ref{tab:survivor})}\\
\code{NNG} & \texttt{net\_normalisation\_gw231123.py} & \parbox[t]{0.64\textwidth}{\raggedright GW231123-like echo with the net-energy normalisation (Sec.~\ref{sec:detect})}\\
\code{RBC} & \texttt{robustness\_checks.py} & \parbox[t]{0.64\textwidth}{\raggedright Tables~\ref{tab:cosmo}, \ref{tab:robust}, Fig.~\ref{fig:robust}}\\
\code{WFP} & \texttt{waterfall\_plot.py} & \parbox[t]{0.64\textwidth}{\raggedright SNR-8 reach (Table~\ref{tab:waterfall})}\\
\code{POP} & \texttt{population\_rates.py} & \parbox[t]{0.64\textwidth}{\raggedright expected numbers at a common Boltzmann reflectivity phase, loudest expected signal, Fig.~\ref{fig:wfpop}, Tables~\ref{tab:rates}, \ref{tab:reformed}, \ref{tab:b23}; run for all scenarios by \code{RRP} (\texttt{run\_reformed\_populations.py})}\\
\code{GBM} & \texttt{gwtc\_bbh\_model.py} & \parbox[t]{0.64\textwidth}{\raggedright Default BBH population model of GWTC-5.0 and GWTC-4.0 from the released posterior samples}\\
\code{PB23} & \texttt{prepare\_barausse23.py} & \parbox[t]{0.64\textwidth}{\raggedright weighted catalogues of Ref.~\cite{Barausse2023}}\\
\code{FK16} & \texttt{fetch\_klein16.py} & \parbox[t]{0.64\textwidth}{\raggedright catalogues of Ref.~\cite{Klein2016} from the GWToolbox data \cite{GWToolboxData}}\\
\code{GWW} & \texttt{gwtc\_waterfall.py} & \parbox[t]{0.64\textwidth}{\raggedright confident LVK binaries (Fig.~\ref{fig:gwtc})}\\
\hline\multicolumn{3}{@{}l}{\emph{Multipoles, numerical checks and text}}\\
\code{RMD} & \texttt{run\_modes.py} & \parbox[t]{0.64\textwidth}{\raggedright single-mode pipeline runs (Sec.~\ref{sec:modes}); \code{MSV} (\texttt{modes\_survivor.py}) and \code{MFG} (\texttt{modes\_figure.py}) give Table~\ref{tab:modes} and Fig.~\ref{fig:modes}}\\
\code{WSE} & \texttt{window\_sensitivity.py} & \parbox[t]{0.64\textwidth}{\raggedright Table~\ref{tab:window}, Fig.~\ref{fig:threshold}; \code{WCO} (\texttt{window\_comparison.py}) Fig.~\ref{fig:window}}\\
\code{PSV} & \texttt{psd\_validation.py} & \parbox[t]{0.64\textwidth}{\raggedright noise curves (Appendix~\ref{app:psd})}\\
\code{MKF} & \texttt{make\_figures.py} & \parbox[t]{0.64\textwidth}{\raggedright Figs.~\ref{fig:packet}, \ref{fig:scan}}\\
\code{CVT} & \texttt{convergence\_table.py} & \parbox[t]{0.64\textwidth}{\raggedright Table~\ref{tab:conv}; \code{GTT} (\texttt{greybody\_td\_test.py}) the time-domain greybody factors (Appendix~\ref{app:td})}\\
\code{R2C} & \texttt{referee2\_checks.py} & \parbox[t]{0.64\textwidth}{\raggedright in-band phase of the reflectivity, absorption time (Appendix~\ref{app:robust})}\\
\code{TDS} & \texttt{checks/check\_td\_solver.py} & \parbox[t]{0.64\textwidth}{\raggedright free ringing of the evolution (Appendix~\ref{app:td}); \code{PLC} (\texttt{checks/pole\_check.py}) the frozen-eigenvalue pole}\\
\code{HLN} & \texttt{checks/hl\_network\_check.py} & \parbox[t]{0.64\textwidth}{\raggedright network response against \texttt{pycbc.detector} (Appendix~\ref{app:psd})}\\
\code{ZPP} & \texttt{checks/zero\_phase\_precursor.py} & \parbox[t]{0.64\textwidth}{\raggedright part of the zero-phase survivor before the echo time (Appendix~\ref{app:robust})}\\
\code{PFC} & \texttt{checks/progenitor\_fold\_check.py} & \parbox[t]{0.64\textwidth}{\raggedright roots of the mapping to the hole before absorption (Table~\ref{tab:progenitor}); \code{STC} (\texttt{checks/spin\_transition\_check.py}) the spin interpolation (Sec.~\ref{sec:detect})}\\
\code{FIL} & \texttt{fill\_v5.py} & \parbox[t]{0.64\textwidth}{\raggedright inserts every number and table of the paper from the run outputs; \code{SUM} (\texttt{summarize\_numbers.py}) prints a summary}\\
\end{tabular}
\end{ruledtabular}
\end{table*}

\begin{acknowledgments}
The author thanks Niayesh Afshordi for useful discussions. The author acknowledges the use of large language models (LLMs) for checking calculations, surveying the literature, coding, and drafting and revising the text. Independent checks of selected analytical relations were performed using SymPy (version 1.14.0) for symbolic algebra and mpmath (version 1.3.0) for arbitrary-precision numerical calculations. These checks included the transformation of the radial Teukolsky equation with fixed separation constant, Kerr geometric and first-law identities, and selected Fourier-transform and filter-response calculations. The idea, guidance, use of data, and implementation of both the linear Kerr analysis and the horizon backreaction are solely those of the author. The author is responsible for the content of the paper, including its interpretation and conclusions. This research has made use of data or software obtained from the Gravitational Wave Open Science Center (gwosc.org), a service of the LIGO Scientific Collaboration, the Virgo Collaboration, and KAGRA. This material is based upon work supported by NSF's LIGO Laboratory which is a major facility fully funded by the National Science Foundation, as well as the Science and Technology Facilities Council (STFC) of the United Kingdom, the Max-Planck-Society (MPS), and the State of Niedersachsen/Germany for support of the construction of Advanced LIGO and construction and operation of the GEO600 detector. Additional support for Advanced LIGO was provided by the Australian Research Council. Virgo is funded, through the European Gravitational Observatory (EGO), by the French Centre National de Recherche Scientifique (CNRS), the Italian Istituto Nazionale di Fisica Nucleare (INFN) and the Dutch Nikhef, with contributions by institutions from Belgium, Germany, Greece, Hungary, Ireland, Japan, Monaco, Poland, Portugal, Spain. KAGRA is supported by Ministry of Education, Culture, Sports, Science and Technology (MEXT), Japan Society for the Promotion of Science (JSPS) in Japan; National Research Foundation (NRF) and Ministry of Science and ICT (MSIT) in Korea; Academia Sinica (AS) and National Science and Technology Council (NSTC) in Taiwan.
\end{acknowledgments}

\bibliography{echo_article_v5}

\begin{thebibliography}{127}
\bibitem [{psd()}]{psds}%
  \BibitemOpen
  \href@noop {} {}\bibinfo {note} {Analytic and tabulated noise curves as
  distributed with LALSimulation~\cite{LALSuite} and PyCBC~\cite{PyCBC}; the
  source of each curve is listed in Appendix~\ref{app:psd}}\BibitemShut
  {NoStop}%
\end{thebibliography}
\end{document}